\documentclass[11pt]{article}
\newcommand{\blind}{0} %

\usepackage[margin=1in]{geometry}
\usepackage[utf8]{inputenc}
\usepackage[T1]{fontenc}

\usepackage{amsmath,amssymb,amsthm}
\usepackage{mathtools}
\usepackage{booktabs}
\usepackage{array}
\usepackage{graphicx}
\graphicspath{{./figs/}}
\usepackage[dvipsnames]{xcolor}
\usepackage{enumitem}
\usepackage{url}
\expandafter\def\expandafter\UrlBreaks\expandafter{\UrlBreaks\do\-\do-}

\usepackage[round]{natbib}

\definecolor{CardinalRed}{cmyk}{0,1,0.65,0.34}
\definecolor{spot}{rgb}{0.6,0,0}  %
\definecolor{OtherDraft}{HTML}{00408C}  %
\usepackage[colorlinks,linkcolor=CardinalRed,citecolor=CardinalRed,urlcolor=CardinalRed,filecolor=OtherDraft,hyperfootnotes=false]{hyperref} %
\usepackage{xurl}  %
\usepackage{cleveref}

\usepackage{titletoc}
\makeatletter
\renewcommand{\@pnumwidth}{2.2em}
\makeatother
\crefname{appendix}{Supplementary Material}{Supplementary Materials}
\Crefname{appendix}{Supplementary Material}{Supplementary Materials}
\crefname{subappendix}{Supplementary Material}{Supplementary Materials}
\Crefname{subappendix}{Supplementary Material}{Supplementary Materials}

\usepackage{setspace}

\usepackage{titlesec}
\titlelabel{\thetitle.\ }
\titleformat{\section}
  {\normalfont\fontsize{15}{15}\bfseries}{\thesection.}{0.5em}{}

\usepackage[labelfont=bf,textfont=bf,labelsep=period]{caption}

\usepackage[flushmargin]{footmisc}
\usepackage{titling}
\usepackage{framed}
\usepackage{tikz}
\usetikzlibrary{arrows.meta,positioning,calc,fit,backgrounds}

\definecolor{vfink}{HTML}{a8003b}    %
\definecolor{vfslate}{HTML}{2b3440}  %
\definecolor{vfband}{HTML}{f4f1ec}   %
\definecolor{vfband2}{HTML}{eef2f4}  %

\theoremstyle{definition}

\theoremstyle{plain}

\newcommand{\floatnote}[2][0.92\textwidth]{%
  \par\vspace{0.4em}%
  {\centering
   \begin{minipage}{#1}%
     \footnotesize\setstretch{1.1}%
     \noindent\textit{Notes:}\ #2\par
   \end{minipage}\par}}

\title{\LARGE\bfseries Replication Certification at Publication:\\ Checkable by Anyone, Anywhere, in Seconds, Forever\ifnum\blind=0\thanks{We thank Danny Ebanks, Stefano Iacus, Chris Lucas, Katalina Toth, Ana Trisovic, Salil Vadhan, and Eddie Yang for helpful comments and suggestions. The authors used Claude Code and Codex as research, coding, and writing assistants. All interpretations, conclusions, and any errors remain the sole responsibility of the authors.}\fi}
\ifnum\blind=1
  \author{}
\else
  \author{Gary King\thanks{Institute for Quantitative Social Science, Harvard
    University; GaryKing.org, \textcolor{spot}{King@Harvard.edu}.}
    \and Yiqing Xu\thanks{Department of Political
    Science, Stanford University; yiqingxu.org,
    \textcolor{spot}{yiqingxu@stanford.edu}.}
    \and Leo Y.\ Yang\thanks{Stanford Center on China's Economy and Institutions, Stanford University, \textcolor{spot}{leoyang@stanford.edu}; currently affiliated with the Department of Accountancy, Economics and Finance, Hong Kong Baptist University;
    LeoYang.org, \textcolor{spot}{leoyang@hkbu.edu.hk}.}}
\fi
\date{\today}
\predate{\begin{center}\large}
\postdate{\par\end{center}}
\usepackage{etoolbox}
\titleformat{\subparagraph}[hang]{\normalfont\normalsize\bfseries}{}{0pt}{}
\titlespacing*{\subparagraph}{0pt}{2.5ex plus .5ex minus .2ex}{0.3ex}

\hypersetup{pdftitle={Replication Certification at Publication: Checkable by Anyone, Anywhere, in Seconds, Forever},pdfauthor={Gary King, Yiqing Xu, Leo Y. Yang}}
\begin{document}
\maketitle

\vfill
\begin{abstract}%
\noindent 
Authors increasingly deposit replication datasets---the data and code behind publications---in archives, and journals increasingly require them to, under progressively onerous rules. But a deposit alone proves little, so authors keep paying: months for the journal's reanalyses, years answering readers' questions, and rancor as failed replications become accusations because no record establishes what actually ran. We introduce a replication certificate: a cryptographically signed record of the deposited files and logged output. It replaces the journal's reanalyses with evidence anyone can check in seconds, which no one, including the archive, can backdate, and no one can fake without risking detection. Certification asks little of anyone and benefits everyone: The author deposits, then reruns the analysis once under a random seed the archive computes from the deposit's hash and a blockchain block; the archive signs, timestamps (in that blockchain), and stores the record---confidential data included---without rerunning anything; the journal merely checks a website for the certificate; and anyone can check or replay any step decades later.
\end{abstract}
\vfill

\clearpage
\baselineskip=1.57\baselineskip
\section{Introduction}
\label{sm:intro}

Ensuring that evidence offered in a scholarly publication stands independently of its author is a goal that echoes through the history of science, back to the 1660 motto of the Royal Society, \emph{Nullius in verba} (``take nobody's word for it'') and even to Heraclitus and Kālāma Sutta more than two thousand years earlier.  We can also see this goal in the now widely accepted ``\emph{replication standard} [which] holds that sufficient information exists with which to understand, evaluate, and build upon a prior work if a third party could replicate the results without any additional information from the author,'' \citep[][p.444]{king1995replication}, and in many efforts to meet the standard.\footnote{We define ``replication'' as most users and data repositories do: running the same code on the same data, producing the same result. Some call this ``reproduction'' and reserve ``replication'' for a study redone with new data \citep{nasem2019reproducibility}; we use the words interchangeably.}

Scholarly publication outlets over four decades have adopted increasingly strict policies designed to convince authors to deposit \emph{replication datasets}, the data and code behind a published result, in professional \emph{archives}, the data repositories such as Dataverse, that store, version, and cite them, and to try to ensure each deposit provides meaningful evidence that results can be reproduced (\Cref{sm:dataverse} describes everything we require of them).  Unfortunately, each of the policy changes has merely moved the burden of verifying replicability from one actor to another, while often increasing costs in money, staffing, publication delays, time investment by authors and future researchers, and misunderstandings between them, while still never quite succeeding. \Cref{sm:problem} explains.

Instead of even more onerous and expensive policies, we propose a technological solution to accomplish the same goals more effectively and at lower cost. By marshalling well-tested cryptographic methods, we show how to create a special replication dataset \emph{certificate} that provides the evidence that the analysis ran on exactly its files and produced exactly the results it records, with its random draws made after the files were fixed. The journal can then substitute its expensive and time-consuming in-house replication runs with a quick check that the certificate exists before final acceptance, and a later reader can check the record in seconds instead of re-running the analysis. A failed replication then has a certified record to be argued from rather than a reputation to be argued about. See \Cref{sm:goal}. 

The rest of the paper states what certification does not cover (\Cref{sm:m-uncovered}) and reviews what our proposal borrows from prior work and what that work leaves open (\Cref{sm:m-related}). The Supplementary Appendix provides all technical details, including what evading certification costs an attacker (\Cref{sm:calculus}) and our contributions (\Cref{sm:related}).

\section{Problems with Replication Policies}\label{sm:problem}

No policy tried thus far lets anyone verify a deposited replication dataset without re-running it, so each future researcher has to verify the same deposit from scratch. We describe here the levels of enforcement journals have escalated through; the evidence indicating how these policies have failed; and the costs imposed by the policies.\footnote{The norm that authors deposit the code and data behind a published result is long-established in the social sciences \citep{dewald1986replication, king1995replication}, and professional archives now supply the storage, versioning, and persistent identifiers this protocol builds on \citep{king2007introduction}. The Transparency and Openness Promotion guidelines extend the same levels across the sciences, with verified re-execution as their top level \citep{nosek2015promoting}.}

\subsection{Policy Enforcement Levels}
Level 1 is the original and least stringent proposal, where journals require authors to ``indicate in their first footnote in which public archive they will deposit the information necessary to replicate their numerical results.'' Editors would be ``responsible only for the existence of the footnote, not for confirming that the data set has been submitted nor for checking whether the results actually can be replicated.'' Any verification ``can and should be left to future researchers'' \citep{king1995replication}.

Level 2 requires authors to go a step further and make data available on request, as in \emph{Science}'s rule from 2011 or by encouraging authors to make data available in some unspecified way. But asked for the materials behind 204 papers published under it, only 44\% of authors supplied them, and the results of only 26\% of the 204 could be reproduced \citep{stodden2018empirical}. 

Level 3 makes replication dataset deposit a condition of publication, as now required by many leading outlets \citep{brodeur2026reproducibility, ferguson2023survey, miske2026reproducibility}. In economics, the requirement reached the \emph{American Economic Review} in 2005 and the \emph{Quarterly Journal of Economics} in 2016 \citep{christensen2018transparency}. In political science, the editors of twenty-seven journals signed the joint statement of the Data Access and Research Transparency initiative, committing to require that cited data be available in a trusted repository \citep{lupia2014openness,dart2015jets}. Among quantitative political-science articles, a working link to the data and code was rare before about 2008, with the share growing approximately exponentially since \citep[][fig.1]{rainey2025data}.

Finally, Level 4 is re-execution of the code by the journal, originally as an option and then a requirement. From July 2009, \emph{Biostatistics} let authors ask it to run their code and marked the article \emph{R} if it reproduced; in two years, five of 125 articles earned the mark \citep{peng2011reproducible}. Journals now increasingly require human-verified re-execution by the journal's data editor \citep{vilhuber2020reproducibility, nust2021codecheck}. The \emph{American Journal of Political Science} has worked with the Odum Institute at the University of North Carolina since 2015 to verify every accepted quantitative article, postponing publication by seventy-five days on average \citep{ajps2018verification}. New proposals for even more stringent practices also regularly appear \citep[e.g.][]{reiss2026improving}.

\subsection{Policy Failures}\label{pfail}
The literature suggests three conclusions. First, \emph{archival deposit alone does not show that the code runs}. Of more than nine thousand R scripts deposited in Dataverse, 74\% did not run to completion without error \citep{trisovic2022large}. Of 15,817 Python notebooks from biomedical publications with declared dependencies, 1,203 of those ran through without error \citep{samuel2024computational}, in line with earlier mass executions of notebooks \citep{pimentel2019large}. When two teams each tried to redo one analysis from each of 18 microarray papers, ten could not be reproduced \citep{ioannidis2009repeatability}.  

Second, \emph{running the code does not guarantee that the materials reproduce the article's results}. For all available replication packages for articles published 2000--2026 in major economics journals, only 11.2\% of 4,452 could be fully reproduced \citep{schwartz2026llm}. Working only from deposited files, one audit reproduced the key result of just 22 of 67 published economics papers \citep{chang2022economics}. At \emph{Cognition}, mandated deposit raised the availability of usable data sharply, while 13 of 35 audited articles still had at least one value that could not be reproduced even with the authors' help \citep{hardwicke2018data}. In the largest journal-run audit, 68\% of attempts succeeded in replicating results \citep{fisar2024reproducibility}.  

And finally, \emph{journal verification catches more failures without catching them all}. Of the twenty-four papers given in-house replication review at the \emph{Quarterly Journal of Political Science}, 14 of 18 produced results that differed from those in the manuscript \citep{eubank2016lessons}. Across 254 papers from the three leading political-science journals whose materials could be obtained, 61 did not reproduce every coefficient exactly and 20 did not reproduce it even approximately \citep{xu2026scaling}.

\subsection{Policy Costs}\label{pcosts}
Finally, journal replication policies impose three categories of costs.

The first cost is \emph{time}, which comes in two parts, delayed acceptance and endless post-publication questions.  In the former, median verification times at the American Economic Association's journals have run three months per round, and some manuscripts need several rounds \citep{vilhuber2024report}; the average manuscript the \emph{AER} published in 2025 spent thirteen weeks in the data editor's hands, up from nine in 2022 \citep{luttmer2026report}. In the latter, authors assemble replication datasets to a data editor's standard but still have to answer questions about them for years. Having access to a deposited replication dataset is far better than no deposit at all, but still considerable work: in one mass-reproduction project, teams averaged thirteen days per paper, and about one article in twelve required more than a hundred days \citep{brodeur2024mass}. None of that work is recorded where the next reader can use it and so every researcher starts from scratch.

The second cost is \emph{money}, paid by journals (and by editors in their time) which hire and manage replicators, in addition to providing cutting edge computational resources used by researchers.  The American Economic Association's Data Editor pays for all student employment and pre-doctoral staff, while Cornell University's Economics Department provides tuition waivers and computing \citep{vilhuber2025impacts}. The Cornell laboratory that replicates received, in one year, 529 requests covering 415 manuscripts. About 15-20 undergraduate research assistants work there per week, supervised by the three staff \citep{vilhuber2022teaching}. At the \emph{American Journal of Political Science}, verification by the Odum Institute took an average of eight person-hours per manuscript, and authors resubmitted materials 1.7 times on average, a ``not trivial'' financial cost \citep{jacoby2017should}. \emph{Management Science} charges \$79 on every original submission since 2025, to fund its reproducibility infrastructure and external verification \citep{informs2025fee}.

The final cost is \emph{rancor}. When a replication fails, the absence of a shared record of what was actually run often turns technical questions into personal ones. The dispute over Reinhart and Rogoff's result turned on what had actually been run---which spreadsheet, which exclusions, which weighting---and was settled only when the authors found and shared the original file \citep{herndon2014does}. The dispute over LaCour and Green's article turned on whether the deposited data had been collected at all \citep{broockman2015irregularities}. Daniel Kahneman, acting as a senior statesman, has gone so far as to propose etiquette rules for how readers attempting replications should approach authors \citep{kahneman2014etiquette}. 

\section{A Technological Solution: The Certificate}\label{sm:goal}

\begin{figure}[p]
\caption{Certification at a Glance}\label{fig:overview}
\centering
\definecolor{ovblue}{HTML}{2f5d9e}\definecolor{ovamber}{HTML}{9a5b00}\definecolor{ovteal}{HTML}{2a6b5f}\definecolor{ovred}{HTML}{9e3b3b}\definecolor{ovgreen}{HTML}{2e7d32}%
\includegraphics[width=\textwidth,height=\dimexpr\textheight-7\baselineskip\relax,keepaspectratio]{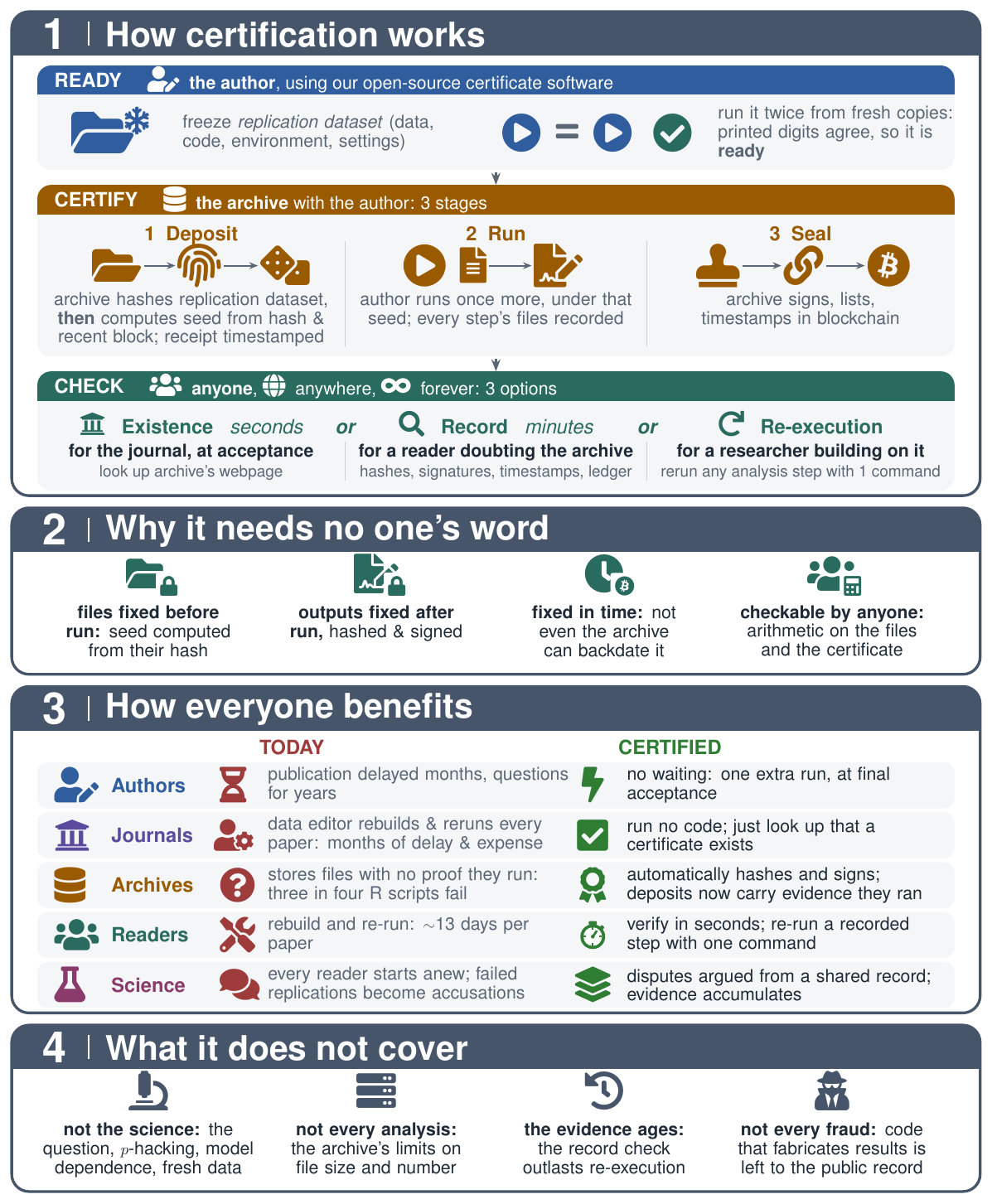}

\floatnote{Panel~1: \textcolor{ovblue}{\textsc{ready}}, the author prepares the replication files until two runs agree. \textcolor{ovamber}{\textsc{certify}}, the archive computes a seed from the deposit's hash and a blockchain block, the author runs the analysis under it, and the archive signs and timestamps the record; draws fall after the seed is computed and before the certificate is issued. \textcolor{ovteal}{\textsc{check}}, three options, depending on need.  Panel~2: why the record is evidence. Panel~3: what each party pays \textcolor{ovred}{\textsc{today}} and gains once \textcolor{ovgreen}{\textsc{certified}}. Panel~4: what certification does not cover.}
\end{figure}

\Cref{fig:overview} gives an overview of the certification protocol. The first three parts of this section follow the figure's first three panels.

\subsection{How It Works}\label{sm:how}

Certification has three parts corresponding to the three bands within Panel~1 of \Cref{fig:overview}. First, the author makes the replication dataset \textsc{ready} by locking down all sources of variability (due to the code, computing environment, or settings), so that it gives the same results if run repeatedly. Second, the author and archive work together to \textsc{certify} the replication dataset. And finally, anyone, at any point in the future, can \textsc{check} the results in seconds without having to rerun the analysis, and each check catches frauds that pass the one before it.

\subsubsection{Ready: Author Prepares the Replication Dataset}\label{sm:ready}

We now formalize the current practice of authors preparing replication datasets for deposit (and thus also prior to our certification protocol). Conceptually, getting ``ready'' in this way involves the author using our certificate software to ensure their replication dataset gives the same results when run at least twice on their machine, which is a minimum definition of ``replicability''. Authors are advised to conduct more than two runs if results may be fragile, since two runs can agree by luck. 

To get ready formally, the software first collects (a) every file in the replication dataset with its hash and the results of a \emph{dependency audit}, a reading of the scripts for the files they open and the seeds they set; (b) the settings to freeze, such as the time zone and the number of processor cores; and (c) the software environment itself (deposited as add-on libraries installed inside the replication dataset, with a ``lockfile'' that lists their exact versions, or, where software licenses allow, as a ``container image'', a packaged copy of the operating system and programs).  From these, the software creates a \emph{closed copy}, which includes the replication dataset plus a few small files for record keeping.  This closed copy enables all the author's analyses to be run with one top-level command, keeps track of each step in the process, and, where possible, has only one input for a random seed.

With the closed copy, the author conducts at least two runs, which we call \emph{closure runs}, each from a fresh temporary copy and under a seed of the author's choosing. The certificate software compares every file the two runs writes and records how closely they agree, as the \emph{determinism tier}: the standard later re-runs must meet. The author cannot declare a tier, only earn one: careful code earns a stricter tier, and unstable code a looser one that the certificate displays. The dataset is \emph{ready} when the two runs agree on every number to a fixed precision, which we recommend setting to at least what articles ordinarily print (such as 2-4 digits), though the goal is byte-identical agreement; the software writes a report, deposited with the replication dataset, that names any file that differed. See \Cref{sm:close} for details.

\subsubsection{Certify: Author Deposits, Receives Seed, Runs Once; Archive Seals}\label{sm:certify}

Certification is designed to eliminate the possibility of \emph{backdating} by anyone including even the archive (in order to surreptitiously change the record after the fact), \emph{seed shopping} (where the author cherry-picks a supposedly random seed to make their results look better), \emph{substitution} (replacing an intermediate file, such as the cleaned data, after the run while claiming the certificate still covers it), and \emph{ambiguity} (in what results the run produced or how to conduct a new one).  Certification occurs after conditional acceptance by the journal; an author may choose to certify earlier, or to certify a revised replication dataset after publication and link it to the earlier one.    

To take a dataset that is ``ready'' and certify it to receive these benefits requires three stages (corresponding to the Certify row of Panel 1 in \Cref{fig:overview}).

First, the author \emph{deposits} the closed copy in the archive. The archive then computes a new random seed from three inputs: a hash of the uploaded files, a cryptographic timestamp, and the private half of the archive's \emph{seed key}, a key pair it uses only for this purpose, so that no one, the author included, could know the seed before it is computed. Version~1 uses Bitcoin's blockchain for the timestamp for both bounds (other options include any public record that no party controls and no one can backdate, with a public source of values no one can predict in advance). The archive's receipt, its signed and timestamped statement that these files arrived, records the hash, the block, the seed, and a short proof with which anyone holding the seed key's public half can verify that the seed was computed from those inputs.

Second, the author \emph{runs} the closed copy once under the seed issued by the archive (producing the ``certified run''). For each step in the run, the certificate software records the ``pre-state'' (every file and its hash present before the step starts), any files it writes, and any output it produces. Because each step begins exactly where the previous one ended, every pre-state already includes the hashes of all files written by earlier steps, so the record of each step is tied to everything that came before it. The software then combines the records of all steps, in order, into a single hash so that changing any intermediate file, or dropping, adding, or reordering a step, would change this hash. The certificate therefore vouches not only for the final numbers but for the entire path from the closed copy to the published results. The published numbers must come from this run. The software produces a ``proposed certificate''.

Finally, the archive \emph{seals} the proposed certificate submitted by the author. To do this, the archive checks the certificate, refuses on any failure, signs the ``witness'' (its statement that it sealed this certificate, using its signing key), timestamps the certificate in the blockchain, and enters the witness in its ledger (a public list of every document it issues including this one, kept so that any later edit is visible publicly). The archive stores the certificate and records it in a metadata field linked to the published dataset.  See \Cref{sm:produce} for details of certification.

\subsubsection{Check: Anyone Can Easily Verify the Record}\label{sm:checking}

A key advantage of a certificate is that anyone can check it without taking anyone else's word for what was run or when. Three types of checks are available, depending on need; after describing them, we show that each catches frauds that pass the one before it; and finally where a certificate sits among higher levels of assurance. \Cref{sm:check} provides details.

\paragraph{Three Types of Checks.}
First, an \emph{existence check} enables a journal to confirm that the replication dataset, the article's citation of it, and the certificate exist (and that the certificate carries no exception mark, the archive's flag indicating that the certificate fails to meet one of the protocol's standards). This requires only a fast lookup on the archive's webpage; no reanalysis is necessary.  An existence check is sufficient for the journal because a sealed certificate enables anyone, without help from the journal, to verify both ``record consistency'' (that the stored files are those the run started from and wrote, and that neither the files nor the record has changed since) and ``replication'' (that rerunning the recorded steps from their stored starting files produces outputs that agree with the certified ones within the certificate's determinism tier).

Second, a \emph{record check} is designed for someone who prefers not to trust the archive's webpage. Here, our open-source certificate software recomputes every hash from the files and checks the archive's signatures, the timestamps, and the ledger, validating record consistency. The record check runs no analysis, typically takes about a minute, needs no outside service when run offline, and, as long as the cryptography it uses stays infeasible to break, works the same way decades from now.

Finally, a \emph{re-execution check} is designed for a future researcher who wishes to build on the research or to test whether the record was faked. Conveniently, our protocol enables a researcher to rerun any single step, without running those that came before or after, whenever all of that step's starting files are stored; the certificate reports which steps qualify.  The researcher can also rerun the entire analysis and see whether their results agree with the certified ones within the determinism tier. A rerun takes as long as the original, uses the deposited libraries or image, and is judged only when the researcher's software environment matches the recorded one; a rerun that agrees gives the researcher a tested starting point for new work.

\paragraph{Each Check Catches Frauds That Evade the One Before.}\phantomsection\label{sm:m-limits}
A certificate is evidence of only record consistency and replication. By all accounts failure to replicate is far more often due to honest error than fraud, but we aim for certification to hold up even against an author acting in bad faith (in security terms, a ``threat actor''). We also consider cases where an author and archive collude to produce a fraudulent result (which is presumably even rarer because we know of no such case). We now summarize how each type of certification check catches frauds that pass the one before it, and the permanent public record reaches what passes all three. \Cref{sm:m-uncovered} shows what certification does not cover.

First is \emph{passing the existence check, caught by the record check.} The existence check relies on the archive's webpage. If it is faked, the record check catches the fraud and any file changed after the certified run, such as edited output tables, swapped data files, or secretly rerun scripts.

Second is \emph{passing the record check, caught by re-execution.} To commit fraud while passing the record check, an author must falsify the record itself: alter the analysis while it runs, such as through hardware control of the computer or a value changed in memory; edit the record before the archive seals it; or hack the certificate software so it still reports the genuine copy's hash. A colluding archive could also sign a falsified record and add it to its ledger. In each case the stored starting files remain untouched, so a reader who reruns a falsified step with a genuine copy of the software, in an environment matching the recorded one, will obtain the honest result and discover the fraud. Each additional researcher who reruns the analysis (which is likely to happen more often with more important scholarly claims) increases the probability of detection.

Last is \emph{passing the re-execution check, caught by the evidence trail.}  Permanent public access protects science from its worst actors, whose frauds reproduce on every rerun or lie outside the run, and so pass all three checks. The certified replication dataset, its certificate, and the archive's ledger stay publicly available, and no one can backdate them. Anyone may read code that fabricates results, read the file that sets a seed missed by the dependency audit, or examine the data (like \citet{broockman2015irregularities}). Anyone can also view the number of deposits to check for seed shopping.

\paragraph{Levels of Assurance.}
Every certificate gives the same assurance on an ordinary computer; later versions of the protocol could allow or require stronger evidence if needed: records of re-executions by others, published on the archive's ledger; an attestation from protected processor hardware, for an author whose analysis runs on it; and, should it become computationally feasible, a cryptographic proof that the computation ran. A journal could then require any of these, but a certificate does not depend on them.

\subsection{Why it Needs No One's Word}\label{sm:claims}

Under the status quo, a replication dataset is useful only if we trust the author's word that re-executing it would produce the published results. The reason re-execution alone is insufficient is that failure to replicate is ambiguous; it may imply any or all of (a) lack of closure with respect to randomness, (b) inaccurate or incomplete documentation, or (c) published results not generated by the data and methods as claimed in the article.

In contrast, an article with a certified replication dataset carries its own evidence, which anyone can check, anywhere, in seconds, forever, without taking anyone's word for what was run or when. ``Evidence'' here means a record anyone can check, which no one, including the archive, can backdate, and no one can fake without risking detection. This is summarized in four features (corresponding to Panel~2 of \Cref{fig:overview}), each of which answers one of the four threats named in \Cref{sm:certify} (seed shopping, substitution, backdating, and ambiguity), even against an author acting in bad faith.

First, \emph{replication dataset files are fixed before the certified run.} The archive hashes the deposit and computes the seed from that hash, so the seeded draws cannot precede the files; this rests on well-tested cryptography, not on trust. Seed shopping therefore needs a new deposit for each seed, and the archive's ledger publicly records every deposit under the depositor's account, so the seeds drawn from each account can be counted; certification is not anonymous.  

Second, \emph{the results are fixed after the certified run.} The certificate records the hash of every file, each step the code started from and produced, and every logged estimate, and the ``witness'' names the certificate's own hash. Thus, the check catches a later change to any file or the record.

Third, \emph{the certified run is fixed in time,} meaning that a researcher trying to correct earlier mistakes will fail or leave a visible trail. The receipt and the certificate are each timestamped, so the certified run's random draws are bounded between two block times.  The later bound is the time of the block that timestamps the certificate. The earlier bound is the time of the block the receipt names, since the seed is computed from that block's hash. Not even the archive can backdate either timestamp, and that earliest date is what no signature, log, or timestamp could supply on its own.

Finally, \emph{the result is checkable by anyone.} No special permission is required, which anyone can easily recompute using only arithmetic starting from the replication dataset and the certificate. Even the seed's derivation can be verified. The archive's word is taken only for the facts only it holds (who deposited and whether a revised deposit and an earlier deposit it names as its predecessor have the same uploader), and for not using its seed key to help an author. That the run happened as recorded rests on the open-source certificate software, which any reader can test by re-running recorded steps.

\subsection{How Everyone Benefits}\label{sm:parties}\label{sm:costs}

Certification lowers the cost of verification by changing who runs the analysis and how often: the author runs it once more, on their machine that already holds their software environment, and everyone afterward checks the record instead of rebuilding and rerunning it from scratch. Panel~3 of \Cref{fig:overview} shows the gains party by party. Because several gains reach more than one party, we take them here by the problem in \Cref{sm:problem} that each one solves, with what certification costs in the same terms.

First, \emph{policy failures}.  Current replication datasets, even when required by journal policies, carry no evidence that their code runs or reproduces the article's results, and journal verification catches only some issues. In contrast, a certified replication dataset carries evidence that its analysis ran on exactly its files and produced exactly the results it records, which the copyeditor or author checks during copy editing to connect the certificate to the publication. Because checking time does not grow with the length of the computation, the evidence extends to analyses too long for a data editor to re-execute, which happen because published articles often represent the computational cutting edge. The closure runs, of the kind authors already make before submission anyway, now leave a record: their report of how closely the two runs agreed, with the estimates each produced, is deposited with the replication dataset and fixed before the archive's seed exists.

Second, \emph{time}.  Data editors presently must rebuild the author's software environment, rerun their analyses, and ask the author for whatever is missing. This takes months per round before acceptance, each later researcher rebuilds it all from scratch, and authors answer questions about their deposits for years. Certification replaces publication delays and later questions with one more run by the author after conditional acceptance, on their machine that already holds the hardware, software, libraries, and data, so that the run comes down to nearly one command. A future reader or researcher attempting a replication confirms that the record is intact with the record check, which is fast and does not require recreating the author's software environment. The record check answers the questions that would otherwise go to the author. Confirming that a step still produces its recorded outputs takes that step's own running time. The author pays by making one extra run.

Third, \emph{money}. Journals today pay for the data editors, staff, and computing that re-execution requires. For a certified paper, journal staff only need an existence check (webpage lookup) at final acceptance with no technical choices or special expertise required. Nor does a journal need a license for proprietary software such as Stata, since the author's own run is the one certified, and readers see its logged estimates without one. Archives pay to operate the software that performs the deposit and seal stages, and for storage of the record, which includes the files the run produces. The archive's institutional work of custody, versions, backups, metadata, terms of use, restricted access, and indefinite storage continues, as before.

Finally, \emph{rancor}. Today a failed replication becomes a dispute about what the author did, because no shared record says what was run. Certification gives the scientific community a public record of precisely what was run for every certified paper, so a dispute over a failed replication turns on what the record shows, which both sides can check, instead of on what the author says was run. The publication and its certified replication dataset can now stand independently of the person who created them. Disputes will continue, but primarily about future research, as should be the case.

\subsection{Extensions to Confidential Data}\label{sm:m-confidential}

We now extend our certification protocol to confidential data, requiring access levels that are not fully public, such as requiring click-through licenses, guestbook signatures, signed contracts, committee approvals, air gapped facilities, and more. We distinguish three types by who is permitted to hold the files: the archive, only a secure facility, or only the author under a license. We assume the archive may publish the data's metadata in any of these categories (otherwise no one would know the data existed in the first place). The guarantees weaken as the files move further from the archive's custody. See \Cref{sm:confidential} for details.

First, \emph{restricted data the archive is allowed to hold} require only trivial changes to the protocol. The certificate includes no data values, and our summaries of the data are hashes of the data mixed with a secret random value, so that no one can confirm a guess about a file's contents; cryptographers call these \emph{hiding commitments}. Before release, the author verifies that the certificate does not include private information. A later reader must then satisfy the archive's access rules for this dataset before checking the files.

Second, \emph{data only accessible inside a secure facility}, such as a federal research data center, is not reachable by the archive, but its notary work needs only the hashes, so the facility takes over the work that uses the files. The author makes the closure runs and the certified run inside the facility, and only hiding commitments, the seed, and a certificate whose estimates pass the facility's review for disclosure leave it. The archive computes the seed from the hiding commitments; facility staff run the record check on the files and sign the result, which the archive verifies before sealing; and the facility keeps the exact files and software environment for a stated period. The certificate is then made public. Anyone can check the signatures, dates, ledger, and deposited code, and a reader admitted to the facility following the same rules as the author did can check the data or rerun the analysis. The cost here is trust in the facility's signed check and how long they keep the data, in addition to cumbersome rules that prohibit installing outside software, accessing familiar materials, or even seeing data or intermediate results.

Finally, \emph{data licensed under terms that forbid archival deposit}, such as from commercial vendors, are analyzed on the author's computer as usual and enter the deposit only as hiding commitments and metadata naming the source and version (such as restricted data used in recent papers in the American Economic Association's journals; \citealt{vilhuber2024report}). Everything but the data is certified in full, and the seed still follows the data's hashes, but only the author checks the data before sealing. A later reader can check them only by obtaining the same license and version of the data, which vendor is trusted to maintain; the archive then supplies the file of secret values needed to check the commitments.

\section{What Certification Does Not Cover}\label{sm:m-uncovered}

What lies outside the certificate's evidence of record consistency and replication forms four groups, which we describe as follows. For details, see \Cref{sm:guarantees}.

First, \emph{a certificate is evidence about a computation, not about the science}: not that the analysis answers the paper's question, avoids $p$-hacking, is free of model dependence, holds in fresh data, or contains genuine data. It locks down one run from the deposited files forward, made after conditional acceptance. It dates the record and the random draws, not the substantive work; binds output files and logged estimates, not the printed numbers which is left to the author and journal copyeditor; and shows which account deposited the files in the archive, not who wrote them.

Second, \emph{practical limits bound which analyses can be certified}. The protocol represents a data file larger than the archive is willing to store only with its hash and where it can be obtained, and so re-executing a step that reads it requires finding a copy with that hash. A folder of millions of separate small files, such as one per social media post, would need millions of hashes, so the protocol hashes it as a single unit. Outputs produced by an API, such as a large language model, that change over time can be dealt with in several ways, none completely satisfactory: (a) if the old version of the API remains available, such as for open source models, the protocol could allow the certified run to make calls to the API, although in this case a future researcher may need to host the then outdated API; (b) the protocol could recompute only a random sample of the items, chosen with the archive's seed after the file was fixed; or (c) intermediate output files from the API can be saved, and the replication dataset certified from that point forward.  We also note that some complicated workflows involving multiple computing environments may require modifications of version 1 of the protocol for certification.

Third, \emph{how evidence ages}. The record check works as long as the archive's files survive and the archive re-timestamps them in time if Bitcoin is ever abandoned; re-execution works only while the recorded software can be run, after which the software would need to be rewritten. The cryptography guarantees that the record is intact and dated, not that it stays available: data availability rests on the archive's custody and backups, as it does today. Deposited add-on libraries, or a container image, hold the installed software, not a list of libraries that may later disappear, but an image runs only on the kind of processor it was built for, such as Intel's or Apple's, or in emulation software. Software whose license forbids deposit cannot be included and so existence and record checks are straightforward, but re-execution requires first obtaining a license to the same software version.

Finally, \emph{what a bad actor can still do}.  Frauds that pass all three checks remain on the permanent public record at risk of being found. This includes results fabricated in the code; code that reads a file or sets a seed the dependency audit misses; and seed shopping paid for with a new, recorded deposit per seed. Re-execution catches a falsified step only when a future researcher reruns it. In addition, an archive colluding with an author, or anyone who steals the private half of the archive's seed key, can compute in advance the seed a planned deposit would receive, which allows seed shopping that leaves no record; and a stolen signing key alone yields certificates the ledger lacks, which the record check reveals against a later copy of the ledger the archive cannot revise.

\section{Prior Work}\label{sm:m-related}

We build on three bodies of prior work: (1) designs that give evidence about a published analysis; (2) the tools the protocol borrows; and (3) methods that date work by drawing public randomness after a commitment.  (\Cref{sm:related} summarizes our contributions.)

First, the designs, in order of increasing assurance. File fingerprints show whether a file changed \citep{dataverse_unf}, as do the file hashes that RO-Crate and Croissant, formats that bundle research files with machine-readable descriptions, can carry \citep{soilandreyes2022rocrate,akhtar2024croissant}; verifiable computational results store each published result under an identifier the article cites \citep{gavish2011universal}. Tools that record a run on the author's machine, by tracing the files it opens so that others can repeat it (ReproZip, SciUnit) or by logging the steps of a Python script (noWorkflow), and hosted platforms that run the analysis in a stored computing environment (Whole Tale, Code Ocean, Renku, Binder) rest on the word of whoever recorded or hosted the run, until someone re-executes it \citep{chirigati2016reprozip,tonthat2017sciunits, murta2015noworkflow, brinckman2019computing, clyburnesherin2019computational,roskar2023renku,jupyter2018binder,trisovic2020advancing}. Signed provenance, graded by the framework SLSA (Supply-chain Levels for Software Artifacts), and signed execution records that authors can neither alter nor disown add a signature but still rest on the word of whoever signed \citep{slsa,keita2026nonrepudiable}. TRACE, whose goal is similar to ours, requires its own platform to run and attest to the workflow, but this lets it reach data usable only inside secure facilities \citep{vilhuber2023trace,li2025trov}. Journal data editors, the named checkers of CODECHECK, and the French certification agency cascad re-run the analysis themselves, at the cost of a qualified person per paper \citep{nust2021codecheck,perignon2019certify}. Trusted hardware signs output produced inside a protected processor, on the word of the chip's maker \citep{costan2016intel, li2026tee}; cryptographic proofs of computation need no such trust but remain far too costly today for general data analysis \citep{goldwasser1989knowledge,walfish2015verifying}. A certificate needs neither a platform that vouches for the run nor a second party's re-run.

Second, the borrowed tools. The cryptography is well-tested: Merkle trees, keyed hashes, verifiable random functions \citep{rfc9381}, digital signatures, and public timestamps (\Cref{sm:tools}), obtained in Bitcoin through OpenTimestamps \citep{opentimestamps}, with a ledger modeled on the append-only logs that Certificate Transparency keeps for website certificates and Sigstore for signed software, an internet standard \citep{laurie2013certificate,sigstore2022,scitt2025architecture}. Blockchain timestamps of research files and pipelines exist, but show only that files existed by some time \citep{gipp2015decentralized,wittek2021blockchain,casas2024topo}. The recording of a run draws on running twice under frozen settings and comparing the outputs \citep{lamb2022reproducible,bizopoulos2020reconciler}, and on signed per-step records of inputs and outputs and the provenance workflow engines keep \citep{torresarias2019intoto}. Re-executing a random sample of recorded steps comes from distributed computing \citep{monrose1999distributed,du2004uncheatable,haeberlen2010avm,jia2021proof}; the deterrence random checks provide has been analyzed as well \citep{golle2001uncheatable,barnett2018randomly}.

Finally, dating work by public randomness drawn after a ``commitment'', a public statement that fixes the materials. Election audits \citep{cordero2006dice,lindeman2012gentle}, the Internet Engineering Task Force's selection of committee members \citep{rfc3797}, randomness beacons \citep{rabin1983transaction,kelsey2019beacons}, and a randomization seeded from a Bitcoin block \citep{saa2019auditable} all draw their randomness after publishing such a commitment, and a signature can postdate a blockchain's latest block, which dates the signature rather than a run's draws \citep{abadi2020timed}. Seeds hashed from the commitment itself \citep{fiat1986prove} or from a preregistration \citep{mazor2019novel}, and blind analysis, which hides results under a perturbation of the data until the analysis is frozen \citep{maccoun2015blind}, leave trust to whoever commits it, a gap that running the whole study once inside protected hardware closes, at the price of ruling out re-execution \citep{meissner2021peqes}. A certificate needs no separately published commitment.

\section{Concluding Remarks}\label{sm:m-concl}

A certificate is designed to move the scientific community closer to the idea of keeping publications based solely on the record and independent of its author.  In place of the author's word, certification provides a permanent public record of exactly which data and code produced each result, dated on the public blockchain so that no one can backdate it, and verifiable by anyone in seconds. It ends publication delays and subsequent disputes over what was done, ensures that every certified result reproduces, puts errors and fraud on a record open to all, and lets anyone question a prominent author without needing their help or their word. Because science advances as researchers check, correct, and build on one another's work, making certification standard practice should rebuild credibility in science and speed the process that produced four centuries of growth in knowledge and human welfare.

{
\renewcommand{\newline}{\unskip\space}
\renewcommand{\bibfont}{\normalsize\setlength{\bibsep}{0.5\baselineskip}}
\bibliographystyle{apsr}
\bibliography{refs,refs_implementation}
}

\clearpage\ifodd\value{page}\else\null\thispagestyle{empty}\clearpage\fi
\renewcommand{\thepage}{S-\arabic{page}}\setcounter{page}{1}
\normalsize
\setcounter{section}{0}\setcounter{figure}{0}\setcounter{table}{0}\setcounter{equation}{0}
\renewcommand{\thesection}{S\arabic{section}}
\renewcommand{\thefigure}{S\arabic{figure}}
\renewcommand{\thetable}{S\arabic{table}}
\renewcommand{\theequation}{S\arabic{equation}}
\renewcommand{\theHsection}{S\arabic{section}}
\renewcommand{\theHfigure}{S\arabic{figure}}
\renewcommand{\theHtable}{S\arabic{table}}
\renewcommand{\theHequation}{S\arabic{equation}}
\begin{center}
{\Large Supplementary Appendix for\\ ``Replication Certification at Publication: Checkable by Anyone, Anywhere, in Seconds, Forever''\par}
\end{center}
\vspace{1em}
\begin{abstract}
\noindent This Supplementary Appendix provides the complete specification of version~1 of the certification protocol proposed in the main paper, together with definitions of its terms and cryptographic tools. Sections, figures, tables, and equations prefixed with S belong to this appendix; unprefixed section numbers and \Cref{fig:overview} refer to the main paper. \Cref{fig:map} presents an overview of the protocol. \Cref{sm:toolkit} introduces the cryptographic tools, and \Cref{sm:close} explains how an author uses the certification software to prepare a replication dataset. \Cref{sm:produce} specifies the protocol for producing a certificate from a prepared dataset, and \Cref{sm:check} explains how anyone can verify the certificate and re-run any recorded step. \Cref{sm:guarantees} expands on the main paper's discussion of the limits of certification (\Cref{sm:m-uncovered}) and quantifies the cost of evading it. \Cref{sm:related} expands on the discussion of prior work (\Cref{sm:m-related}) and states our contributions. Finally, \Cref{sm:tools} formally defines each tool, \Cref{sm:dataverse} specifies the requirements for a participating archive, and \Cref{sm:bytes} provides the byte-level conventions needed for implementation.
\end{abstract}
\clearpage
\setcounter{tocdepth}{3}
\makeatletter
\patchcmd{\ttl@savel@section}{1.0em}{0.5em}{}{\PackageWarning{cert-supp}{contents spacing patch failed}}
\makeatother
\startcontents[supp]
{\noindent\Large\bfseries Contents\par}\vspace{1em}
\printcontents[supp]{}{1}{}
\clearpage

\section{The toolkit, in brief}\label{sm:toolkit}

\begin{figure}[p]
\caption{The Protocol on One Page}\label{fig:map}
\footnotesize
\setlength{\fboxsep}{2.5pt}%
\setlength{\tabcolsep}{0pt}%
\renewcommand{\arraystretch}{0.92}%
\definecolor{vfanyone}{HTML}{2a6b5f}%
\definecolor{vfmute}{HTML}{5b6675}%
\newcommand{\bandhead}[3]{{\color{vfink}\textbf{#1}}\ \
{\scriptsize\itshape
#2}\hfill\colorbox{#3}{\color{white}\strut\scshape\scriptsize\,#1\relax}\par\vspace{1pt}}
\newcommand{\bhead}[4]{{\color{vfink}\textbf{#1}}\ \
{\fontsize{9}{10.5}\selectfont\itshape
#2}\hfill\colorbox{#4}{\color{white}\strut\scshape\scriptsize\,#3\,}\par\vspace{1.5pt}}
\newcommand{\mapstripe}[2]{\noindent\colorbox{#1}{\begin{minipage}{\dimexpr\textwidth-2\fboxsep\relax}#2\end{minipage}}\par}
\newcommand{\mapgap}{\par\vspace{5pt}\par}
\newcommand{\maparrow}[1]{\par\vspace{0.2pt}\centerline{{\color{vfink}$\boldsymbol\downarrow$}\
{\footnotesize\itshape #1}}\vspace{0.4pt}\par}
\newenvironment{symgrid}{\fontsize{9}{10.5}\selectfont\begin{tabular}{@{}p{0.132\textwidth}@{\,}p{0.353\textwidth}@{\
\ }p{0.11\textwidth}@{\,}p{0.371\textwidth}@{}}}{\end{tabular}}
\newcommand{\srow}[4]{\textbf{#1}&#2&\textbf{#3}&#4\\}
\newcommand{\lrow}[2]{\textbf{#1}&\multicolumn{3}{@{}p{\dimexpr 0.834\textwidth}@{}}{#2}\\}

\centering
\scalebox{1.0}{\begin{minipage}{\textwidth}%
\mapstripe{vfband}{%
\bhead{READY (\S\ref{sm:close})}{the author's work, before the
protocol: freeze, measure, name by hash}{author's machine}{vfslate}%
\begin{symgrid}
\srow{$P$}{the replication dataset as supplied; $P^\ast$'s files form
four groups: code $c$, data $d$, other inputs $u$, environment
$e$}{$P^\ast$}{closed copy: $P$ +
capture code, capsule $\kappa$ (software), target list; run
frozen; at most $C_{\mathrm{size}}$ (1 GB), $C_{\mathrm{files}}$ entries;
outputs $o$ arrive at the run}
\lrow{capture}{added files that deliver the seed and copy listed-command estimates (in Stata, those
its log shows), and otherwise only read, print, and write capture files;
$h_c^{-}$ hashes the code as supplied}
\srow{$h_G$}{group hash: one fingerprint per file group}{$x$,\,$h_x$}
{commitment $x=(h_c,h_d,h_u,h_e,h_o)$; $h_x=H(\mathrm{enc}(x))$ names
the certified dataset}
\srow{$s_0$,\,$\tau$}{two closure runs under author's $s_0$, compared
after normalization, earn $\tau$}{tiers}{\textsf{T0} bytes;
\textsf{T0-meta}; \textsf{T1} $10^{-9}$; \textsf{T1-reported} 4 significant figures;
\textsf{T2} disclosed; \textsf{TF} failed}
\srow{audits}{dependency audit (no undeclared reads); source state;
target list, in order}{evidence}{$P^\ast$ + evidence set stored with
$\rho$, or no certificate}
\end{symgrid}}
\maparrow{$P^\ast$ uploaded; $\mathrm{Run}(e,c,d,u,s)=o$ is now well defined}

\mapstripe{vfband}{%
\bhead{1\ \ DEPOSIT (\S\ref{sm:deposit})}{hash first, then seed: the
seed cannot precede the bytes it is bound to}{archive}{vfink}%
\begin{symgrid}
\srow{$\mathit{id}$}{deposit identifier $H(\mathrm{enc}(H(P^\ast),v))$;
same bytes replay it}{$\gamma$,\,$\beta$}{nonce context: $\mathit{id}$,
$H(P^\ast)$, key id, and $\beta$, a public random value
$D$ cannot predate}
\srow{$\nu\to s$}{$(\nu,\phi)=\mathrm{VRF}(k_{\mathrm{seed,priv}},\mathrm{enc}(\gamma))$;
the public rule gives $s$}{$D$}{signed receipt: $\mathit{id}$,
$H(P^\ast)$, group hashes, closure hash, depositor, predecessor,
$t_{\mathrm{arr}}$, ledger position, $\beta$, $\nu$, $s$, and more;
timestamped}
\end{symgrid}}
\maparrow{author holds $D$ and a seed no one could have known earlier}

\mapstripe{vfband}{%
\bhead{2\ \ RUN (\S\ref{sm:run})}{the one certified run, every step
written down}{author's machine, once, under $s$}{vfslate}%
\begin{symgrid}
\srow{$i$,\,$M_i$}{per step: pre-state manifest, every file hashed
(root $R_{M_i}$)}{$L_i$,\,$\Delta$}{console transcript; processes
sampled every $\Delta$ (50 ms)}
\srow{$r_i$}{step record: script, exit, times, written-file hashes,
$H(L_i)$, processes}{$Q$}{result capture: every logged
estimate, saved to a file or not}
\srow{$T$,\,$R_T$}{trace $(r_0,\ldots,r_{n-1})$, Merkle root $R_T$}
{$C_{\mathrm{blob}}$}{blobs kept to 200 MB; re-execution coverage
$n_c/n$}
\end{symgrid}}
\maparrow{one run $\to$ one certificate}

\mapstripe{vfband}{%
\bhead{3\ \ SEAL (\S\ref{sm:seal})}{sign, timestamp, list, link; then
freeze the published version}{archive}{vfink}%
\begin{symgrid}
\srow{$\rho$}{the certificate: $x$, $h_x$, $\kappa$, $\tau$, $D$,
$\nu$, $s$, $T$, $R_T$, $H(Q)$, audits, coverage, $v$}{$W$,\,$\pi$}{witness:
archive-signed statement naming $H(\mathrm{enc}(\rho))$; timestamp on $\rho$}
\srow{$E_j$}{entry $j$ of the ledger, a Merkle log of receipts, witnesses, key-log entries,
and publications, its head signed and timestamped hourly; proofs of
$D$'s and $W$'s entries are stored with $\rho$}{DOI}{stored by hash,
linked from the published version; cite
it, ``Certified.''}
\lrow{$\mathit{id}^-$}{a revision is a new deposit naming its
predecessor; per-uploader version chain}
\end{symgrid}}
\maparrow{published; from here on, anyone can check}

\mapstripe{vfband2}{%
\bhead{CHECK (\S\ref{sm:check})}
{$\mathrm{Verify}(\rho,P^\ast,\theta,k,\omega)$: no lenient setting;
every yes-or-no finding that runs must pass}{anyone}{vfanyone}%
\begin{symgrid}
\srow{record}{every $k$: recompute every hash, root, seed, tier, audit;
check signatures, dates, and that $D$ names the tree; ten findings,
seconds}{existence}{journal looks up: $P^\ast$
present, one link to $\rho$,
no exception mark}
\srow{\mbox{re-execution}}{$k$ random steps from stored pre-states, at $\tau$;
$\omega$: re-query pending timestamps}{chance}{$1-\tbinom{n_c-m}{k}/\tbinom{n_c}{k}$: chance $k$ steps
hit one of $m$ faked among $n_c$ replayable}
\lrow{not covered}{(\S\ref{sm:limits}) fabricated or honest-but-wrong
code, $p$-hacking, confinement; seed shopping made visible}
\end{symgrid}}
\end{minipage}}

\floatnote{Chips name who acts (slate = author's machine, red = archive, teal = anyone, gray = no one); arrow text names what passes on; each band's section defines its bold symbols, and \Cref{sm:toolkit} names the tools whose symbols the bands use, such as $H$, $\mathrm{enc}$, and $\pi$.}
\end{figure}

In this section, we introduce the cryptographic tools and the notation used throughout the protocol. Every stage of the protocol relies on the same small set of standard tools and follows two conventions. The tools fall into three groups: three tools name a run of bytes, four bind a person or a time to bytes, and one folds an ordered list into a single hash. Every security claim rests on five standard assumptions---about SHA-256, HMAC (hash-based message authentication code, a keyed hash that the protocol uses only for data under restricted access, where it gives each file the secret random value mixed into its hash), the verifiable random function of RFC~9381, Ed25519, and Bitcoin's blockchain---together with two operational premises and one assumption of trust in the archive (\Cref{sm:tools} states them). We describe each tool only in enough detail to read the protocol. \Cref{sm:tools} defines each tool formally, and \Cref{sm:m-related} cites the sources of the cryptography.

The two conventions concern the schema and the protocol version. A \emph{schema} is the machine-readable half of this specification. It consists of a small tree of files that fixes every byte-level spelling, including key names, grammars, and patterns, whereas this specification explains what each spelling means (\Cref{sm:schema}). The protocol version $v$ identifies the edition of the rules, and hence of the schema, under which each signed object was built. This specification describes version~1. In version~1, the constants are binding rules and the lists are exhaustive (\Cref{sm:version}).

First, three tools give a run of bytes a name. The hash $H$ is SHA-256, which maps any run of bytes to a 32-byte fingerprint. No one can feasibly find two inputs with the same fingerprint, and the protocol names every file, list, and signed object by its hash (\Cref{sm:naming}). To ensure that every implementation hashes the same record to the same value, the canonical encoding $\mathrm{enc}(\cdot)$ writes a record with named fields as one exact run of bytes. It adopts a single spelling in JSON, a standard text format for records, with fields sorted by key and no floating-point values (\Cref{sm:naming}). Similarly, $\mathrm{pack}(\cdot)$ lays out a whole file tree as one run of bytes, in a form that every implementation reproduces, so that a tree also has a single hash (\Cref{sm:pack}).

Second, four tools bind a record to a person or to a point in time. First, a signature $\mathrm{Sign}(k_{\mathrm{priv}}, m)$ is a mark on the bytes $m$ that only the holder of the private key $k_{\mathrm{priv}}$ can make and that anyone holding the matching public key $k_{\mathrm{pub}}$ can verify. The protocol uses Ed25519, a standard signature scheme (\Cref{sm:binding}). Second, the nonce $\nu$ is a number that the author could not have known in advance. The archive computes it with a verifiable random function (VRF), a keyed function whose output only the holder of a private key can compute. The VRF also returns a short proof, the seed proof $\phi$, which anyone holding the matching public key can verify. The archive uses a dedicated key pair for this purpose, which we term its \emph{seed key}, with private half $k_{\mathrm{seed,priv}}$ and public half $k_{\mathrm{seed,pub}}$. The input is a short record of facts about the deposit, and $(\nu,\phi)=\mathrm{VRF}(k_{\mathrm{seed,priv}},\text{that record})$. These facts include the hash of the deposited files and the newest Bitcoin block that already had five later blocks after it when the files arrived (\Cref{sm:pulse}). As a result, the nonce cannot exist before the bytes to which it is bound, nor before a moment fixed in a public record.

Third, the seed $s$ is derived from $\nu$ by a public rule. The author therefore cannot choose the seed, and anyone can verify it using $\phi$. Finally, the timestamp proof $\pi$ records the certificate's hash in Bitcoin's blockchain. Its block sets the later bound of the window around the certified run (\Cref{sm:window}), and no party, including the archive, can backdate it.

The last tool folds an ordered list into a single hash. A Merkle root is a single hash computed over an ordered list by hashing its items in pairs, level by level. Changing any item, or the order of the items, changes the root. Moreover, anyone can prove that an item belongs to the list by presenting a few hashes rather than the whole list. The protocol names the certified run's list of per-step records in this way, as $R_T$ (\Cref{sm:merkle}).

We now state the whole protocol in this notation, with the section that defines each object. Closing (\Cref{sm:close}) turns the author's replication dataset $P$ into the closed copy $P^\ast$ and measures its tier $\tau$ from two closure runs. Certification then proceeds in three stages.
\begin{enumerate}[leftmargin=1.6em,itemsep=2pt]
  \item Deposit (\Cref{sm:deposit}). The archive computes $H(P^\ast)$ and names the public random value $\beta$. It then forms the nonce context $\gamma$ from $\mathit{id}$, $H(P^\ast)$, the key id, and $\beta$, computes the nonce and seed proof $(\nu,\phi)=\mathrm{VRF}(k_{\mathrm{seed,priv}},\allowbreak\mathrm{enc}(\gamma))$, and derives the seed $s$ from $\nu$. Finally, it returns the receipt $D$, which holds these values and is signed under $k_{\mathrm{priv}}$, timestamped, and entered in its ledger.
  \item Run (\Cref{sm:run}). The author runs $P^\ast$ once under $s$. The run yields the outputs $o$, the trace $T$ of step records with Merkle root $R_T$, and the logged estimates $Q$. The certificate $\rho$ collects these together with $D$.
  \item Seal (\Cref{sm:seal}). The archive checks $\rho$, signs the witness $W$ naming $H(\mathrm{enc}(\rho))$, obtains the timestamp proof $\pi$ over $\mathrm{enc}(\rho)$, and enters $W$ in its ledger.
\end{enumerate}
Anyone can then run the check $\mathrm{Verify}(\rho,P^\ast,\theta,k,\omega)$ (\Cref{sm:check}). Here $\theta$ denotes the pair of values that the reader pins, that is, obtains independently and holds fixed: the archive's trust root (the key under which the archive's key log is signed) and one Bitcoin block hash. In addition, $k$ is the number of steps to re-execute, and $\omega$ indicates whether to consult live sources. The check recomputes every hash and root, verifies $\phi$ under $k_{\mathrm{seed,pub}}$ and the signatures under the keys that the archive's key log names, checks the timestamps and the ledger, and re-executes $k$ sampled steps.

\section{Help for authors preparing replication datasets}\label{sm:close}

A replication dataset is ready (the Ready part of \Cref{sm:how}) when the two closure runs of its closed copy agree to within half a unit in the fourth significant figure of every number (tier $\mathsf{T1}$-reported of \Cref{tab:tiers}, or a stricter tier). This section gives the procedure for getting there, \emph{closing}: the certificate software checks that the dataset can run, runs the dependency audit, fixes the list of scripts to run, freezes whatever could make a second run differ, adds its capture code, and runs the closed copy twice, measuring how closely the two runs agree. Readiness is the author's own work, on the author's machine and at the author's pace, and it is finished before the protocol of \Cref{sm:produce} begins. \Cref{sm:objects} defines the objects closing works on and the hashes that name them; \Cref{sm:grading} gives the closing procedure, which ends by measuring the determinism tier that later re-execution is judged against; and \Cref{sm:records} gives the two records closing makes for the certificate to carry. Byte-level spellings only an implementer needs are in \Cref{sm:bytes}.

\subsection{The replication dataset and its closed copy}\label{sm:objects}

Closing exists so that one function is well defined: the function from a closed copy and a seed to the files a run writes. This subsection states that function and then defines, in order, the four groups of files; the outputs; how a file is named; the folders whose files are hashed as one entry; the hash of a group of files; the record of the software environment, and the optional container image that carries the environment itself; the closed copy (introduced in the Ready part of \Cref{sm:how}); the commitment, which names a certified dataset by one hash; and the packed form, which gives the closed copy one run of bytes.

\paragraph{Run.}
Write
\begin{equation}
\mathrm{Run}(e,c,d,u,s)=o
\end{equation}
for the outputs $o$, the files that a run creates or changes, given the four groups of a closed copy---code $c$, data $d$, other inputs $u$, and the record of the software environment $e$ (\Cref{sm:dataset})---under the seed $s$, in the environment $e$ describes. The function is \emph{well defined} when every run with the same arguments writes outputs that agree at the determinism tier $\tau$ the two closure runs measured (\Cref{sm:cruns}).

\paragraph{Replication dataset $P$.}\phantomsection\label{sm:dataset}
A replication dataset $P$ is a \emph{file tree}: a folder and every file inside it, including nested folders. The files of its closed copy $P^\ast$ (\Cref{sm:pstar}) are divided, with no overlap and no remainder, into four groups, by a rule an implementation applies without judgment. Code $c$ is the \emph{executable scripts}, the files the computer runs, recognized by the version-1 schema's list of extensions (\texttt{.do}, \texttt{.R}, \texttt{.py}, and the others the schema lists), except those under the \emph{library root}, the folder of add-on libraries installed inside the dataset. The list of extensions is complete, so an analysis written in a language not on it cannot be certified under version~1 (\Cref{sm:limits}). Data $d$ is the non-code files the scripts read, as found by the dependency audit, which searches the scripts' text for file reads (\Cref{sm:audit}). Environment $e$ is exactly one file, the environment capsule $\kappa$, which closing writes to record the software the run used (\Cref{sm:capsule}). Other inputs $u$ are all remaining files. They include, whatever their extensions, the files under the library root, whose path the schema fixes, since such a library is not the author's analysis (\Cref{sm:audit}). A packed folder, a folder whose files are hashed as one entry (\Cref{sm:unit}), belongs to one group whole. A file or folder whose path or name the schema fixes is \emph{schema-named}.

\paragraph{Outputs $o$.}
The \emph{outputs} $o$ are the files a run creates or changes, such as tables, figures, logs, and derived data. They are not the article: a paper is typeset separately, often with numbers moved into it by hand, and the certificate binds the output files, not what the article prints (\Cref{sm:article}). Nor are they a fifth group, because the author does not deposit them: the certified outputs do not exist until the certified run writes them (\Cref{sm:run}), after the deposit has been hashed (\Cref{sm:deposit}). An output file left over from the author's own runs belongs to one of the four groups, and it is part of the certified $o$ only if the certified run changes it (\Cref{sm:run}).

\paragraph{Paths.}\phantomsection\label{sm:paths}
A path is a file's name inside the replication dataset, written in one fixed style so that two computers spell it the same way. Paths are \emph{relative}, starting from the dataset's top folder rather than the top of the disk (\path{code/estimate.R}, not \path{/Users/smith/paper/code/estimate.R}), and they use \texttt{/} as the separator, with no leading \texttt{./} and no trailing slash (a packed folder's entry adds one, \Cref{sm:unit}). The remaining rules fix how characters are encoded and forbid a few spellings that operating systems treat differently (\Cref{sm:bytes-paths}). They belong to the protocol, not to an implementation's choice, because a path goes into a hash, and two implementations that spell one file differently compute different hashes for the same dataset. A symbolic link, an entry that only points to another file, is not a file of the dataset: packing neither follows nor lists it. A link that points outside the dataset is an undeclared external reference, for which the dependency audit stops certification (\Cref{sm:audit}).

\paragraph{Packed folders.}\phantomsection\label{sm:unit}
A folder of millions of small files, such as one per social-media post, would need millions of recorded hashes, one per file in the group hashes and in every step's manifest, so the protocol instead hashes such a folder as one entry (\Cref{sm:m-uncovered}). A \emph{packed folder} is a folder $U$ of $P^\ast$ hashed this way, as $H(\mathrm{pack}(U))$: the folder packed as a tree of its own, with paths starting from $U$ (\Cref{sm:pack}). The author lists the packed folders in the \emph{packed-folder list}, a schema-named file inside $P^\ast$ beside the target list, spelled as the schema fixes, so that $h_u$ and the receipt bind the choice before the seed exists. The list is fixed at preflight, the first part of closing (\Cref{sm:preflight}): the software proposes the folders that hold the most files, and the author confirms or edits the proposal, as for the targets (\Cref{sm:entry}); only the list is bound, so the proposal is not part of the protocol. A listed folder must be a folder of $P^\ast$; it may not be, contain, or lie under the library root or the dependency root; and it may contain no code file, no schema-named file, and no other listed folder. A packed folder belongs to one group whole: to the data group $d$ when the dependency audit finds a read of the folder or of any path under it, and otherwise to the other inputs $u$ (\Cref{sm:audit}).

Version~1 hashes at most $C_{\mathrm{files}}$ entries of $P^\ast$ separately, a ceiling fixed at 100,000, counting each file outside a packed folder as one entry and each packed folder as one. The ceiling counts the finished closed copy, not the files a run writes or fetches, which enter the later manifests one by one. Preflight refuses a tree already over the ceiling, and the archive refuses at deposit a $P^\ast$ over it, or one whose packed-folder list breaks the rules of this unit (\Cref{sm:dataverse}); finding~1 recomputes both (\Cref{sm:verify}). Wherever the protocol lists files with their hashes, in a group hash (\Cref{sm:group}) and in a step's manifest (\Cref{sm:run}), a packed folder appears once, as one entry: its path, which obeys the rules of \Cref{sm:paths}, followed by one \texttt{/}, a spelling no file's path can have, together with $H(\mathrm{pack}(U))$; both lists sort an entry by that spelling, the \texttt{/} included. The run must not create, change, or delete anything under a packed folder: closing stops when a closure run does (refusal~ix, \Cref{sm:choices}), and finding~2 (\Cref{sm:verify}) fails a certificate in which a step's write-set, the files it wrote (\Cref{sm:run}), names one. A packed folder is therefore unchanged since deposit at every step, and re-execution rebuilds it from $P^\ast$ (\Cref{sm:reexec}).

\paragraph{Group hash $h_G$.}\phantomsection\label{sm:group}
A \emph{group hash} is one hash that stands for a set of files, such as all the code. For a group $G$ of files, write one line per file, giving the path and then the hexadecimal hash of the file's bytes, and one line per packed folder in the group, as \Cref{sm:unit} spells it; sort the lines by path (a packed folder's line by its entry, the trailing \texttt{/} included), compared as bytes; join them; and set $h_G=H(\text{those lines})$. The schema spells the separators, the join, and the value for an empty group (\Cref{sm:schema}). Because each line carries a content hash, changing one byte of any file changes $h_G$; because the lines carry the paths, renaming, adding, or removing a file changes it too. The subscript names the group: $h_c$ for code, $h_d$ for data, $h_u$ for other inputs, $h_e$ for the environment.

\paragraph{Environment capsule $\kappa$.}\phantomsection\label{sm:capsule}
The \emph{environment capsule} $\kappa$ answers the question ``what software was this run made with,'' and its fields are fixed so that every implementation records the same things. It is encoded with $\mathrm{enc}$ and stored as the one file of the environment group, at the fixed path \texttt{environment.json}. It records: the operating system and the processor architecture; each language \emph{runtime}, a program that executes analysis code, such as R, Stata, or Python, with its exact version; each add-on library the dependency audit finds the scripts loading (\Cref{sm:audit}), with its version; the numerical library that does the runtime's matrix arithmetic, and the C mathematics library that computes its exponentials, powers, and logarithms, the two sources of last-digit disagreement between machines; the frozen settings in force (\Cref{sm:frozen}); when the author declares at closing that the analysis uses a graphics processor, that processor's model, its driver's version, and the version of its numerical software, such as NVIDIA's CUDA; the container image's digest, when one is deposited (\Cref{sm:container}); and the certificate software's version and the hash of its installed source code, $(v_{\mathrm{tool}},\,h_{\mathrm{tool}})$ (\Cref{sm:deposit}). \Cref{sm:bytes-capsule} lists the fields exactly. Closing fills $\kappa$ before the closure runs (\Cref{sm:cruns}), from the versions the machine reports, without running the analysis. Reading a version does start the runtime, through commands the schema names so that two implementations fill $\kappa$ identically.

$\kappa$ is \emph{declared evidence}: a claim the certificate records and the check cannot recompute. Recomputing $h_e$ shows that the capsule in hand is the capsule certified, but nothing confirms that the author's machine held that software, and re-execution rebuilds nothing from $\kappa$: it restores files from a recorded pre-state and runs them under the frozen settings (\Cref{sm:reexec}). $\kappa$ tells a later reader what to install to attempt a fresh replication, and which environments may judge a re-executed step (\Cref{sm:reexec}).

\paragraph{Container image.}\phantomsection\label{sm:container}
A ready replication dataset comes with its software environment, not only a description of it: its add-on libraries are vendored under the library root (\Cref{sm:audit}), usually with the lockfile that lists their exact versions, and, where licensing allows, the author may deposit a container image (defined in the Ready part of \Cref{sm:how}), and the closure runs and the certified run execute inside it. $\kappa$ records the image's digest in the Open Container Initiative (OCI) format, a hash of the image's bytes. The image is not a file of the closed copy: the archive stores it alongside, addressed by that digest and outside the one-gigabyte limit on the closed copy (\Cref{sm:preflight}), and serves it to readers who re-execute (\Cref{sm:dataverse}). The image also fixes the operating system and the numerical libraries, and lets a later reader re-run a step with one command (\Cref{sm:reexec}); where a license forbids depositing it, as for a proprietary runtime, $\kappa$'s exact versions are the list of what to install instead.

\paragraph{Closed copy $P^\ast$.}\phantomsection\label{sm:pstar}
The closed copy $P^\ast$ is a copy of $P$ plus the files closing writes: the capture code (\Cref{sm:instr}), the environment capsule $\kappa$ (\Cref{sm:capsule}), the target list (\Cref{sm:entry}), when there are external files, the dependency declaration (\Cref{sm:audit}), when there are packed folders, the packed-folder list (\Cref{sm:unit}), and, when there are saved service calls or declared item computations, the service recording and the item-computation declaration (\Cref{sm:spot}). For a restricted deposit (\Cref{sm:confidential}), it also holds a file of 32 random bytes from which the salts, the secret values mixed into each of its published per-file hashes, are derived, so that no published hash of its files confirms a guess (\Cref{sm:bytes-salt}); and closing deletes from it any runtime cache, a file a runtime writes to speed itself up, such as compiled bytecode (\Cref{sm:reexec}). Every run of $P^\ast$ happens under the frozen settings (\Cref{sm:frozen}), which fix the usual sources of ``it came out different this time''. \emph{Capture code} is a small amount of added code that copies estimates out of the statistical runtime as it runs and delivers the run's seed to each script (\Cref{sm:instr}). The author's original $P$ is never modified.

Because the copy carries the capture code, its code-group hash $h_c$ covers that code, a reader who re-runs the deposit runs the capture code too, and the deposit no longer matches the author's scripts byte for byte. The certificate therefore also carries a \emph{capture-code record}: the code-group hash of the code as the author supplied it, written $h_c^{-}$, with the before-and-after hashes of every file the capture code touched and the hashes of the files it added. A reader can then check the author's original code against a code repository or a working paper and see exactly which files the capture code changed or added. Like $\kappa$, the record is declared evidence: the check recomputes $h_c$ from the deposit, but $h_c^{-}$ names code that was not deposited.

\paragraph{Commitment $x$ and commitment hash $h_x$.}
The five hashes, for code, data, other inputs, environment, and outputs, are published together as a \emph{commitment}, a bundle of hashes stored inside a later signed or hashed object so that none can be swapped afterward (\Cref{sm:naming}):
\begin{equation}
x=(h_c,h_d,h_u,h_e,h_o),
\end{equation}
where the first four are the group hashes of $P^\ast$'s four groups and $h_o$ is the group hash of the certified outputs over their \emph{raw} bytes, exactly as written. The certificate $\rho$ (\Cref{fig:cert}) also stores a normalized output hash, computed after normalization, which strips marks that file formats embed without changing content (\Cref{sm:norm}); only the raw $h_o$ enters $x$. The normalized hash covers only outputs stored in full: a file above the per-file storage ceiling $C_{\mathrm{blob}}$ (\Cref{sm:run}) is kept only as its hash, and no reader could renormalize bytes it does not hold. $x$ is encoded with $\mathrm{enc}$ as a JSON array of the five hashes in that order, each as 64 hexadecimal characters. The \emph{commitment hash}
\begin{equation}
h_x=H(\mathrm{enc}(x))
\end{equation}
names the \emph{certified replication dataset}: $P^\ast$ together with the outputs $o$ the certified run wrote from it, everything the run consumed and produced. The receipt $D$ (\Cref{sm:receipt}) binds the four input hashes at deposit, and the witness $W$ (\Cref{sm:documents}) binds all five, through $h_x$, after the run.

\paragraph{Packed replication dataset $\mathrm{pack}(P^\ast)$.}\phantomsection\label{sm:pack}
Packing lays a file tree end to end as one run of bytes, so the tree has one hash, as a zip file does; unlike a zip file, the packed form is defined so that every implementation produces the same bytes. This unit has three parts: first, the form; second, why the archive serves the packed bytes themselves; and finally, the two hashes that now name a replication dataset.

First, the form. A tree is many runs of bytes plus the names that tell them apart, so packing must carry the names and lengths, or two different trees could flatten to the same bytes. $\mathrm{pack}(P^\ast)$ writes the files in path-sorted order, each as a short header line followed by the file's raw bytes. The header holds the path, the byte count, and the permission bits, written as the constant \texttt{644} for every file. A constant suffices because an interpreter, not the operating system, runs every recognized script, so an executable bit would do no work; and a constant makes $\mathrm{pack}$ a function of paths and bytes alone, so two closings of one tree write identical headers on any filesystem. No file owner, owning group of users, or modification time appears, and the schema fixes the header's spelling (\Cref{sm:schema}). Two identical trees must produce identical packed bytes on every implementation.

Second, transport. An archive's ordinary handling of uploads can change bytes: converting tables to its own format, converting line endings or file names, reassembling downloads. So the archive serves $P^\ast$ as the packed bytes themselves, one opaque file exempt from such processing, stored under the \emph{content-addressed} name $H(\mathrm{pack}(P^\ast))$, a name that is the object's own hash (\Cref{sm:dataverse}). Unpacking is defined as the inverse of $\mathrm{pack}$, so unpacking the packed bytes returns exactly the original files.

Finally, the two hashes. $H(P^\ast)$, which abbreviates $H(\mathrm{pack}(P^\ast))$ throughout, is taken over the closed copy's bytes; the archive computes it when the files arrive, and it names what was sent. $h_x$ is taken over the commitment $x$, which includes the outputs' hash $h_o$, so it covers the outputs too, although they did not exist at deposit; it names what was certified.

\subsection{The closing procedure and its tier}\label{sm:grading}

Closing proceeds in the order of this subsection (summarized in the Ready part of \Cref{sm:how}): check that the dataset can run at all, and fix the packed folders (preflight); fix the scripts the run starts (targets), which the capture code must know; freeze the settings that would otherwise vary between runs; add the capture code; and run twice, comparing the two runs after normalization and grading their agreement. The dependency audit and the source state, which closing also records, are described in \Cref{sm:records}; the audit runs after preflight and before the targets are fixed, because the division of the files into groups uses its result (\Cref{sm:dataset}), and the targets are found with the invocation patterns it also matches (\Cref{sm:entry}).

\paragraph{Preflight.}\phantomsection\label{sm:preflight}
Preflight is a quick check before any real work: can this dataset be run, and can the archive be reached? The software rejects the tree on three grounds. First, a tree that cannot be run or packed: no code file, a missing runtime, two paths that differ only in letter case, an empty directory, no writable place for the run's fresh temporary copy, or a packed-folder list that breaks the rules of \Cref{sm:unit}. Second, a tree too large: one larger than the size ceiling $C_{\mathrm{size}}$, fixed at one \emph{gigabyte} ($10^9$ bytes) before encoding, which keeps the check's hashing to about a second (\Cref{sm:verify}) and bounds what the archive must store, larger materials entering as declared external dependencies (\Cref{sm:audit}); or one with more than $C_{\mathrm{files}}$ entries once each folder the author lists as packed counts as one (\Cref{sm:unit}). Finally, an archive that cannot be reached. \Cref{sm:bytes-preflight} gives the list, with the reason for each ground.

\paragraph{Targets.}\phantomsection\label{sm:entry}
The \emph{targets} are the scripts the run starts itself, in a fixed order: the master \texttt{.do} file, say, and not the files that file runs with \texttt{do}. The run executes each target as one step. The targets are found from a \emph{directed graph} of scripts, in which an arrow from $A$ to $B$ means that script $A$ calls script $B$. The arrows are exactly the matches of the schema's per-language \emph{invocation patterns}, fixed text patterns for one script calling another: \texttt{source} in R, \texttt{do} in Stata, \texttt{import} in Python, a command that starts a separate program naming a deposited code file, and the others the schema lists. The dependency audit uses the same patterns (\Cref{sm:audit}). The capture code's added files are never targets and contribute no arrows; they are loaded only through the line inserted at the top of each target (\Cref{sm:instr}). A script's \emph{in-degree} is the number of arrows pointing at it, and the default targets are the scripts with in-degree zero, in path order. If every script is called by another, every code script is a target. Files under the library root contribute no arrows and are never targets (\Cref{sm:audit}).

Closing always writes the target list, in execution order, to a schema-named file inside $P^\ast$. The software proposes the list by the in-degree rule and shows it to the author to confirm or edit, since real trees hold stale scripts that nothing calls, and path order can put an analysis before its cleaning step. Because the list is inside $P^\ast$, $h_u$ covers the order, and the check compares the run's record of steps with that file rather than with a declared value (\Cref{sm:verify}). Some projects' top level is a file the patterns do not read---a Makefile, an RMarkdown or Quarto document, a pipeline's configuration---and their authors edit the list to name, in order, the scripts that top level runs.

\paragraph{Frozen settings.}\phantomsection\label{sm:frozen}
The \emph{frozen settings} (introduced in the Ready part of \Cref{sm:how}) fix the time zone, the formatting of dates and numbers, the build date, the number of processor cores, and the order in which a runtime lists the items of a set, so that none of them changes the numbers. Most are \emph{process settings}, the named values (environment variables) that the operating system hands every program it starts. This unit has three parts: first, clocks and formatting; second, cores; and finally, item order.

First, clocks and formatting. The settings do not change the time the computer reports: code that writes the current time into an output is the author's to remove at closing, where the closure runs expose it. The reproducible-builds community, software developers who make compiled programs come out byte-identical, intercepts the clock itself \citep{reprotest}; this design does not, because no tool that does so runs without administrator privileges on every desktop system. The software sets the process time zone to UTC (Coordinated Universal Time) and the \emph{locale}, the settings that choose the date format and decimal mark, to the plain C-language default, with a period for decimals and dates in one fixed spelling. It also freezes the ``date of this build'' to the schema's fixed value, so output files do not embed today's date; build tools honor that setting, and where statistical software ignores it, as most does, normalization does the same work afterward (\Cref{sm:norm}). \Cref{sm:bytes-frozen} names each setting.

Second, cores. Numerical libraries read process settings to decide how many processor cores to use when adding numbers, and two machines that differ only in core count can then disagree in the last digits, since adding numbers in a different order is a different floating-point computation. The software sets each setting on the schema's list of core-count settings to $1$ and, where a numerical library runs different code for different processors, sets its reproducibility switch to its fixed code path (\Cref{sm:bytes-frozen}), so these libraries' sums run on one core, in one order. That is the protocol's largest cost in running time: an analysis tuned for many cores runs correspondingly longer in each of the three runs, the two closure runs and the certified run. A frozen core count above one, measured by the closure runs like everything else, is a possible relaxation for version~2. Code on graphics processors adds instability of its own, since the same container can give different results on other hardware or drivers \citep{heil2021reproducibility}. Code that runs in parallel by other means is untouched by these settings, and the closure runs measure whatever instability remains.

Finally, item order. Some runtimes vary from run to run the order in which they store the items of a set, so two runs can list the same items in different orders; Python's sets are the usual case. The software fixes that order by a seed: in each step of the closure runs, a seed derived from the \emph{closure seed} $s_0$, a starting number of the author's own used only there (\Cref{sm:cruns}); in each step of the certified run and of re-execution, the step's seed $s_i$ (\Cref{sm:run}). A seed rather than a constant is used so that output that depends on item order shows up as sensitivity to the seed rather than being hidden.

\paragraph{Capture code.}\phantomsection\label{sm:instr}
Closing adds capture code so that the certificate can bind the numbers a paper reports, not only the files that produced them. This unit has three parts: first, why files alone are not enough; second, what capture produces, and how it differs from the outputs; and finally, in brief, how the code arrives.

First, why files are not enough. A paper's coefficients and standard errors are held in the runtime's memory while it runs, and only what the author's code chooses to write reaches a file. A certificate built from files alone would bind only the estimates the author's code happened to write, often in forms hard to compare. Numbers a paper states in its text, such as a coefficient quoted in a sentence or a $p$-value in a footnote, often reach no file. And the numbers that do reach a file are often hard to compare: a figure is a picture, and a table may be a typeset fragment from which a number must be read back by eye (\Cref{sm:article}).

Second, what capture produces. Capture code writes, into a per-step \emph{capture file}, one record for every estimation made by a command on the schema's list of estimation commands, except, in Stata, one whose output the log does not show (\Cref{sm:run}). The capture files are outputs like any other, so $h_o$ covers them; the certificate also binds the \emph{result capture} $Q$, the sequence of their records in run order, so that a reader can compare an estimate without parsing a table. The estimates these records hold are the \emph{logged estimates} of the main paper (\Cref{sm:claims}). The outputs other than the capture files hold what the author's code chose to write, a figure for example; $Q$ holds every listed estimation, including one the author discarded. $Q$'s exact form, and how the check rebuilds it, are given with the certified run, when $Q$ comes into being (\Cref{sm:run}).

Finally, how the code arrives. Capture code arrives as files added to the code group, which in R and Python register \emph{hooks}, small routines the language calls at set moments, to fire after each estimation command (Stata's estimates are read from its log, \Cref{sm:run}); and as one \emph{loading line} inserted at the top of each target. The loading line does four things before any of the author's code runs. It \emph{seats} the step's seed, setting the runtime's random-number generator to it. In R and Python it \emph{masks} the language's set-seed call, replacing it with one that seats a seed derived from the run's seed ($s$, or $s_0$ in the closure runs); Stata's cannot be masked (\Cref{sm:audit}). It sets the run-to-run settings that no process setting reaches (in Stata, the number of processors and the order in which \texttt{sort} breaks ties). And it prints a \emph{seed banner}, a fixed line naming the seed, into the step's transcript, the record of everything the script prints. The files and the line are fixed in the version-1 schema, and after those duties capture code may only read, print, and write its capture files. \Cref{sm:bytes-capture} gives the mechanism by language, the protections on the capture files, and the \emph{capture command}, which the author may call to bind an estimate that capture code would otherwise miss, such as one from a hand-written estimator or a Stata estimation run under \texttt{quietly}.

\paragraph{Normalization.}\phantomsection\label{sm:norm}
Normalization removes marks that file formats embed without changing content, so that two honest runs are not called different for cosmetic reasons. Before comparison, the software strips this \emph{generation noise}: creation dates, the names of the programs that wrote a file, log timestamps, and metadata padding. Its categories are adapted from those of the reproducible-builds tool strip-nondeterminism \citep{stripnondeterminism}, and \Cref{sm:bytes-norm} lists every category and how each is stripped. The list is complete for version~1, because normalized bytes are hashed: an implementation that stripped one mark more than another would compute a different normalized output hash and reject an honest certificate. Every byte not on the list is left exactly as written. Normalization decides only whether two runs agreed, and because both the raw and the normalized output hashes are stored, a reader can see whether normalization, rather than identical bytes, made two runs agree.

\paragraph{The two closure runs and tier $\tau$.}\phantomsection\label{sm:cruns}
The closure runs (the Ready part of \Cref{sm:how}) are two runs of $P^\ast$, each from a fresh temporary copy, under the same closure seed $s_0$ and frozen settings. The software compares what they wrote and grades the agreement as the dataset's determinism tier. Two runs are the least that can reveal instability, and so the protocol's minimum; the author may run pairs privately as often as desired, and should when results look fragile, and deposits the report of one, the \emph{kept pair}. This unit has three parts: first, how the tier is earned; second, the closure report that records it; and finally, the extra disclosure a $\mathsf{T2}$ certificate carries, and the boundary cases of the numeric tiers.

First, the tier. The software compares every file by its raw hash, for tier $\mathsf{T0}$, and by its hash after normalization, for $\mathsf{T0}$-meta. For formats the schema parses for numbers, it also pairs the two runs' numbers, by rules the schema fixes (\Cref{sm:schema}), for the \emph{numeric tiers}, $\mathsf{T1}$ and $\mathsf{T1}$-reported, which compare numbers within a tolerance. A structure no rule of the schema reads is \emph{unparseable}, and for it only hash equality counts. The dataset's tier $\tau$ is the strictest row of \Cref{tab:tiers} that holds for every file, so one unstable file sets the tier for the whole dataset. Two runs can reveal instability but not rule it out: a source of difference that acts only sometimes can leave two runs in agreement by chance. The tier records what was observed. It also measures agreement under \emph{one} seed and says nothing about how much an estimate moves \emph{across} seeds, which is Monte Carlo variation, a property of the analysis rather than of its determinism (\Cref{sm:sensitivity}). Even the tier could in principle depend on the seed, as when one seed produces a degenerate bootstrap resample and another does not. Closing also stops, naming the path, if a run wrote a path that spells the current date or time in one of the schema's formats (refusal~vii, \Cref{sm:choices}): a re-execution on a later day would write the file under another name, and re-execution fails a step whose written files differ from the record (\Cref{sm:reexec}). The fix is one line of the author's code.

\begin{table}[htbp]
\centering
\caption{Determinism Tiers, from Strictest to Loosest}\label{tab:tiers}
\begin{tabular}{p{0.16\textwidth} p{0.74\textwidth}}
\toprule
\textbf{Tier $\tau$} & \textbf{What held for every output} \\
\midrule
$\mathsf{T0}$ & byte-for-byte identical (the two files have the same
hash) \\
$\mathsf{T0}$-meta & identical after removing generation noise
(\Cref{sm:norm}) \\
$\mathsf{T1}$ & numerically equal within relative tolerance $10^{-9}$:
for each pair of numbers,
$\lvert a-b\rvert/\max(\lvert a\rvert,\lvert b\rvert)\le 10^{-9}$;
both zero match; one zero and one nonzero do not; non-finite and
missing values match only themselves \\
$\mathsf{T1}$-reported & equal to within half a unit in the fourth
significant figure: $\lvert a-b\rvert\le\tfrac12\cdot10^{\ell-3}$, where
$10^{\ell}\le\max(\lvert a\rvert,\lvert b\rvert)<10^{\ell+1}$ (both zero
match; non-finite and missing values match only themselves); every
pair within $10^{-9}$ passes, so the tiers nest \\
$\mathsf{T2}$ & a numeric or structural difference, not cosmetic \\
$\mathsf{TF}$ & execution did not complete; nothing is certified \\
\bottomrule
\end{tabular}
\floatnote{A replication dataset's tier is the strictest row that held for every output of its two closure runs; it is measured, never declared by the author, and later re-execution is judged against it. In the formulas, $a$ and $b$ are a pair of corresponding numbers from the two runs, and $\ell$ is the power of ten of the larger. ``-meta'' means identical except for metadata that normalization removes; ``-reported'' means to the precision papers report.}
\end{table}

Second, the closure report. The kept pair's measurements are written into the \emph{closure report}, a document in the schema's format (\Cref{sm:schema}). For each file it holds the two raw hashes, the two normalized hashes, and, for a parsed format, its \emph{relative gap} for $\mathsf{T1}$ and its \emph{reported gap} for $\mathsf{T1}$-reported (\Cref{tab:tiers}), each the largest difference over the file's paired numbers and stored with the pair that produced it, from which the file's own tier follows. It also holds both runs' logged estimates, the deposit's $H(P^\ast)$ and $s_0$, and the number of pairs of runs the software saw attempted, a self-report the author can reset. The report is deposited beside $P^\ast$, never inside it, because it names $H(P^\ast)$. The archive signs the report's hash, the \emph{closure hash}, into the receipt (\Cref{sm:receipt}). The kept pair is therefore fixed before the seed exists: an author cannot run more pairs after seeing the certified numbers and deposit one whose tier or logged estimates suit better. No certificate is issued without closure runs, since without them there is no tier for the check to re-derive or for re-execution to be judged against (\Cref{sm:verify,sm:evidence}).

Finally, tier $\mathsf{T2}$ and the boundary cases of the numeric tiers. A $\mathsf{T2}$ dataset has at least one file whose two closure runs disagreed, so no single tolerance can judge a later re-execution of it: one tight enough to mean something would fail honest re-runs. On the ordinary route a $\mathsf{T2}$ result stops certification (refusal~ii, \Cref{sm:choices}); on the deliberate disclosed-$\mathsf{T2}$ route the certificate carries an exception mark (\Cref{sm:exception}) and the \emph{per-output disclosure}: the closure report's measurement of each file, its two hashes or its two gaps. For each output on which the closure pair disagreed, re-execution reports its new measurement against that disclosure and gives no pass-or-fail verdict (\Cref{sm:reexec}). A run that did not finish, $\mathsf{TF}$, never yields a certificate. The four significant figures of $\mathsf{T1}$-reported are a version-1 constant chosen to cover every digit a paper ordinarily prints (published tables commonly report two to four), so agreement at that tier holds every value printed to four or fewer significant figures within one unit in its last place. Significant figures rather than decimal places are used because they do not depend on scale, holding a coefficient near $10^{-3}$ and one near $10^{3}$ to the same relative standard. One boundary case is deliberate: under the numeric tiers' rule for zeros, an exact zero in one run against a residue such as $10^{-17}$ in the other fails, so the dataset falls to $\mathsf{T2}$ at closing, where the author can fix it, rather than at a later reader's re-execution.

\subsection{What closing records}\label{sm:records}

Closing also makes two records for the certificate: the dependency audit, of the files the scripts read and the seed resets they make, and the source state, of the version-control state of the author's files. Which choices belong to the author and which to the protocol, and when certification refuses, are collected with the production stages (\Cref{sm:choices}).

\paragraph{Dependency audit.}\phantomsection\label{sm:audit}
The \emph{dependency audit} reads the text of the scripts, without running them, and matches fixed text patterns the schema lists, to find the files the scripts read, inside the replication dataset or outside it, and every addition to a language's search path; it finds only the reads its patterns match (\Cref{sm:limits}). It also searches every text file for calls that reset a seed and for mentions of a capture file. This unit has four parts: first, how the audit classifies what the scripts read; second, how an external file and an installed library are bound; third, how it treats code that resets the seed; and finally, what else it reads or cannot read, and why the schema fixes its patterns. \Cref{sm:bytes-audit} gives the mechanism.

First, classification. Before deposit, the audit classifies every file-read reference, and every addition to a language's \emph{search path}, the places the language looks for add-on files (\texttt{.libPaths()} in R, \texttt{adopath} in Stata, \texttt{sys.path} in Python). A read of a non-code file inside $P^\ast$, outside the library root, places that file in the data group $d$, and a read of a packed folder or of any path under it places the whole folder there (\Cref{sm:dataset,sm:unit}); a read that matches a declared external dependency is bound as the second part describes; and an undeclared reference outside $P^\ast$ stops certification. A search-path addition naming anything but the library root, or the deposited image's own, also stops certification (\Cref{sm:container}). The patterns that find reads, search-path additions, and calls to other scripts form the audit's \emph{read family}. The protocol forbids reading an undeclared file during the run but does not detect it: the software does not monitor which files the run opens, and restricting the run's access is outside the protocol (\Cref{sm:limits}), so a read the patterns miss goes uncaught. Such a read is a \emph{residual risk}: a way, left open by the protocol, in which what happened can differ from what the certificate records (\Cref{sm:limits} lists them). Tracing every system call, as ReproZip and SciUnit do on Linux \citep{chirigati2016reprozip,tonthat2017sciunits}, would detect such reads, but at the price of machinery specific to each platform and, on macOS and Windows, of administrator rights, the price that also rules out restricting access; version~1 keeps the static audit and states this limit.

Second, external files and installed libraries. An external data file is allowed only if the \emph{dependency declaration file}, a schema-named file inside $P^\ast$ beside the target list, lists it with three things: a \emph{locator}, an address that says where to fetch it (a URL, or a library name and version); the hash of its bytes; and the path, under the schema-named \emph{dependency root}, at which the author's code reads it. The declaration has one entry per external file the run reads, and because it is inside $P^\ast$, $h_u$ and the receipt bind it before the seed exists (\Cref{sm:entry}). Before the run, the \emph{fetch layer}, the part of the certificate software that resolves declared dependencies, fetches each entry, requires its bytes to match the declared hash, places it at its declared path in the fresh temporary copy, and, if it is at most $C_{\mathrm{blob}}$, stores it as a \emph{blob}, a raw copy named by its hash, which the check reads (\Cref{sm:run,sm:evidence}). The fetched files then appear in every pre-state, and re-execution rebuilds them like any other file. The author's code reads each file at that path, never at the locator, so the run needs no network beyond the certificate software's own spot-check (\Cref{sm:spot}); a script that reads from the locator itself reaches outside $P^\ast$ and makes an undeclared reference, while a request to a remote service sent through the schema's listed web libraries is answered from the service recording instead (\Cref{sm:spot}). A locator can stop resolving even when its hash is recorded, as when a library version is withdrawn or a repository branch deleted, which is why the fetch layer stores a dependency as a blob; one larger than $C_{\mathrm{blob}}$ is bound only by its hash and locator, and re-executing a step that reads it requires finding a copy with that hash (\Cref{sm:m-uncovered,sm:reexec}). For source code, a locator into Software Heritage, the content-addressed public archive of software, remains available after the original repository is deleted \citep{dicosmo2020archiving}. The fetch layer serves data files, whose declared hash names exactly the bytes the run reads. Fetching and installing a library from a package repository is deferred to version~2, because a library's locator names its source archive while the run loads installed files built from it, which the declared hash would not cover. A ready dataset therefore carries its own add-on libraries: in version~1 they are \emph{vendored}, installed inside $P^\ast$ under the library root as a project-local library, or they live in a deposited container image (\Cref{sm:container}), and either way the language's search path points only there.

Third, resetting the seed. The audit's \emph{reseed family} matches the calls that reset the seed---\texttt{set.seed} in R, \texttt{set seed} in Stata, \texttt{numpy.random.seed} and the like in Python---and the constructors that draw fresh entropy from the operating system instead, bypassing the certified seed rather than overriding it (Python's unseeded \texttt{default\_rng()} is the common case). It reports every match by file and line, and $\rho$ records the list; the check recomputes it by re-running the audit (\Cref{sm:verify}). Legitimate code resets seeds: nearly every stochastic script opens with a literal seed, and a separate stream per bootstrap replicate is ordinary, so a match is disclosed, not refused. But disclosure alone would tell a reader little. A reset to a fixed seed after the loading line would make the outputs independent of the archive's $s$, so an author could try seeds privately and fix a favorable one in the code: seed shopping without a new deposit (\Cref{sm:claims,sm:limits}). In R and Python the capture code therefore \emph{re-maps} the author's resets: the masked set-seed call, asked for seed $j$, seats the seed derived from $(s,j)$ by the rule of \Cref{sm:binding}, under a schema-fixed \emph{domain string}, a fixed label hashed with the input so that this use of the hash cannot collide with any other. The derivation starts from the run's seed $s$ rather than the step's seed $s_i$, so that one literal seed in two scripts yields one random stream, as in the author's own runs. The author's stream structure is preserved, every draw depends on the archive's seed, and the audit still records each match. Stata's built-in \texttt{set seed} cannot be masked, so the capture code provides a command, \texttt{certified\_seed} $j$, that seats the same derived seed, and closing names the one-line replacement for each reset. A reset that the capture code cannot re-map and that remains after the loading line, such as a Stata \texttt{set seed} left unreplaced, is disclosed and marks the certificate (\Cref{sm:exception}). The rule covers \texttt{set seed} and the \texttt{seed()} and \texttt{rseed()} options of Stata's resampling commands, as a literal or a macro, at top level or inside a program, in a target or in any file the run reaches from one.

Finally, what else the audit reads or cannot read, and why the schema fixes its patterns. The reseed family and the \emph{capture-path family}, which matches any deposited file that names a capture file's path (\Cref{sm:instr}), run over every text file of $P^\ast$ and of the container image, whatever its extension---a shell script, a Julia file, an \texttt{.ado} program---so no text file inside $P^\ast$ or the image that the run executes carries a reset the audit did not read. A capture-path match is disclosed and closing stops (\Cref{sm:choices}), so a counterfeit record of $Q$ cannot come from text the audit read. The author's startup files on the machine (Python's user site, R's user environment file, Stata's personal profile) lie outside both, so every target and launched process runs with the interpreter flags that skip them (\Cref{sm:bytes-capture}); a startup file those flags do not skip could run code the audit never read, a risk \Cref{sm:falsified} treats with the other ways of changing the one certified run while it runs. A compiled file the audit cannot read is unaudited, and $\rho$ carries instead an \emph{artifact manifest}, naming for each such file the digest in a public package registry that it matches, or none; the check recomputes the manifest (\Cref{sm:verify}), and the count of unmatched files appears on the metadata field. Code that a data file carries into a run, such as a serialized object a script loads, is a residual risk (\Cref{sm:limits}). A reference by author code to the process settings that carry the run's seed to each step (\Cref{sm:run}) would let the code tell the certified run from the closure runs; the read family matches such references, and a match marks the certificate (\Cref{sm:bytes-capture}). The audit's patterns are fixed in the version-1 schema (\Cref{sm:schema}), not chosen by each implementation, because the check re-runs the audit on $P^\ast$ and requires the same verdict (\Cref{sm:verify}), and an implementation that searched for one pattern more than another would refuse a dataset the other had certified. A version-1 check uses the version-1 patterns, and $\rho$ names both the protocol version $v$ (\Cref{sm:version}) and the build of the certificate software that produced it (\Cref{sm:deposit}).

\paragraph{Source state.}\phantomsection\label{sm:source}
The \emph{source state} records which version of the files the author had saved, when the author uses a version-control system such as Git, which records a history of changes to files, so that a reader can match the deposit to a public repository and see whether it held edits made after the last saved snapshot. Closing records the \emph{commit}, the saved snapshot, named by its own hash; the \emph{branch}, the line of history it belongs to; the number of \emph{dirty} files, tracked files edited since the last commit; and the number of \emph{untracked} files, which version control does not record. It binds that record by hash. The source state is declared evidence: the check does not reconstruct Git history from the files.

\section{Producing certificates}\label{sm:produce}

Once a replication dataset is ready (\Cref{sm:close}), the protocol produces a certificate in the three stages of the Certify part of \Cref{sm:how}, whose order is what gives the certificate its dates: the files are hashed before the seed exists, and the certificate is timestamped after the run. In Stage~1 the author deposits the closed copy and receives a seed the author could not have known (\Cref{sm:deposit}). In Stage~2 the author runs the dataset once under that seed, and every step is recorded (\Cref{sm:run}). In Stage~3 the archive seals the result: it checks the certificate, signs and timestamps it, stores it by its hash, and links it from the published replication dataset (\Cref{sm:seal}). Three subsections follow the stages. A revision is a new deposit that names its predecessor and, because its bytes differ, receives a new seed (\Cref{sm:rev}). A restricted deposit (\Cref{sm:m-confidential}) is certified like any other; what changes is who may run the parts of the check that read the files, and which values are published, and data the archive may not hold are kept by a custodian instead (\Cref{sm:confidential}). And the author sets a short list of things, the protocol fixes the rest, and certification stops only for named causes (\Cref{sm:choices}).

\subsection{Stage 1: deposit the replication dataset, receive the seed}\label{sm:deposit}

Depositing, the first stage of the Certify part of \Cref{sm:how}, hands the closed copy $P^\ast$ to the archive, which hashes the bytes and only then returns a signed, timestamped receipt that includes the seed $s$, so that the seed cannot exist before the files do. This subsection defines, in order, the deposit identifier; the nonce context, the record from which the nonce, and then the seed, are computed; the public random value, the part of that record that gives the deposit a public earliest date; the receipt and its fields; how the receipt is timestamped; what the archive must do when a timestamp lags; the order these events must follow; why the seed has two sources and the protocol two documents; and the archive's check on the build of the certificate software.

\paragraph{Deposit identifier $\mathit{id}$.}
The deposit identifier names a deposit by its bytes and its edition of the rules, so that sending the same bytes again, with the same declared predecessor, returns the same receipt and a later edition gets a new identifier: $\mathit{id}=H(\mathrm{enc}(H(P^\ast),\,v))$, spelled as the schema fixes. When the same bytes arrive again under the same $v$ with the same declared \emph{predecessor}, the earlier deposit, if any, that this one revises (\Cref{sm:rev}), the archive serves the receipt it stored instead of issuing a new one, which would carry a new arrival time and public random value and so be a different receipt. The one exception is a receipt superseded for cause, whose replacement is served instead (\Cref{sm:receipt}). The stored receipt names the account that first uploaded the bytes, by a stable pseudonym the archive assigns to the account (\Cref{sm:bytes-receipt}); another account can submit a certificate for it only with the uploader's authorization (\Cref{sm:dataverse}). Declaring a different predecessor for the same bytes draws a new receipt under the same $\mathit{id}$; the predecessor, a signed field of each receipt, tells the two apart (\Cref{sm:rev}).

\paragraph{Nonce context $\gamma$.}
The nonce context $\gamma$ is the four facts to which the nonce $\nu$ is tied, so that a nonce cannot be reused for another replication dataset. Its fields are exactly these: $\mathit{id}$; $H(P^\ast)$; the \emph{archive key id}, the hash $H(k_{\mathrm{pub}})$ of the archive's current public signing key; and the public random value $\beta$ (\Cref{sm:pulse}). The arrival time $t_{\mathrm{arr}}$ is deliberately excluded; it stays in the receipt, where the check bounds it (\Cref{sm:verify}). The archive's one free choice in $\gamma$ is therefore which recent block the public random value names, a choice the check limits to about one day (\Cref{sm:pulse}); an arrival time in $\gamma$ would instead have let it choose among the day's 86,400 seconds. Because the archive holds $k_{\mathrm{seed,priv}}$, it could compute the seed for each candidate block and name the one it prefers, or compute in advance, for an author, the seeds of deposits not yet made; the one assumption of trust in the archive (\Cref{sm:claims,sm:tools}) is that it does neither (\Cref{sm:calculus}). $\gamma$ is encoded with $\mathrm{enc}$ as an array of the four values in that order, each spelled as the schema fixes (\Cref{sm:schema}); then
\begin{equation}
(\nu,\,\phi)=\mathrm{VRF}(k_{\mathrm{seed,priv}},\,\mathrm{enc}(\gamma)),
\end{equation}
where $\phi$ is the seed proof, which anyone holding $k_{\mathrm{seed,pub}}$ can check, and $s$ is computed from $\nu$ by the public rule of \Cref{sm:binding}.

\paragraph{Public random value $\beta$.}\phantomsection\label{sm:pulse}
The \emph{public random value} $\beta$ gives a deposit an earliest possible date that does not depend on the archive's clock. The receipt's arrival time says when the files arrived, but only as the archive's own assertion, which would leave the archive free to backdate the receipt; the public random value replaces that assertion with a value no one could have known before a public moment. Its three fields name a public source of unpredictable values, identify one of its values, and give that value. In version~1 the source is Bitcoin's blockchain; the identifier is the height (the position in the chain) of the newest block that is six \emph{confirmations} deep when the files arrive, meaning that the block and five later blocks are in the chain; and the value is that block's hash. Six confirmations protect the named block against a \emph{reorganization}, the network's occasional replacement of its newest few blocks by a competing sequence, at the cost of an earliest date about an hour before the newest block. That block did not exist until it was \emph{mined}, that is, added to the chain by the network, so no receipt containing it can have been built earlier, by anyone. The archive chooses which recent block to name, so the check limits that choice. It rejects a receipt unless the receipt's timestamp proof confirms within the \emph{coherence window}, the $C_{\mathrm{time}}=24$ hours that follow the named block's time (\Cref{sm:verify}): the block that records the proof must be six confirmations deep and carry a time inside the window (\Cref{sm:receipt}). The coherence window leaves the archive about 144 blocks to choose among; \Cref{sm:window} derives from it the earliest date of the run's seeded draws.

The public random value's unpredictability dates the deposit, but it does not make the seed unpredictable: the archive's seed key does, and that is why a miner cannot choose the seed. A block hash can be influenced: a miner who dislikes a block's hash can discard the block and mine another, so a seed computed from a block hash alone could be steered. Here influencing the block gains a miner nothing, because the miner cannot compute $\nu$ without $k_{\mathrm{seed,priv}}$ (\citealp{bonneau2015bitcoin} quantify the economics of manipulating seeds taken from block hashes). Reading the public random value writes nothing to any public record and so costs no fee; its one cost is that the source must be reachable at the moment of deposit. The protocol specifies only what the source must provide: unpredictable values published on a schedule under stable identifiers. Version~1 fixes Bitcoin; a later version may substitute another source that meets the same requirements, such as a \emph{randomness beacon}, a public service that publishes unpredictable values on a schedule, like the one the National Institute of Standards and Technology runs (\Cref{sm:related}). If the source cannot be reached, certification stops (refusal~v, \Cref{sm:choices}).

\paragraph{Receipt $D$.}\phantomsection\label{sm:receipt}
The receipt is the archive's one signed document for the deposit: these files arrived, from this account, under this edition of the rules, and this is the seed. The archive signs $\mathrm{enc}(D)$. It carries the statement type \texttt{deposit}; every signed object carries its statement type, so that a signature on one kind of object cannot be passed off as one on another, the discipline the domain strings apply to hashes (\Cref{sm:audit}). Its other fields, listed in \Cref{sm:bytes-receipt}, name the deposit ($\mathit{id}$, $H(P^\ast)$, the four group hashes, and the predecessor $\mathit{id}^-$ or its recorded absence); the uploader's account and \emph{identity assurance}, which kind of login sent the files; the archive's name and key id, $v$, and $h_{\mathrm{schema}}$, the hash of the version-1 schema (\Cref{sm:schema}); the recognized build of the certificate software (\emph{Certificate-software build}, below); the position the receipt's entry will take in the ledger (\Cref{sm:ledger}); the arrival time $t_{\mathrm{arr}}$; $\beta$, $\nu$, the seed proof $\phi$, and $s$; the closure hash (\Cref{sm:cruns}); and, for a receipt reissued for cause, the supersedes field (\emph{Attestation deadline and reissue for cause}, below).

\paragraph{Timestamping the receipt.}
The archive timestamps the receipt at once, so that the deposit has a latest possible date as well as an earliest one. As soon as it signs $D$, the archive sends $H(\mathrm{enc}(D))$ to the timestamp service, keeps the proof the service returns, and appends $D$'s entry to the ledger (\Cref{sm:ledger}). The \emph{timestamp service} is a public service whose servers, called \emph{calendars}, collect many users' hashes, record one hash of them all in a single Bitcoin transaction, and return to each user a proof linking its hash to that transaction (\Cref{sm:binding}). A timestamp proof \emph{confirms} when the Bitcoin block that records its hash is six confirmations deep, and its confirmed time is that block's recorded time. The check compares both bounds, the block named in $\beta$ and the receipt's confirmed time, with the arrival time they bracket (\Cref{sm:verify}).

\paragraph{Attestation deadline and reissue for cause.}
The coherence window obliges the archive to act in time. The calendars commit on their own schedule, and an honest deposit whose proof confirmed late would fail the check permanently. So a receipt whose proof would confirm outside the coherence window must be recorded at the archive's own expense rather than left to lag: if twelve hours after issuing a receipt no Bitcoin block yet records its hash, the archive pays to record it in a Bitcoin transaction of its own, or in one it buys. Those twelve hours are the \emph{attestation deadline}, a version-1 constant that leaves the rest of the window for confirmation. If a proof nevertheless confirms outside the coherence window, the archive \emph{reissues for cause}: it signs a replacement receipt for the same bytes under a new public random value, and so with a new seed. The replacement's signed \emph{supersedes} field names the \emph{superseded receipt} by the hash of its document as the ledger stores it (\Cref{sm:ledger}) and by its confirmed time, and the superseded receipt and its proof are stored with the replacement, so that the check can confirm the delay (\Cref{sm:verify}). The archive signs the witness only after the receipt's proof has confirmed inside the coherence window (\Cref{sm:dataverse}), so the author has not yet been certified and repeats the certified run under the new seed. The \emph{landing page}, the archive's webpage for a published version, which carries the certification metadata field that the existence check reads (\Cref{sm:checking,sm:dataverse}), counts the superseded receipt as the archive's fault, not as a receipt the author left unused, which the page counts separately as a possible sign of seed shopping (\Cref{sm:limits}).

\paragraph{Order.}
The receipt comes first and the certified run second: the run needs $s$, which exists only in $D$, and each target's seed banner shows that the run used it (\Cref{sm:verify}). The run needs no archive: once the author holds $D$, the archive need not answer again until the certificate is submitted, and no deadline applies between the two. The check detects a receipt from another deposit, or a seed that $D$'s $\nu$ does not yield, by recomputation (\Cref{sm:verify}).

\paragraph{Two sources for the seed, and two documents.}
The seed combines the archive's seed key with a public random value because each does a job the other cannot: the key makes the seed unpredictable and, through its proof, checkable, and the public random value dates the deposit. A seed taken from a public randomness beacon alone, with no archive key, would need a rule tying each deposit to one beacon value that the archive could not pick by misstating its clock, would still need the archive's ledger to count each author's draws, and would add the beacon's operators as parties to trust. A stronger design would also mix in a public value published only after the receipt is recorded, so that not even an archive colluding with an author could compute a seed in advance. Version~1 does not adopt it, to keep the author's path short: the seed arrives with the receipt rather than after a wait for the later value, and the protocol needs one rule fewer. The choice leaves one gap, stated as a residual risk with its cost: an archive colluding with an author can search privately among versions of a deposit, at one seeded run per candidate, and no check detects it (\Cref{sm:calculus}). The stronger design is a direction for version~2. The protocol issues two documents rather than one for a related reason: the deposit's hash is computed before the seed, and the receipt records both, which fixes the files before the draws, while the witness fixes the results after the run (\Cref{sm:seal}); neither can stand in for the other.

\paragraph{Certificate-software build.}
The archive's comparison of builds stops a stale, damaged, or foreign build of the certificate software, where a \emph{build} is one packaged version of the software. The software hashes its own installed source as a group hash is computed, over sorted relative paths and file contents, excluding the intermediate files the language runtime generates while running; the schema, not the implementation, lists the excluded files. It sends the pair $(v_{\mathrm{tool}},\,h_{\mathrm{tool}})$, its version and source hash, with the deposit, and $\kappa$ carries the same pair (\Cref{sm:capsule}). The archive compares the pair with the builds it publishes, refuses a hash it does not recognize or a build whose recognition has lapsed, and signs a recognized pair into $D$. A build stays recognized for a fixed overlap after its successor is published, ninety days in version~1, so that a dataset closed shortly before a new release can still be deposited. Which builds are recognized is a matter of governance rather than protocol, and it is the closest thing in this design to a central authority. Version~1 requires the recognized builds to be listed in the archive's \emph{key log}: the record, signed with the private half of its trust root's key pair, of the archive's signing keys, each with its interval of validity, and of the recognized builds (\Cref{sm:binding}). The key log is public and dated by the ledger's timestamps. Version~1 also requires every recognized build to be a \emph{reproducible build}: anyone who builds and installs it from the published source obtains the same installed files, and so the same $h_{\mathrm{tool}}$. At least two independent parties must build it separately and publish that value. Archives may maintain the list of recognized builds jointly. The comparison cannot stop a modified build of the certificate software that reports the expected hash (\Cref{sm:m-limits,sm:falsified}), since a program cannot establish its own integrity when the modification can also change the code that reports it.

\subsection{Stage 2: run once under the archive's seed}\label{sm:run}

The certified run (the second stage of the Certify part of \Cref{sm:how}) is the one official execution of the analysis, once, under the archive's seed $s$, with every step written down; the numbers a paper reports must be taken from it. This subsection defines, in order, the steps and the pre-state taken before each; how a step is launched and observed; what happens when the author attempts the certified run more than once under one receipt; the record each step leaves; the capture of estimates, and its limits in Stata; the seed sensitivity, which compares the certified run's estimates with the first closure run's; the trace that binds the step records; the outputs and write-sets, both derived from the pre-states; the stored files that decide which steps can be re-executed; and finally, the spot-checks of saved intermediate files.

\paragraph{Step $i$ and pre-state $M_i$.}
A step is one target (\Cref{sm:entry}), run as its own \emph{process}, one launched program, and steps therefore follow exactly the author's top-level structure: a dataset driven by one master script is one step, re-executable only whole, while one split into several top-level scripts can be spot-checked script by script. $P^\ast$ runs in a fresh temporary copy, never in the stored original. Any script a target calls runs inside its caller's step; the software records nothing finer than a step. A script that a target launches as a separate process also belongs to the caller's step: the certificate software installs a \emph{startup hook}, a routine the language runs whenever a process starts. The software installs it outside the tree and makes it reachable through the process settings it gives the run, and the flags that skip the author's own startup files leave it active (\Cref{sm:audit,sm:bytes-capture}). The hook gives the launched process the caller's step seed, masks its set-seed call, and captures its estimates, as the loading line does for the target (\Cref{sm:bytes-capture}). A launch whose flags skip that routine or clear the process settings, such as R's \texttt{--vanilla}, is a refusal at closing (refusal~viii, \Cref{sm:choices}). A process the hook cannot reach, such as a compiled program, marks the certificate when the software's sampling of running programs observes it (\Cref{sm:exception}); a process that starts and exits between two samples goes unobserved (\Cref{sm:limits}).

Write $i$ for the step's \emph{index}, its place in the order, starting at $0$. The step's pre-state, the file tree just before the step starts, is recorded as its \emph{manifest} $M_i$, the list of $(\mathrm{path},\,H(\mathrm{bytes}))$ pairs, one per file, with one entry for each packed folder (\Cref{sm:unit}). The manifest is summarized by a Merkle root $R_{M_i}$, computed by the rule of \Cref{sm:merkle} under the domain string \texttt{certified-manifest-v1|}, with the entries sorted by path and each pair a \emph{leaf}, one item of the Merkle tree, spelled as the schema fixes (\Cref{sm:schema}). Each $M_i$ is kept in full, paths and hashes only, so it is small, and is stored under its root; re-execution rebuilds a step's tree from it (\Cref{sm:reexec}). $M_0$ must equal the manifest of a fresh temporary copy of $P^\ast$ with the declared dependencies placed at their declared paths (\Cref{sm:audit}), and no other entry, so the recorded run began from the deposited bytes, a starting point the check recomputes rather than trusts (\Cref{sm:verify}).

\paragraph{Launching a step.}
The software launches each step's process under the frozen settings (\Cref{sm:frozen}), with the fresh temporary copy's top folder as the folder its relative paths start from. Each step runs under its \emph{step seed} $s_i=\mathrm{derive}(s,i)$, computed from $s$ and the step's index by the rule of \Cref{sm:binding} under the schema's per-step domain string, so that two steps that both simulate draw independent random streams while every draw depends on $s$. The software passes $s$ and $i$ to the process as two schema-named process settings; the loading line computes $s_i$ from them and seats it with the language's own set-seed call, and the masked set-seed call computes every re-mapped seed (\Cref{sm:audit}). The closure runs do the same with $s_0$ in place of $s$. While the step runs, the software records the \emph{console transcript} $L_i$, everything the script printed, and every $\Delta=50$ milliseconds samples the list of running programs descended from the step's process, never the whole machine. When the step exits, it records the next manifest, from which the files the step wrote are derived (\emph{Outputs and write-sets}, below). A failed step stops certification: a non-zero \emph{exit code}, the number a script returns on finishing (0 means success, by Unix convention), or, for a runtime whose unattended mode returns zero regardless (Stata), a failure of the schema's per-language error test on the log the runtime writes.

\paragraph{One certificate per receipt.}
The author may attempt the certified run more than once under one receipt; the certificate software does not choose among the attempts, and the archive accepts one certificate per receipt (\Cref{sm:dataverse}); the author's choice among attempts is a residual risk (\Cref{sm:limits}). The run uses the build of the certificate software that $\kappa$ records; using a later build means closing again, which changes $P^\ast$ and so requires a new deposit.

\paragraph{Step record $r_i$.}
A \emph{step record} is the written log of one step: what ran, what it wrote, and which programs were seen running. Its fields are the index $i$; the script's path; the language; the exit code; the start and end times in Unix seconds and the duration in milliseconds; $R_{M_i}$; the before-and-after hashes of every file written, read off the adjacent manifests; $H(L_i)$; the hash of the step's capture file (\Cref{sm:instr}); the \emph{observed} processes, each with its process id (the number the operating system assigns to a running program), its parent's id, its executable's path and hash $H(\mathrm{executable})$, its command line, and when it was first and last seen; and the pair $(\mathrm{sampler},\,\Delta)$, naming the sampler and its interval. Executable hashes do not cover \emph{dynamic libraries}, code a program loads while it runs, or add-on libraries (R packages, Python modules, Stata \texttt{.ado} files).

\paragraph{Result capture $Q$.}
Result capture copies the numbers out of the statistical runtime itself, through the capture code (\Cref{sm:instr}). After each \emph{estimation command}, a command on the schema's per-language list of commands that compute a statistical result (\texttt{regress} in Stata, \texttt{lm()} in R), the capture code reads the command's results---the coefficients, their standard errors, the sample size, and the command line---from the runtime's stored results in R and Python and from the log in Stata (\emph{Capture in Stata}, below), and writes them to the step's capture file. The capture file is a schema-named path inside the fresh temporary copy, and so a member of $o$ (\Cref{sm:instr}). Its form is one object per estimation in JSON, each number spelled exactly rather than rounded by a formatter (\Cref{sm:bytes-q}) wherever the stored results are read from memory.

From the capture files the software writes the sequence $Q$ of those objects, in run order, in the form the schema fixes; an empty $Q$ is the empty run of bytes. The check rebuilds $Q$ from the stored capture files and requires a byte-for-byte match. Capture is required for every language the dataset uses. The schema fixes the lists of estimation commands per language (\Cref{sm:schema}), since $\rho$ records how many records $Q$ holds and implementations working from different lists would disagree on that count. The lists include the standard post-estimation commands (marginal effects, predictions, linear and nonlinear combinations of coefficients), since those give the numbers tables print; any other quantity reaches $Q$ only through the capture command. Figures left in memory are exported as files and enter $o$.

\paragraph{Capture in Stata.}
Stata offers no way to run code automatically after each command, so there the capture code parses the log the run writes, and it reaches only the estimations the log shows, at the digits the log prints. An estimation run under \texttt{quietly}, or one whose stored results a later command overwrote before the parse, is not captured; closing names each such command so that the author can route it through the capture command, which reads the stored results exactly (\Cref{sm:instr}). A record from inside a loop carries the loop's echoed line, so its command line does not distinguish iterations; its dependent variable and coefficient names do, but only when they differ across iterations (\Cref{sm:sensitivity}). And a \texttt{bootstrap} prefix hides its replicates from the log, while R's capture code records an R \texttt{boot} call once per replicate, so the count of records in $\rho$ cannot be compared across the two languages.

\paragraph{Seed sensitivity.}\phantomsection\label{sm:sensitivity}
Because the archive's seed $s$ differs from the closure seed $s_0$, simulation-based numbers in the final tables can legitimately differ from any computed before deposit, and the certificate records by how much. The \emph{baseline} is the first closure run (\Cref{sm:cruns}). For each coefficient among the logged estimates that can be uniquely paired between the baseline and the certified run, by the grouping rules of \Cref{sm:bytes-sensitivity}, the certificate records its value under $s_0$ and under $s$; the absolute movement between them; the movement divided by the baseline standard error, when that is positive and finite; the two standard errors; and the absolute change in standard error. The record is an observed contrast between two seeds, not an estimate of Monte Carlo error, and no threshold of movement across seeds refuses certification. The accompanying \emph{sensitivity summary} reports how completely the estimates paired before it reports how far they moved. It gives its state and the reason for it, whether a comparison was made and, if not, why; its scope; the two seeds; and its \emph{pairing counts}: the numbers of records from each run, of paired coefficients, of coefficients scaled by a standard error, of unpaired records, of \emph{ambiguous records}, whose identifying fields (\Cref{sm:bytes-sensitivity}) occur more than once, and of invalid coefficients (\Cref{sm:bytes-sensitivity}). Its largest movement is the maximum movement in baseline standard errors among coefficients for which that scale is positive and finite. An invalid coefficient, an unpaired or ambiguous record, an absent baseline, or a missing standard error is never recoded as zero movement; it is disclosed as unavailable, or as incomplete pairing. The certificate software computes the movements from the closure report's logged estimates and the certified run's $Q$ and writes them, with the summary, into $\rho$, and the check re-derives them (\Cref{sm:verify}).

The tier and the seed sensitivity can disagree. A Markov chain Monte Carlo run with a fixed seed can be byte-identical across the closure runs ($\mathsf{T0}$) while its posterior summaries move in the third significant figure across seeds; another certificate can show little movement but incomplete pairing, because its estimates could not be paired. So the metadata field a journal reads reports the sensitivity state, the pairing counts, and the largest standard-error-scaled movement beside the tier (\Cref{sm:dataverse}).

\paragraph{Trace $T$.}
The \emph{trace} is the ordered list of all step records,
\begin{equation}
T=(r_0,\ldots,r_{n-1}),
\end{equation}
where $n$ is the number of steps, with Merkle root $R_T$ (\Cref{sm:merkle}), one hash for the whole log of the run. Altering any record, any recorded output hash, or the shape of the Merkle tree (how many leaves, and which pairs were hashed together) changes $R_T$. After the last step the software records a final manifest $M_n$ by the same rule, with root $R_{M_n}$.

\paragraph{Outputs and write-sets.}
The outputs and each step's \emph{write-set}, the files it wrote, are derived from the manifests, not reported by the run. $o$ is the set of files in $M_n$ that are new or changed relative to $M_0$; step $i$'s write-set is the set of differences between $M_i$ and $M_{i+1}$; and a write that leaves a file's bytes unchanged counts as no write. A file written and then deleted is absent from $o$; it appears in the intermediate manifests if it survived a step boundary, and nowhere if it came and went within one step.

\paragraph{Blobs and re-execution coverage.}
Written files are stored as blobs (\Cref{sm:audit}) up to the per-file ceiling $C_{\mathrm{blob}}$, fixed at 200 megabytes ($2\times10^8$ bytes). Above the ceiling only the hash is kept, except for capture files, which are always kept because the check rebuilds $Q$ from them (\Cref{sm:verify}). A step's pre-state is \emph{complete} when every entry in its manifest is available as bytes, as a stored blob or, for a file or packed folder unchanged since deposit, in $P^\ast$ itself, and only then can a reader rebuild the pre-state and re-execute the step (\Cref{sm:reexec}). A declared dependency counts toward a complete pre-state only when it is stored as a blob, that is, when it is at most $C_{\mathrm{blob}}$ (\Cref{sm:audit}), since its locator can stop resolving. If $n_c$ of the $n$ steps have complete pre-states, the \emph{re-execution coverage} is $n_c/n$, stored as the two integers and computed when the certificate is issued.

\paragraph{Spot-checks of saved intermediate files.}\phantomsection\label{sm:spot}
Some computations cannot be repeated at certification: calls to a remote service, such as a language model, and computations too long to rerun in three runs. Their results are saved in an intermediate file inside $P^\ast$, hashed at deposit like any file, and after the certified run the certificate software checks a random sample of the saved items, chosen from $s$, which no one knew when the file was fixed. This unit has four parts: first, service calls; second, item computations; third, the sample; and finally, the spot-check record and its check.

First, service calls. When the author runs the analysis before closing with recording turned on, the certificate software records each request the code sends through the schema's listed web libraries for R and Python, and each response. The \emph{service recording}, a schema-named file inside $P^\ast$, holds each request, its response, and the service's reported model name, version, and settings, with credentials removed. In the closure runs and the certified run, the software answers each such request from the recording, matched by the schema's rule over method, address, and body, with repeated identical requests receiving the recorded responses in the order recorded; a request the recording does not hold fails the step (\Cref{sm:run}). These runs therefore send no request through those libraries to the network and get the same answers every time, as does every later re-execution, so the analysis stays reproducible after the service changes its models or closes. A request sent by any other means reaches the service live, like a file read the audit misses (\Cref{sm:limits}).

Second, item computations. An \emph{item computation} is a computation, run before deposit, that wrote one result per item into a saved file, such as a classification of each document, and is too long to rerun at certification. The author names each in the schema-named \emph{item-computation declaration}: the saved file, the field that names each item, the seed the saved results were computed under, and a script in $P^\ast$ that recomputes any listed items from the deposited files. The script is code like any other, audited and never a target (\Cref{sm:audit,sm:entry}).

Third, the sample. Separately for each service recording and each item computation, the software lists the items in the order they appear (for a service, the requests the certified run replayed), and draws 100 of them, or all when there are fewer, by a partial Fisher--Yates shuffle: its $g$th draw, counting from 0, takes the first eight bytes of $H(\Lambda_{\mathrm{spot}}\,\|\,\mathrm{BE}_{64}(s)\,\|\,\mathrm{BE}_{64}(g))$ modulo the number of items not yet drawn, where $\Lambda_{\mathrm{spot}}$ is the schema's spot-check domain string and $\mathrm{BE}_{64}$ is as in \Cref{sm:binding}. For a service, the software sends each sampled request to the live service, with the author's credential supplied at the time and never recorded, and counts the fresh responses equal to the recorded ones after the schema's normalization of the response body, and the requests that return no answer. For an item computation, it runs the declared script on the sampled items, under the frozen settings and the declared seed, in a fresh temporary copy of $P^\ast$ with the declared dependencies placed, and compares each result with the saved one field by field, numbers at $\mathsf{T1}$-reported (\Cref{tab:tiers}) and everything else for equality; any difference stops certification. The closure runs check item computations the same way, so one that does not reproduce stops at closing, where the author can fix it; they send no live requests. A service can answer differently between calls even when nothing on the author's side changes, and can retire a model, so its counts are a disclosure, not a verdict: the certificate and the metadata field report them, and no threshold refuses certification. The live requests are skipped, and reported as skipped, for a restricted deposit or held files (\Cref{sm:confidential}), since they would send restricted data to a third party. For an item computation, if a share $f$ of the items a sample is drawn from was altered, the chance that the sample includes none of them is at most $(1-f)^{100}$, under 0.6\% when $f$ is 5\%. This 0.6\% figure does not apply to a service. There the sample is drawn from the requests in the order the certified run replayed them, and the run's code receives $s$. Code written to cheat could therefore work out which places will be checked and replay the requests with altered responses in other places.

Finally, the spot-check record and its check. The spot-check is not a step: it runs after the final manifest $M_n$ is taken and writes nothing into the run's tree, so the outputs $o$, their hash $h_o$, the trace, the tier, and the re-execution coverage are unchanged by it. It leaves its own record in $\rho$: for each sample, the sampled items' identifiers, the counts, and the hashes of the fresh responses and recomputed results, which are stored as blobs in the evidence set (\Cref{sm:evidence}). The record check recomputes each sample from $s$, requires the record to name it, and recomputes the counts and comparisons from the stored responses and results (\Cref{sm:verify}). A reader can rerun an item computation's check as a re-execution; a service's live requests cannot be repeated, and their record rests on the certificate software being unmodified (\Cref{sm:falsified}).

\subsection{Stage 3: sign, timestamp, and store the certificate}\label{sm:seal}

Sealing (the last stage of the Certify part of \Cref{sm:how}) makes the record tamper-evident and publicly dated: the archive checks the proposed certificate $\rho$ that the author submits, refusing on any failure, signs the witness $W$ for it, timestamps $\rho$ with the proof $\pi$ (\Cref{sm:binding}), and appends $W$'s entry to the ledger beside the entry $D$ received at deposit (\Cref{sm:deposit}). The archive stores beside $\rho$ the \emph{certificate bundle}, the documents of this certificate that a check reads besides $\rho$ and $P^\ast$: $\sigma_D$, the archive's signature on the receipt, and the receipt's timestamp proof ($D$ itself is a field of $\rho$); $W$ with its signature $\sigma_W$; $\pi$; the \emph{ledger proofs}, which show that $D$'s and $W$'s entries are in the archive's public ledger under a signed, timestamped summary of it (\Cref{sm:anchor}); the key-log entries for the keys that signed $D$, $W$, and the ledger heads covering their entries; the evidence set (\Cref{sm:evidence}); and, for a receipt reissued for cause, the superseded receipt with its timestamp proof (\Cref{sm:receipt}). This subsection has three parts: the documents stored with the run (\Cref{sm:documents}); the window those documents fix around the run, the argument on which the design rests (\Cref{sm:window}); and the public ledger and the timestamp that fixes its latest state (\Cref{sm:ledger}). How the archive stores the documents and recovers from a crash is in its requirements (\Cref{sm:dataverse}).

\subsubsection{The documents}\label{sm:documents}

This unit has four parts: one for each of the three documents sealing produces (the certificate $\rho$, the witness $W$, and the timestamp proof $\pi$), and finally one on who signs what.

\paragraph{Certificate $\rho$.}
The certificate is the finished record: every hash and list from closing, deposit, and the certified run, in one document. Each field is defined where its object was introduced (\Cref{sm:tools,sm:close,sm:deposit,sm:run}), and \Cref{fig:cert} shows an example with every top-level field annotated. Three fields are first introduced here. The first is the full receipt $D$, exactly as signed. The second is the output \emph{inventory}, the counts of output files by kind under the schema's \emph{kind map}, its rule from a file's extension to a kind (figure, table, log, or data). The last is the \emph{oversize list}, the written files that exceeded $C_{\mathrm{blob}}$ and are kept only as hashes. $\rho$ cannot prove on its own that it is genuine; the witness supplies that, and $\pi$ fixes only when the bytes existed.

\begin{figure}[p]
\caption{A Certificate $\rho$, Annotated}\label{fig:cert}
\scriptsize\setstretch{0.72}
\begin{verbatim}
{                                     <- line breaks and readable key order
                                         for display; enc(rho) has neither
 "protocol_version": "1",             <- v: which edition of the rules
 "schema": "9e4c...77",               <- h_schema: the v1 schema file
 "commitment": [                      <- x: the five group hashes
   "ab40...9e",                       <-   h_c: code
   "17c9...0d",                       <-   h_d: data
   "e803...77",                       <-   h_u: other inputs
   "9c41...5b",                       <-   h_e: environment capsule kappa
   "6f2d...c4"],                      <-   h_o: outputs, raw bytes
 "commitment_hash": "20fa...91",      <- h_x = H(enc(x)): one name for it all
 "final_manifest_root": "aa31...9d",  <- R_{M_n} (Stage 2)
 "predecessor": null,                 <- id^- as signed in D; null = none
 "environment_capsule": {...},        <- kappa; h_e is the group hash over it
 "source_state": {...},               <- commit, branch, dirty/untracked counts
 "capture_code": {
   "code_hash_pre": "3d18...7a",      <- h_c^-: code as the author supplied it
   "touched": {"code/clean.R":
     ["9b02...c1","4e77...30"], ...}, <- each target: before, after
   "added":
   ["code/certified_capture.R"]},     <- files the capture code created
 "closure": {
   "report_hash": "5e21...aa",        <- closure hash: H of the closure
                                         report; always present
   "tier": "T0"},                     <- tau, measured (tier table)
 "exception_mark": [],                <- marks that hold, in schema order:
                                         disclosed-T2, unmasked-reseed, unhooked-process
 "seed_sensitivity": {...},           <- per-estimate movement, two seeds
 "sensitivity_summary": {...},        <- state, pairing, max SE-scaled movement
 "deposit_receipt": {...},            <- D in full, archive-signed (Stage 1):
                                         id, H(P*), group hashes, predecessor,
                                         depositor, t_arr, ledger position,
                                         closure hash, beta, nu, phi, seed, ...
 "nonce": "77b0...13",                <- nu, as issued in D
 "seed": 1583296741,                  <- s, derived from nu by the public rule
 "targets": ["code/clean.R",
   "code/estimate.R","code/figs.R"],  <- targets, in execution order
 "timeouts_ms": {"code/clean.R": 3600000, ...}, <- per-script time limit
 "trace": {                           <- T: one record per step (Stage 2)
   "n": 3,                            <- step count
   "root": "c9d4...3e",               <- R_T: Merkle root of the records
   "steps": [
     {"index": 0,
      "script": "code/clean.R",
      "language": "R",
      "exit": 0,                      <- 0 = success; anything else stops
      "start": 1789000000,            <- Unix seconds
      "end": 1789000041,
      "duration_ms": 41250,
      "pre_state_root": "b2ac...77",  <- R_{M_0}: every file, hashed, before
      "wrote": {"data/clean.csv":
        ["9f31...c8","40cc...19"]},   <- before- and after-hash
      "transcript_hash": "88d1...b5", <- H(L_0): console output
      "capture_file_hash": "1c0a...e2",<- H of this step's capture file
      "processes": [...]},            <- observed programs; sampler, Delta
     {...}, {...}]},
 "outputs": {
   "normalized_hash": "b7aa...02",    <- h_o after stripping generation noise
   "files": {"figs/fig1.pdf": "d0e7...4f", ...},  <- per-file output hashes
   "inventory": {"figure": 2,
                "table": 3}},         <- counts of outputs by kind
 "estimates": {
   "count": 4,                        <- number of records in Q
   "hash": "3af2...96"},              <- H(Q); Q is rebuilt at check time
 "dependency_audit": {
   "external": [],                    <- declared (locator, hash) pairs
   "reached": [],                     <- what the fetch layer fetched
   "undeclared": 0,                   <- must be 0
   "reseeds": [],                     <- matched seed resets, by file and line
   "artifacts": []},                  <- compiled files: registry digest or none
 "reexecution_coverage": [3, 3],      <- n_c and n: every step replayable
 "spot_check": null,                  <- samples and counts; null = none
 "oversize": []                       <- files kept only as hashes (> C_blob)
}
\end{verbatim}

\floatnote{Hashes are truncated; \texttt{\{...\}} elides contents defined in the text. Field names are the version-1 schema's (\Cref{sm:schema}); $\mathrm{enc}(\rho)$ has no line breaks and sorts its keys. A $\mathsf{T2}$ certificate adds the per-output disclosure. $W$ and $\pi$ are issued over $\mathrm{enc}(\rho)$, so they cannot appear in it.}
\end{figure}

\paragraph{Witness $W$.}
The witness $W$ is the archive's signed statement that it sealed the finished certificate, and $\sigma_W$ is the archive's signature on it. $W$ is a statement of type \texttt{witness} whose other six fields are $H(\mathrm{enc}(\rho))$; the archive's identity; the protocol context, the pair $(v,h_{\mathrm{schema}})$ and the key id (\Cref{sm:version}); the ledger position $W$'s entry takes; the account that submitted $\rho$, which is the uploader's or one the uploader authorized (\Cref{sm:dataverse}); and the time of signing. Because the first field is a hash of all of $\rho$, $W$ binds everything in $\rho$, the receipt included, and nothing $\rho$ holds is restated. $W$ attests none of what \Cref{sm:m-uncovered,sm:limits} exclude, such as the correctness of the outputs, the honesty of the code, or what happened between process samples. $W$ is dated through the ledger (\Cref{sm:ledger}).

\paragraph{Timestamp proof $\pi$.}
The timestamp proof puts a public mark on the certificate's exact bytes, so that no one, the archive included, need be trusted about when $\rho$ came into existence. The archive obtains $\pi$ from the timestamp service (\Cref{sm:binding}) over $\mathrm{enc}(\rho)$ when it seals $\rho$. Once $\pi$ confirms (\Cref{sm:receipt}), a later pass by the archive completes the stored proof with its path to the confirming block (\Cref{sm:dataverse}); until then $\pi$ may be $\mathsf{pending}$, which is not a failure.

\paragraph{Who signs.}
The author signs nothing. For each certificate the archive signs its receipt $D$, its witness $W$, and the publication document that the ledger records when the version is published (\Cref{sm:ledger}), and it signs the ledger's head every hour; all are checked under keys listed in the key log, whose signature is checked under the trust root (\Cref{sm:binding}). $D$ records who deposited, as an account, and how that account signed in, as the identity assurance: a shared login, a person registered with the archive, or a person signed in through ORCID, the Open Researcher and Contributor ID, a researcher's persistent public identifier (\Cref{sm:bytes-receipt}). Version~1 has no author signature: a key that arrives with the files binds nothing the deposit does not already bind, and authors do not, in practice, publish a key separately, so author signatures are left as a candidate for version~2. An author who wants to sign may place a signature file in the tree, where it is hashed and bound like any other file, and the protocol does not check it.

\subsubsection{The window around the certified run}\label{sm:window}
The seed and the two timestamp proofs, the receipt's and the certificate's, together place the certified run between two moments, neither controlled by the author (the third feature of \Cref{sm:claims}), and this is the argument on which the whole design rests. It has three parts: first, why fixing a date takes a public record no party controls; second, the window itself; and finally, the window's two limits.

First, the public record. The timestamp proof settles when the certificate existed without trusting anyone's clock---the author's, the archive's, the journal's, or a later reader's---because the time a reader obtains is a Bitcoin block's, from a public record no party to the certificate controls. Some such outside source is needed, because every date the archive writes into its own signed documents, such as the receipt's arrival time, is its own assertion: a stolen signing key, or a cooperating archive, could produce a complete, internally consistent certificate at any later date and give it whatever dates suited. The proof rules out backdating. A proof confirmed in a 2027 block means that the certificate's exact bytes existed in 2027, and no cooperation between an author and an archive produces such a proof in 2029, since that would mean inserting data into a block mined two years earlier. A certificate can therefore be shown to be at least as old as it claims to be. A confirmed proof also remains valid after the archive is gone: if the archive's signing key is later compromised, or the archive disappears, the proof still shows that the certificate existed while the key was good.

Second, the window. Its later bound is the latest time by which the certificate existed, the moment its timestamp proof names, which matters when a question arises years later: whether the analysis predates an objection, a failed replication, or a competing paper it stands accused of having followed. The earlier bound comes from the seed, and it is not a clock reading. The archive produces the nonce only after receiving the deposit, and the seed comes from the nonce, so an analysis that ran under the seed ran after the deposit. The public random value gives that order a date (\Cref{sm:pulse}): the run is no earlier than the time of the block the receipt names, because that block did not exist before. The archive chooses which block to name, so the protocol guarantees only that this time is at most one day before the receipt's own timestamp proof, which, like the later bound, is the time of a block. So the certified run happened no earlier than a day before one public moment and no later than another; neither moment can be revised, and neither the author nor the archive controls either.

Finally, the window's two limits. The first is that the earlier bound dates the run's outputs (\Cref{sm:run}) only insofar as they depend on the seed. An analysis with no random draws could have been run before deposit and simply run again under the certificate software, and an author who computed every deterministic part before deposit and only the seeded remainder after it holds a certificate that dates the record and the seeded draws, not the substantive work (\Cref{sm:limits}). The second and last is that the later bound dates the certificate's bytes, not the moment the analysis finished.

\subsubsection{The public ledger and its anchor}\label{sm:ledger}

The ledger (introduced in the Certify part of \Cref{sm:how}) lets anyone see everything the archive has issued in a form the archive cannot edit without detection. Every hour the archive signs and timestamps a \emph{head}, a summary of the whole ledger (\Cref{sm:anchor}). A certificate whose entries the ledger does not hold fails the check whenever the reader holds a later head from a source the archive cannot revise. This subsection defines the ledger's entries and then the head.

\paragraph{Ledger entry $E_j$.}
The ledger is an \emph{append-only} public list, one that can only grow, of the archive's receipts, witnesses, key-log entries, and publications, whose entries are the leaves of one Merkle tree (\Cref{sm:merkle}) under its own domain string, \texttt{certified-ledger-v1|}. The tree's root at any size therefore names the whole list up to that point. Write $E_j$ for entry number $j$, starting at $0$ (the letter is reused: elsewhere $j$ is a seed an author's code requests), and $t_j$ for the time of its append, in integer Unix seconds. Write $\mathrm{kind}$ for which issuance the entry records: a receipt; a witness; an entry of the key log (\Cref{sm:binding}); or a publication, the publication of a version that carries a certification field, whose document is that field's contents (\Cref{sm:dataverse}). Write $\mathrm{doc}$ for the entry's document, which the archive stores addressed by its hash; for every kind, the document is the statement together with its signature, so that the head covering the entry, once timestamped (\Cref{sm:anchor}), dates the signature and not only the statement. Entry $j$ is
\begin{equation}
(j,\,\mathrm{kind},\,H(\mathrm{enc}(\mathrm{doc})),\,t_j),
\end{equation}
encoded with $\mathrm{enc}$ as an array of those four fields in that order, spelled as the schema fixes; it carries the document's hash, not the document, so it is under a hundred bytes. Each certification appends three entries: $D$'s at deposit, $W$'s at seal, and its publication's when the version is published. $D$ and $W$ each name, before they are appended, the ledger position their entry will take, so appends are made under a \emph{lock}, which lets only one writer append at a time, and no other entry can take that position between the moment a document names it and the moment its entry is appended. The ledger makes the archive's history \emph{tamper-evident}, since an edit changes every later root. Its tree supports two short proofs, both defined by RFC~6962, the Internet standard for Certificate Transparency logs (\Cref{sm:merkle}), and each a handful of hashes: an \emph{inclusion proof} that a named entry is in the tree of a given root, and a \emph{consistency proof} that one root's tree is a prefix of another's. The ledger dates $W$, and so $\rho$, indirectly, through the head covering $W$'s entry, which the check uses in place of $\pi$ when $\pi$ remains pending because a calendar failed (\Cref{sm:verify}); $\pi$ is the direct mark.

\paragraph{Anchoring the ledger head.}\phantomsection\label{sm:anchor}
A Merkle tree shows that the archive has not edited its own past, but not that everyone is being shown the same past. This unit has three parts: first, the attack in which an archive shows different readers different ledgers, a \emph{split view}; second, how signed, timestamped heads and the proofs each certificate stores make such an attack provable; and finally, the two limits that remain.

First, the split view. An append-only Merkle tree does not stop an archive from keeping two ledgers, each internally flawless, and serving one to some readers and the other to others. A reader who holds only what the archive serves cannot detect such a split view, because every proof within each ledger verifies. Catching it means comparing one's own copy of the latest state with someone else's, which is what it means for observers to \emph{gossip}: to exchange the latest statements of the tree's state they have seen. This protocol does not require anyone to do that.

Second, the head and the proofs, which turn conflicting heads into proof of misbehavior whenever two certificates are compared. At a fixed interval, one hour in version~1, the archive signs a head: a statement of type \texttt{head} naming the tree's size; its Merkle root; its top hash $t$, the RFC~6962 tree hash from which this protocol's root is computed and over which the standard's consistency proofs are stated (\Cref{sm:merkle}); the time; the archive's identity, and the protocol context. It sends $H$ of the head's encoding to the timestamp service and keeps the proof; a head so timestamped is \emph{anchored}. Each certificate's ledger proofs (\Cref{sm:seal}) locate its two entries in the ledger: for each of its $D$ and $W$ entries, an inclusion proof under the first signed, anchored head at or after the entry, with that head and the head's timestamp proof. The archive attaches them once the head's own proof has confirmed (\Cref{sm:dataverse}), and until then the check reports them pending, as it does $\pi$. A reader then checks the certificate's membership in the ledger from the stored documents alone, with no archive to ask. Two signed heads of the same size with different roots prove, from the two documents alone, that whoever holds the archive's key signed two conflicting versions of its ledger. For two heads of different sizes, the archive must serve a consistency proof that the larger extends the smaller, and a proof that fails, or a refusal to produce one, is the demonstration. An archive can still anchor two diverging versions of its ledger, since the timestamp service accepts any hash from anyone, but any two certificates whose stored heads conflict now prove the split whenever they are compared.

Finally, the two limits. The first is that a reader who compares with no one detects a split only by holding a head from a source the archive cannot later revise: in version~1, the ledger snapshots, complete copies of the ledger the archive deposits at least daily under their own DOIs (digital object identifiers, permanent citable addresses), and the copies other archives keep of them (\Cref{sm:dataverse}). A head may also carry cosignatures, signatures by independent parties, distinct from the witness $W$; such cosigning is the standard defense against split views. A cosignature under a key the reader has not pinned (\Cref{sm:binding}) is ignored, never failed; the place for cosignatures is optional and costs nothing when empty, and it lets a consortium add this defense without a new version. The second and last limit is that anchoring is periodic, so a split that opens and is abandoned within one interval is caught only when a certificate issued on it is later checked against the surviving tree, which the certificate then fails. The heads describe the whole ledger; what ties a certificate to them is its own inclusion proofs, and the check requires both of the certificate's entries to be under signed, anchored heads and, when a reference head, a later head the reader holds (\Cref{sm:check}), is available, consistent with it (\Cref{sm:verify}).

\subsection{Revising a certified replication dataset}\label{sm:rev}

A \emph{revision} is a later, different replication dataset that the author wants linked to an earlier certificate. Unchanged files under the same declared predecessor are not a revision: depositing them again returns the stored receipt (\Cref{sm:deposit}). Declaring a different predecessor for the same bytes draws a new receipt, whose seed is new only if the public random value names a later block, since $\gamma$ omits the predecessor. Changed bytes produce a new $\mathit{id}$ and a new $s$. Because $\mathit{id}$, $H(P^\ast)$, and $h_x$ all change across a revision, the author declares the link, and, when the predecessor is in the same archive, the archive confirms that both deposits have the same uploader, a fact taken on the archive's word (\Cref{sm:claims}), and records the link in the version chain.

\paragraph{Predecessor $\mathit{id}^-$.}
The predecessor $\mathit{id}^-$ is the identifier of the earlier deposit the new certification revises, as the author names it. Before signing $D$, the archive checks that a predecessor in this archive belongs to the same uploader, a fact a reader cannot recompute, and refuses the deposit otherwise. A predecessor not in this archive is signed as an \emph{unresolved link}, which the landing page shows as named by the author and unconfirmed. $\rho$ carries the predecessor as the archive signed it, not as the author sent it, and a certificate whose predecessor disagrees with $D$'s fails the check (\Cref{sm:verify}).

\paragraph{Version chain.}
The \emph{version chain} is an uploader's growing list of deposits, each entry tied to the one before. For each uploader, an append-only version chain records
\begin{equation}
(H(P^\ast),\,\mathit{id},\,H(\text{previous entry})).
\end{equation}
It is a display the archive derives from the receipts on the ledger, not a signed object, and its length appears on the landing page as the count of certified versions. Published history is derived only from predecessor links the archive signed; titles and display names do not decide which certificates belong to the same work. Two certificates that name the same predecessor are both listed as current. The archive adds this history to the landing page when the page is displayed and never writes it into a published version, so a published version never changes and a paper that cites it still cites the same bytes.

\subsection{Confidential data: the certificate public, the files restricted}\label{sm:confidential}

Confidentiality restricts who may hold and read the files, not whether the run can be certified. A \emph{restricted deposit}, a deposit of the main paper's restricted data the archive is allowed to hold (\Cref{sm:m-confidential}), is stored and hashed by the archive but served only to readers it has approved; it is a deposit in every respect (\Cref{sm:deposit,sm:seal}), and its certificate is public. Data the archive may not hold at all are certified too, with the archive's work on the files taken over by a secure facility or left to the author. This subsection has five parts: first, what is published and what stays restricted; second, two rules for what a restricted deposit publishes; third, who can run which part of the check; fourth, the author's duty of disclosure review; and finally, data the archive may not hold.

First, what is published. For a restricted deposit the certificate carries no data values: it holds names, hashes, counts, times, seeds, versions, settings, and command lines (\Cref{fig:cert}), with the sensitivity summary in place of the seed sensitivity's per-estimate values. So $\rho$, $D$, $W$, $\pi$, and the ledger entries are published and linked as usual, while $P^\ast$ and the evidence the check reads (\Cref{sm:evidence}), whose transcripts and blobs can contain data, stay restricted and are never published with $\rho$.

Second, the two rules. The first protects the restricted bytes: every per-file hash a public document carries is, for a restricted deposit, a \emph{hiding commitment}: a hash computed over the file's bytes preceded by a secret random value, the salt (\Cref{sm:m-confidential}). An approved reader, who holds the salts, still verifies, while an outsider cannot confirm a guess about a file by hashing it, a leak long known from storage services that keep one copy of identical files \citep{harnik2010side}. The salts are derived from one random file added to the other-inputs group $u$; because that file is inside $P^\ast$, no one can reproduce from a guess the whole-object hashes the protocol computes without a salt, such as the group hashes and $H(P^\ast)$ (\Cref{sm:bytes-salt}). The second rule lets any reader compare the article's numbers with the logged estimates, the number comparison of \Cref{sm:article}, without access to the restricted capture files: those estimates are published as a standalone $Q$ stored beside $\rho$, bound by the $H(Q)$ inside it and released only after the author's disclosure review (the last part of this subsection). Where the rules that govern release forbid a complete release, a \emph{redacted} $Q$ is published instead, each unreleased value replaced by its hiding commitment, and the full $Q$ stays with the restricted evidence. The standalone $Q$ is committed to like any other file, so $H(Q)$ confirms no guess. The per-estimate rows of the seed sensitivity (\Cref{sm:sensitivity}), which would republish withheld values, stay with the restricted evidence, where the check recomputes them for an approved reader.

Third, the check, whose individual tests are its \emph{findings} (\Cref{sm:verify}). The findings that read only the signed documents and the public record (findings~7 through~10, on the witness, the certificate's timestamp proof, the dates, and the ledger, and the signature and seed checks of finding~6) run for anyone. The findings that recompute hashes over the files, findings~1 through~5 and the rest of finding~6, and any re-execution, run only for a reader with access; a reader without access reports them as unevaluated, never as passed (\Cref{sm:check}). The journal's existence check needs no files and is unchanged (\Cref{sm:article}).

Fourth, the author's duty. Every estimate the capture code records, including estimates the article never reports, is a candidate for release in the standalone $Q$, and so gets the same \emph{disclosure review} as the article's tables: a review, before release, that no published number or name identifies a person or organization in the restricted files (\Cref{sm:m-confidential}). The author must anticipate that review at closing; an intermediate regression on restricted microdata, for example, is captured like any other estimate and released only if the review allows. The file paths and dependency locators $\rho$ publishes are part of the review, since a path or locator can name a dataset; a restricted deposit may use pseudonymous paths of the form the schema fixes.

Finally, data the archive may not hold. Their files, the \emph{held files}, are kept instead by a \emph{custodian}: either a secure facility, such as a federal research data center or a national statistics office, where data may be analyzed but not copied out, or the author, under a license that forbids deposit with a third party. Any other file the protocol would deposit whose release the custodian's terms forbid, such as a transcript, blob, capture file, or closure report that reproduces held values, is held on the same terms. Held files lie outside $P^\ast$ and are declared in it like external dependencies (\Cref{sm:audit}), each by its path, its hiding commitment, and a locator naming the custodian, the source, and the version. A held folder may be declared as one entry, under the rules a packed folder obeys (\Cref{sm:unit}): its path, one hiding commitment over the folder packed as a tree, taken with the salt derived from its entry (\Cref{sm:bytes-salt}), and one locator, so that a corpus of millions of items needs no record per item; a read of any path under it matches that entry. Re-executing a step that reads held files needs the same files again, and a corpus whose items can be deleted at their source, such as social-media posts, may no longer be complete. The archive never fetches them and accepts their hiding commitments without recomputing them, the one hash it takes from the depositor (\Cref{sm:dataverse}). Because the declarations are inside $P^\ast$, the seed still follows the held files' hashes; because held files are never stored as blobs, every step that reads one is outside re-execution coverage (\Cref{sm:run}). Held files carry no exception mark; the metadata field names their custodian instead (\Cref{sm:dataverse}).

A secure facility does the archive's work on the held files, and the author makes the closure runs and the certified run inside it (\Cref{sm:m-confidential}). The facility keeps the salt file with the held files (\Cref{sm:bytes-salt}), and only the hiding commitments, the rest of $P^\ast$, and a $\rho$ and standalone or redacted $Q$ that pass its review for disclosure leave it. Before the archive signs $W$, facility staff, or an accredited verifier working there \citep{perignon2019certify}, run findings~1 through~5 over the held files and sign a \emph{custodian statement}, of statement type \texttt{custody}, naming $H(\mathrm{enc}(\rho))$, each finding's state with its reason code, and the date until which the facility keeps the held files and the software environment. The archive verifies the statement under the key its key log lists for that facility and refuses on a failed finding (\Cref{sm:dataverse}); the statement joins the certificate bundle, and finding~7 checks it again (\Cref{sm:verify}). Until that date, a reader the facility admits on the terms the author met can run every finding and re-execute inside the facility.

For licensed data the author holds, which only the author checks before sealing (\Cref{sm:m-confidential}), the findings over the held files stay unevaluated for the archive, as for any reader without the files, and the metadata field names the author as custodian. The archive stores the salt file under restricted access and serves it to a reader who shows the same license, who can then check the held files of the declared version and re-execute, for as long as the vendor still supplies that exact version.

The share of papers each kind of confidential data covers differs by discipline. At the American Economic Association's journals, a majority of recent papers use restricted data held under agreements that forbid deposit with a third party, most of it licensed rather than held in a secure facility \citep{vilhuber2024report}. At the leading political-science journals most quantitative articles use public data.

\subsection{The author's choices, the protocol's requirements, and its refusals}\label{sm:choices}

Every choice in production belongs either to the author or to the protocol, except the archive's choice of the public random value (\Cref{sm:pulse}), and the protocol says in advance when it will stop. This subsection gives, in order, what the author may set; what the author may not set; the evidence every certificate must carry; the exception mark (introduced in the Check part of \Cref{sm:how}), the archive's flag that a certificate falls short of one of the protocol's standards; and finally, the refusals.

\paragraph{What the author may set.}
The author controls the following, and every such choice is recorded in the certificate or in $P^\ast$, which the certificate binds: the dataset's contents; the closure seed $s_0$; each script's \emph{timeout}, how long it may run before it is stopped; the target order (\Cref{sm:entry}); the packed folders (\Cref{sm:unit}); whether the analysis uses a graphics processor (\Cref{sm:capsule}); and the item computations to spot-check (\Cref{sm:spot}). A stopped script counts as a failed step and stops certification (\Cref{sm:run}); the default timeout is three times the longer of the two closure runs, and never less than an hour.

\paragraph{What the author may not set.}
The protocol, not the author, sets the following, and that is what makes them evidence: the receipt, and therefore $s$; the arrival time, which the archive records (\Cref{sm:pulse}); the tier $\tau$, which is measured rather than declared, the author choosing only which measured pair to deposit (\Cref{sm:cruns}); the result of the dependency audit; whether the archive signs the finished run; and the trace $T$ (\Cref{sm:run}), once the run has started. Where the runtime searches for add-on libraries is not a choice either: it is the schema-named library root inside $P^\ast$, or the deposited image's (\Cref{sm:container}). Nor are the ceilings $C_{\mathrm{size}}$, $C_{\mathrm{blob}}$, and $C_{\mathrm{files}}$, which the protocol version fixes so that no author's tree is judged by a looser rule than another's.

\paragraph{The evidence set.}\phantomsection\label{sm:evidence}
The \emph{evidence set} is everything the check reads other than $P^\ast$, the signed documents, their timestamp and ledger proofs, and the key-log entries (\Cref{sm:seal}). It comprises the closure report; every transcript $L_i$; every manifest, $M_0$ through $M_n$; the blobs recorded as stored, the capture files and fetched dependencies among them (\Cref{sm:run}), and the spot-check's stored responses and results (\Cref{sm:spot}); the startup hook's record of the processes it seeded (\Cref{sm:bytes-capture}); the container image, when $\kappa$ names one (\Cref{sm:container}); and, for a restricted deposit, the salts (\Cref{sm:bytes-salt}). Without it the check could not re-derive $\tau$ or the closure hash, and signatures and a timestamp on a record nobody can check would not be a certificate. So the archive signs the witness (\Cref{sm:seal}) only when it holds $P^\ast$ and the evidence set, and a deposit without the closure report receives no receipt either (\Cref{sm:receipt}). A journal therefore has nothing to require here: every certificate carries its evidence.

\paragraph{Exception mark.}\phantomsection\label{sm:exception}
The protocol's standards, those a certificate must meet for a journal to accept it by default, are three: agreement at $\mathsf{T1}$-reported or better (\Cref{tab:tiers}); outputs that depend on $s$ wherever the audit sees a reset of the seed; and no process outside the startup hook's reach. The protocol fixes these in advance so that no journal has to choose a tier (\Cref{sm:parties}). A certificate that fails any of them carries an \emph{exception mark}, disclosed in $\rho$ and shown on the metadata field, which names the failure; a journal's existence check refuses it (\Cref{sm:article}), and a journal accepts it only by an exception its policy states in advance. There are three marks. The first is \emph{disclosed-$\mathsf{T2}$}: a $\mathsf{T2}$ certificate issued on the disclosed-$\mathsf{T2}$ route (refusal~ii), the one route around the ordinary stop at $\mathsf{T2}$. The second is \emph{unmasked-reseed}: the audit matched, in code the certified run executes, a reset of the seed after the loading line that the capture code could not re-map, such as a Stata \texttt{set seed} left unreplaced (\Cref{sm:audit}), the one disclosed way for the outputs not to depend on $s$. The third and last is \emph{unhooked-process}: the certified run launched a process the startup hook could not reach (\Cref{sm:run}), the one disclosed way for a process to run outside the hook. The mark is the schema's function of the tier, the audit record, and the process observations, and a certificate with more than one failure carries the list. The certificate software writes the mark into $\rho$, the archive recomputes it before signing $W$ and writes its own value onto the metadata field at publication (\Cref{sm:dataverse}), and the check recomputes it again (\Cref{sm:verify}), so what a journal reads is a value any reader re-derives.

\paragraph{Refusals.}
Production refuses in nine cases, each with a named cause. (i)~An undeclared external reference exists. (ii)~The two closure runs agree only at $\mathsf{T2}$ on the \emph{ordinary route}, in which the author closes the replication dataset and then deposits and certifies it, because the author is better placed to fix the cause than a reader is to discount it later. On the deliberate \emph{disclosed-$\mathsf{T2}$ route}, certifying a dataset closed earlier without treating $\mathsf{T2}$ as a stop, a $\mathsf{T2}$ certificate may be issued, carrying an exception mark (\Cref{sm:exception}). (iii)~A deposited script names a capture file's path (\Cref{sm:instr}). (iv)~The archive cannot be reached at deposit or when $\rho$ is submitted; no seed can be obtained without it, because a seed the author made would give the certified run no more assurance than the closure runs have. (v)~The public source of the public random value cannot be reached (\Cref{sm:pulse}). (vi)~The closure report, $P^\ast$, the rest of the evidence set, or the container image $\kappa$ names is neither deposited nor, when held, declared (\Cref{sm:evidence,sm:confidential}). (vii)~A closure run wrote a path that spells the current date or time (\Cref{sm:cruns}). (viii)~A target launches a process with a flag or process setting that drops the runtime's startup hook (\Cref{sm:run}). (ix)~A closure run created, changed, or deleted something under a packed folder (\Cref{sm:unit}). Certification also stops at six points named where they occur: at preflight (\Cref{sm:preflight}); at a failed step during the run (\Cref{sm:run}); at a spot-checked item computation whose result differs (\Cref{sm:spot}); at an unrecognized software build (\Cref{sm:deposit}); at a storage mismatch (\Cref{sm:dataverse}); and at the archive's own run of the check before it signs $W$, which stops a second $\rho$ for one receipt, a $\rho$ from an account the uploader did not authorize, and a $\rho$ that fails any part of the check the archive can evaluate at that moment (\Cref{sm:verify,sm:dataverse}).

\section{Checking certificates}\label{sm:check}

This section first defines the check and what a pass establishes, and then gives the three types of checks of the Check part of \Cref{sm:how}: the record check, which recomputes the certificate from the files without running the analysis (\Cref{sm:verify}); the re-execution check, which optionally re-runs some recorded steps (\Cref{sm:reexec}); and finally, the existence check, the journal's look-ups, which are not computations (\Cref{sm:article}). The existence check comes last here rather than first, because it is sufficient only when anyone can run the other two, which are the same for everyone and open to any reader with access to the files.

\paragraph{The check $\mathrm{Verify}$.}
The check recomputes the hashes from the files and compares them with the certificate, the way one recomputes a table from the data rather than trusting the printed numbers. Write it $\mathrm{Verify}(\rho,\,P^\ast,\,\theta,\,k,\,\omega)$. Its arguments are the certificate $\rho$, read together with its certificate bundle (\Cref{sm:seal}); the closed copy $P^\ast$; the \emph{pinned pair} $\theta$, the archive's trust root and the hash of one Bitcoin block, which the reader obtains apart from the files being checked (\Cref{sm:binding}); the number $k\in\mathbb{N}_0=\{0,1,2,\ldots\}$ of steps to re-execute (a bare $k$ always means this count, and keys always carry subscripts, as in $k_{\mathrm{pub}}$); and $\omega\in\{\mathsf{offline},\mathsf{online}\}$, whether to consult live sources. Offline, every part of the check is evaluated from stored material: the stored chain of Bitcoin block headers, the 80-byte summaries of each block, each naming the block before it (\Cref{sm:obsolete}); the stored timestamp proofs; the certificate's ledger proofs; and the newest ledger snapshot the reader holds. Online, pending proofs are looked up again, and the archive's current head can serve as the \emph{reference head}, the later ledger head against which finding~10 checks the certificate's heads.

The check has no lenient setting: every finding that runs must pass. A finding ends passed, failed, or unevaluated, and a finding that cannot be evaluated is never reported as passed. A finding is unevaluated when something it needs is not in the reader's hands: a timestamp proof still awaiting public confirmation, reported as \emph{pending}; a reference head, when no source supplies one; a piece of the bundle the archive serves but the reader has not downloaded, such as a container image; or, for a restricted deposit, the files themselves, when the reader lacks access (\Cref{sm:confidential}). A finding with several parts takes the worst state among them (failed, then unevaluated, then passed). Every finding reports its state with a \emph{reason code} naming the part responsible, so two implementations print the same verdict. A failure names its cause, a missing piece or a mismatch, and which file, so that a half-downloaded evidence set is never reported as tampering. A piece the archive fails to serve to a reader entitled to it, $P^\ast$ or any piece of the evidence set, is a failure, not an unevaluated finding. The certificate software fetches the whole bundle, with $\rho$, by the DOI and version the article cites (\Cref{sm:dataverse}), and fetches it again before reporting a mismatch.

\paragraph{What a pass establishes.}
What a pass establishes depends on the arguments chosen. Without the pinned pair $\theta$ the check fails; with it, a pass establishes record consistency (defined in the Check part of \Cref{sm:how}). With $k>0$ it also establishes that each drawn step replicated; at $\mathsf{T2}$, that each replicated on the outputs the closure runs agreed on, with a fresh measurement reported beside the per-output disclosure (\Cref{sm:cruns}) for the others. A drawn step unevaluated because the reader's environment does not match is neither a replication nor a failure (\Cref{sm:reexec}). With $\omega=\mathsf{online}$, pending proofs are looked up again, so the report can say whether $\pi$ is now confirmed.

\subsection{The record check: recompute the certificate}\label{sm:verify}

The record check recomputes every value that the deposited files and the public record determine, checks the signatures, and asks whether each matches; declared evidence is checked only for consistency. It runs no analysis, covers every step rather than a sample, and is fast enough to be routine (the Check part of \Cref{sm:how}). This subsection gives its cost and then its ten findings.

\paragraph{Cost.}
The record check reads each file once and hashes it. Hashing runs at roughly a gigabyte per second on an ordinary laptop, so a replication dataset at the one-gigabyte ceiling of \Cref{sm:preflight} is hashed in about a second, and checking the handful of Ed25519 signatures takes well under a millisecond. Rebuilding $Q$ and the Merkle roots, and reading the evidence set, whose stored blobs can exceed the deposit itself, take time proportional to what was stored, which the one-gigabyte ceiling does not bound. A complete record check typically takes from seconds to a few minutes, mostly spent reading the disk, while the analysis it certifies may have taken hours or weeks. Re-execution (\Cref{sm:reexec}) is the expensive part, and it is optional: $k=0$ still settles record consistency for every step.

\paragraph{The record check ($k=0$).}
No value that can be recomputed is read off $\rho$. The first five findings recompute from the deposited files: do the files match the certificate? The last five check the archive's signatures and the public record: did the archive bind this certificate, do its dates cohere with the public record, and does the ledger hold it? The numbering orders the report, not the execution: an implementation runs finding~5's audit first, because finding~1's division of the files into groups uses its result (\Cref{sm:dataset}). The last five use these public times: $t_\beta$, the recorded time of the block the public random value names; $t_D$, the time at which the receipt's timestamp proof confirms; $t_\pi$, the time at which the certificate's does; and $t_W$, the time at which the proof of the head covering $W$'s ledger entry confirms; finding~10 also reads the confirmed time of the head covering $D$'s entry. They also use the archive's key log (\Cref{sm:receipt}).

\subparagraph{From the deposited files.}
\begin{enumerate}[leftmargin=1.6em,itemsep=2pt]
  \item Recompute the group hashes, including the environment capsule's, and the commitment hash, and confirm that the capture code is the schema's. $h_c,h_d,h_u,h_e$ are recomputed from $P^\ast$, and $h_x$ from $x$, taking $h_o$ from $\rho$ until finding~3 recomputes it; $h_e$ covers $\kappa$ (\Cref{sm:capsule}). $P^\ast$ is no larger than $C_{\mathrm{size}}$ and has no more than $C_{\mathrm{files}}$ entries, and its packed-folder list obeys \Cref{sm:unit}. The capture-code record coheres: its after-hashes match the deposited files, and $h_c^-$ recomputed from its before-hashes matches its recorded value. And every file in the record's list of added files is byte-identical to the schema's canonical file of capture code for its language, while removing the schema's loading line, at the schema's point of insertion, from each touched target yields bytes whose hash is the recorded before-hash. Closing's changes are thus recomputed rather than trusted; without this, altered capture code could pass every other finding and write the same false records again on re-execution.

  \item Recompute the trace's Merkle root, and check that every target finished and received the seed. $R_T$ is recomputed from the step records, and each stored manifest, $M_0$ through $M_n$, reproduces its recorded root. $M_0$ equals, entry for entry, the manifest recomputed from the $P^\ast$ in hand plus the declared dependencies' $(\mathrm{path},H(\mathrm{bytes}))$ pairs from the declaration file inside $P^\ast$, with no other entry (\Cref{sm:run}). Each $r_i$'s list of written files equals the difference between $M_i$ and $M_{i+1}$, no write-set names a packed folder (\Cref{sm:unit}), and no manifest or write-set names a path on the schema's list of \emph{runtime caches}, files a language runtime writes for its own speed, such as compiled bytecode (\Cref{sm:reexec}). $n$ equals the number of records; every step passed the schema's per-language test of success (a zero exit code, or for Stata the log's error test, \Cref{sm:run}); and every target appears in $T$. The seed $s$ arrived at every target: each target's transcript carries the seed banner (\Cref{sm:instr}) exactly once between the schema-fixed delimiters that only the loading line prints; a banner printed by a launched process carries other delimiters (\Cref{sm:bytes-capture}) and is not counted; and the counted banner names that step's $s_i$, derived from the $s$ of finding~6 (\Cref{sm:run}); a runtime's own log header may precede it. Re-execution coverage, recomputed from which manifests have all their entries available as bytes (\Cref{sm:run}), equals the recorded $(n_c,\,n)$. When there is a spot-check record, each sample in it is the one recomputed from $s$, and its counts and comparisons recompute from the stored responses and results (\Cref{sm:spot}).

  \item Recompute the outputs from the manifests, their raw and normalized hashes, and their agreement with the trace. $o$ is recomputed from $M_0$ and $M_n$ (\Cref{sm:run}); the raw $h_o$ is over all of $o$, and the normalized output hash over its members stored as blobs (\Cref{sm:run}). Every stored output file hashes to the value its manifest records; each output's hash in $M_n$ equals the last after-hash any step recorded for that path; and the output inventory is recomputed by the schema's kind map.

  \item Rebuild the logged estimates from the capture files, and require an exact match. $Q$ rebuilt from the stored capture files (\Cref{sm:run}) hashes to the $H(Q)$ in $\rho$, and a certificate that records no logged estimates must rebuild to an empty $Q$.

  \item Recompute the tier and the seed sensitivity, re-run the dependency audit on $P^\ast$ and on the image, check the list of fetched dependencies and the artifact manifest, confirm that each step records its observed processes, and recompute the exception mark. $\tau$ is recomputed from the closure report, whose hash is the closure hash $D$ signs (finding~6) and which names this deposit's $H(P^\ast)$ and $s_0$, equal to the deposit's. The seed sensitivity and its summary, including the state, the pairing counts, the availability of standard errors, and the largest standard-error-scaled movement, are recomputed from the report's values under $s_0$ and the $Q$ of finding~4, and equal those in $\rho$ (\Cref{sm:sensitivity}). An absent closure report fails the finding, since $\tau$ is then unmeasured. The dependency audit, re-run on $P^\ast$, finds no undeclared external reference, and its list of seed resets equals $\rho$'s. The list of fetched dependencies the certificate software recorded (\Cref{sm:audit}) holds exactly the pairs of the declaration file, with matching hashes. Each step record carries its observed processes, labeled as observed rather than complete, and its $(\mathrm{sampler},\,\Delta)$ pair (\Cref{sm:run}). When $\kappa$ names a container image, the audit re-run covers the image in hand, and the artifact manifest recomputed from it equals $\rho$'s; the parts of this finding that read the image are unevaluated when the image is not in hand. And the exception mark in $\rho$ equals the schema's function of the recomputed tier, audit record, and process observations (\Cref{sm:exception,sm:bytes-capture}).
\end{enumerate}

\subparagraph{From the archive's signatures and the public record.}
\begin{enumerate}[leftmargin=1.6em,itemsep=2pt,resume]
  \item Recompute the seed from the nonce, check the receipt's signature, and check that $D$, $\rho$, and the files name the same deposit. The seed proof $\phi$ in $D$ verifies, under the public seed key in the key-log entry that $D$'s key id names, valid at $t_D$, and with that key validated as RFC~9381 provides, that $\nu$ is the one output of the verifiable random function for $\mathrm{enc}(\gamma)$, with $\gamma$ rebuilt from $D$'s fields; that part is unevaluated while the receipt's proof is pending. $s$ is recomputed from $\nu$ by the public rule (\Cref{sm:binding}). $\mathrm{Vrfy}$ holds on $D$ and its signature $\sigma_D$ from the bundle, under the key its key id names, valid at $t_D$ under the key log; that part is unevaluated while the receipt's proof is pending. $D$'s recognized build $(v_{\mathrm{tool}},h_{\mathrm{tool}})$ equals $\kappa$'s and is recognized at $t_D$ by the key log. $H(P^\ast)$, recomputed by packing the $P^\ast$ in hand (\Cref{sm:pack}), equals $D$'s; the deposit identifier recomputed from $(H(P^\ast),v)$ equals $D$'s $\mathit{id}$; $D$'s four group hashes equal those of finding~1; and $D$'s closure hash equals $H$ of the closure report in hand. $\rho$'s $\nu$ and $s$ are $D$'s. $D$ and $\rho$ agree on $v$ and on $h_{\mathrm{schema}}$, which equals the printed version-1 hash (\Cref{sm:schema}), and $\rho$'s predecessor equals $D$'s, or both record its absence.

  \item Check the witness $W$ under the key the archive's key log names, a log the pinned trust root signs, and never under a key that arrived with the files. The finding requires $\mathrm{Vrfy}(k_{\mathrm{pub}},\,\mathrm{enc}(W),\,\sigma_W)=1$. Here $k_{\mathrm{pub}}$ is the key the key log names as valid at $t_W$, since the head covering $W$'s ledger entry, not $\pi$, dates the signature (\Cref{sm:ledger}); that part is unevaluated while that head's proof is pending. $W$ binds $H(\mathrm{enc}(\rho))$, and through it everything in $\rho$; $W$'s $(v,h_{\mathrm{schema}})$ equals $D$'s and $\rho$'s, while its key id may differ, since keys may rotate between deposit and seal; and $W$ records the account that submitted $\rho$. An absent trust root is a failure; that the root is genuine is the reader's duty (\Cref{sm:binding}), which no finding can test. When the bundle holds a custodian statement (\Cref{sm:confidential}), the finding also requires its signature under the key the key log lists for that custodian as valid at $t_W$, and that it name the same $H(\mathrm{enc}(\rho))$.

  \item Check the public time mark on the certificate's bytes. $\pi$ must be for the exact bytes of $\mathrm{enc}(\rho)$: a proof for other bytes fails, and a pending proof leaves the finding pending. With $\omega=\mathsf{online}$, confirmation is looked up again. With several calendars (\Cref{sm:binding}), $t_\pi$, like $t_D$, is the earliest confirmed time. While $\pi$ is pending, $t_W$, once confirmed, serves as $t_\pi$.

  \item Check that the dates cohere with the public record. The public random value $\beta$ is real: the block at the named height carries the named hash in the stored chain of headers. That chain passes through the block hash pinned beside the trust root and, after it, each header carries the proof of work Bitcoin's rules require (\Cref{sm:binding,sm:obsolete}); for blocks after the pinned one, a pass carries a reason code marking it provisional until the online check or a later pin confirms them (\Cref{sm:binding}). The receipt's timestamp proof is present. Once it confirms at $t_D$, the arrival time must lie between the two block times, widened by two hours ($\mathrm{h}$) at each end for the imprecision of Bitcoin's block times (\Cref{sm:binding}): $t_\beta - 2\,\mathrm{h} \le t_{\mathrm{arr}} \le t_D + 2\,\mathrm{h}$. And $t_D - t_\beta \le C_{\mathrm{time}}$, the coherence window, with no two-hour widening (\Cref{sm:receipt}). When $D$'s supersedes field is set, the finding fails unless the superseded receipt and its proof are in hand, its entry is in the ledger, its $\mathit{id}$, uploader, and predecessor equal the replacement's, no other replacement names it (a part tested against a ledger snapshot, and otherwise unevaluated), and its own $t_D - t_\beta$ exceeds $C_{\mathrm{time}}$.

  \item Check that the archive's ledger holds this certificate: the one finding that someone who has stolen the signing key cannot satisfy, since a holder of the key can satisfy every other finding with genuine signatures and timestamp proofs obtained at the real time. The finding has two halves. \emph{Inclusion}: from the stored documents, $D$'s entry and $W$'s entry each carry an inclusion proof under a signed head whose timestamp proof is confirmed, the head's signature being checked as in finding~7 at the head's own confirmed time; the entries' document hashes are those of the $D$ and $W$ in hand; and each entry's position is the one its document names (\Cref{sm:anchor}). \emph{Consistency}: the certificate's heads, those under which the two entries are included, are compared with the reference head (\Cref{sm:check}) in one of two ways. Against a ledger snapshot, a source the archive cannot revise (the newest snapshot the reader holds, obtained through its DOI or from a copy another archive keeps, rather than from the archive being checked; \Cref{sm:dataverse}), the reader rebuilds the tree at each of the certificate's heads' sizes from the snapshot's entries and compares roots; a mismatch is a split view and fails. With $\omega=\mathsf{online}$, against the archive's current signed head, the archive's consistency proof shows that the certificate's heads are in the archive's current ledger, but it cannot reveal a split view. Either way, the half passes only against a reference at least as large as the certificate's heads: a smaller one leaves it unevaluated, and one of the same size with a different root is a split view and fails. Without any reference head the consistency half is unevaluated, and proofs the archive has not yet attached are pending, as $\pi$ is. A certificate whose entries the ledger does not hold, one produced with a stolen signing key, say (\Cref{sm:limits}), fails here whenever a reference head newer than its own is available, and nowhere else.
\end{enumerate}

\subsection{The re-execution check: re-execute sampled steps}\label{sm:reexec}

After publication anyone may run recorded steps again and see whether they still produce the recorded outputs. Re-execution tests the replication claim (defined in the Check part of \Cref{sm:how}), and it is the reader's work alone: the author's closure runs happened before deposit and measured $\tau$ (\Cref{sm:cruns}), and the journal only checks that the certificate exists (\Cref{sm:article}). The record check already covered every step without running anything, so re-execution adds evidence that the certificate describes real computation. Its cost to the reader is the step's own running time, plus installing the recorded software where no container image was deposited (\Cref{sm:parties,sm:container}). The reader chooses how many steps to run, trading time for confidence, and because re-execution samples, it deters fabrication and spot-checks for it rather than proving honesty. This subsection gives the procedure and then the probability that a sample detects a faked step.

\paragraph{Sampled re-execution.}
Re-executing a step means rebuilding the exact file tree it started from, running it again under the same settings and seed, and comparing what it wrote and captured with the record, at the tier the author earned. The procedure has five parts: first, the sample; second, the rebuild; third, the run and its comparison; fourth, the precondition on a verdict; and finally, an optional re-run under the closure seed.

First, the sample. A step with a complete pre-state (\Cref{sm:run}) is \emph{replayable}. The reader draws $k$ distinct steps from the $n_c$ replayable ones, \emph{uniformly without replacement}: every remaining step is equally likely and none is drawn twice, and if $k\ge n_c$ all are drawn. The draw uses the reader's own randomness, never anything in the certificate, so different readers' draws are independent. A step that is not replayable is outside the sample (the second part, the rebuild, says when it can still be re-executed). A drawn step the reader declines to run is reported unevaluated and not replaced, since redrawing would make the sample non-uniform. The per-step durations in $T$ tell an author which steps a reader on a budget is least likely to run, so the probability of \Cref{eq:hyper} is that of a reader who draws before looking at them. Each declined step counts as a miss, so that probability, computed with the nominal $k$, is an upper bound.

Second, the rebuild. For each drawn step $i$, the reader reconstructs the tree from $M_i$, placing at every path in the manifest the bytes that hash to the recorded value (for a packed folder, its files from $P^\ast$) and creating folders as needed, including the parent folder of every path in the step's recorded write-set, so that a folder an earlier step created and left empty is present. The bytes come from the stored blobs; from $P^\ast$, for files unchanged since deposit; or, for a declared external dependency, from fetching its locator again and requiring the bytes to hash as declared. The hash, not the locator, is the file's identity, so a matching fetch rebuilds the step as a blob would, though only a stored blob counts toward coverage (\Cref{sm:run}). A step whose only missing bytes are a declared dependency too large to store is outside the sample, and coverage says so, but a reader can still re-execute it by fetching the dependency.

Third, the run and its comparison. The reader runs the recorded script as at certification, in its own process and under the frozen settings, with $s$ and $i$ passed through the two schema-named process settings, and compares every written file with the recorded output under $\tau$. The recorded output is the stored blob's bytes, and at the numeric tiers the comparison parses them, which a hash alone could not support. The set of paths the re-run wrote must equal the recorded write-set, $M_{i+1}$ against $M_i$, deletions included. A recorded path the re-run did not write, or a written path the record lacks, fails the step; so a build that falsely recorded a write by step $i$ is exposed when step $i$ is re-run. A path on the schema's list of runtime caches is treated asymmetrically. If the re-run writes it, it is ignored, since a reader's environment may write such caches; if the record holds it, finding~2 fails, since the recognized build's frozen settings write none, and closing deletes any listed path (a vendored library's cache of compiled bytecode, say) from the closed copy before hashing. A written file above $C_{\mathrm{blob}}$ is compared by hash only: equality passes at any tier, and inequality fails at $\mathsf{T0}$ and is otherwise unevaluated for that file, since the bytes a looser tier would need were never stored.

The re-run also fires the capture code, and its records are compared with $Q$'s records for that step, matched by position within the step, at the same tier. A differing count of records fails the step, as does a record that differs at the tier, even when the step's other outputs agree. At $\mathsf{T0}$ only byte identity passes. At $\mathsf{T2}$, the outputs and records the closure runs agreed on are judged at $\mathsf{T1}$-reported where the schema parses them and by normalized hash otherwise, because instability in one output of a $\mathsf{T2}$ deposit suggests that the others may vary more than one pair of runs showed; the write-set and record-count rules still apply. Only for an output the pair disagreed on does the check return neither a pass nor a failure. A disclosed gap is one draw of a random quantity, and a fresh draw, measured against the certified output, exceeds any fixed multiple of it at a computable rate (about 30\% for twice the gap, under Gaussian noise), so any threshold set at a fixed multiple of the disclosed gap would fail honest outputs at a known rate. The check instead reports the fresh gap beside the disclosed gap, and, for a file compared by hash, the fresh hash beside the two disclosed hashes. A drawn step ends in the worst state among its outputs and records, in this order from worst: \emph{failed}; \emph{unevaluated}, with a reason code such as environment mismatch (the precondition that follows), a declined draw, or an oversize file at a looser tier than $\mathsf{T0}$; \emph{re-executed against the disclosure}, at $\mathsf{T2}$, neither a pass nor a failure; and \emph{replicated}, which is a pass. If $\tau$ is unmeasured, no replication check is possible, and the check has already failed (finding~5).

Fourth, the precondition. The tier was measured as two runs on the author's machine, so it describes one environment: the numerical and mathematics libraries, and the processor they run on, not the analysis, are what usually move last digits between machines (\Cref{sm:frozen}). A re-executed step is therefore judged at $\tau$ only when the re-executing environment matches $\kappa$ on the schema's \emph{numeric-critical fields} (\Cref{sm:capsule}): the operating-system family, the processor architecture, the language runtime's version, the numerical library's implementation and version, the C mathematics library, the numerical library's reproducibility switch (\Cref{sm:frozen}), where it has one, the container image's digest, when one is deposited, and the graphics processor's model, driver version, and numerical-software version, when $\kappa$ records them. The architecture, the reproducibility switch, and the graphics processor's fields are included because numerical libraries run different code on different processors, so identical versions can still differ in the last digits. For an output the schema does not parse, compared by hash at any tier, the precondition also requires the versions $\kappa$ records for every add-on library the writing step loaded, and the runtime's exact build; otherwise that output is unevaluated (environment mismatch), never failed. The step's parsed outputs and $Q$ records are still judged under the numeric-critical fields alone. A step whose environment fails the numeric-critical fields is unevaluated (environment mismatch) as a whole, never failed, like every finding that cannot be evaluated (\Cref{sm:check}). Without this precondition, an honest $\mathsf{T0}$ certificate re-run on a different numerical library would be declared not replicated for a reason the tier never measured, creating the false disputes the protocol exists to end.

The precondition is met most easily by a deposited container image (\Cref{sm:container}), which is what keeps re-execution possible years later; without one, the reader runs the vendored libraries and must match the other numeric-critical fields on their own machine. When one is deposited, the closure runs and the certified run execute inside it, and $\kappa$'s numeric-critical fields also record the image's content digest, so matching also means running the same image. A build recipe that lists every package and its exact source, as the Guix system writes, goes further: it lets a future reader rebuild the environment from source instead of relying on a stored image \citep{vallet2022toward}. Support for using such a recipe instead of vendored libraries is deferred to version~2, along with library fetching (\Cref{sm:audit}). A proprietary runtime, Stata, cannot be included in a deposited image, so matching $\kappa$ there depends on the reader's own license.

Finally, a re-run under the closure seed, which tests the closure report. A \emph{seeding idiom} is any construction in code that sets a seed; one the reseed patterns do not match can make the outputs independent of the archive's seed (\Cref{sm:limits}), and its first clue is an estimate that did not move between $s_0$ and $s$. The $s_0$ values, however, come from the author's own closure report, so the clue is evidence only if that report is a run of these files. A reader tests whether it is, and so obtains the second clue of \Cref{sm:limits}, by re-executing a drawn step under the step seed derived from $s_0$ and comparing its records with the closure report's. The test is valid when every entry of $M_i$ is an entry of $M_0$, so that no earlier step wrote the step's inputs, and the step then runs alone in the tree rebuilt from $M_i$ (the rebuild, above), without any file of $P^\ast$ that an earlier step deleted; otherwise the reader first re-runs the earlier steps under $s_0$ from a fresh copy of $P^\ast$. A mismatch then shows that the closure report is not a run of these files. The exception is a file that a later step rewrote: the report holds its value at the end of the run, not after step $i$, so that file is reported unevaluated.

\paragraph{Detection probability.}
How likely a sample is to catch a fake depends on how much was faked, so a reader can choose $k$ knowingly. The chance that a random sample includes a faked step follows the hypergeometric distribution, which describes sampling without replacement, as in survey sampling. Write $m$ for the number of \emph{faked} steps among the $n_c$ replayable ones, steps whose re-execution would fail a comparison the tier can judge. (The letter is reused: elsewhere $m$ is a signed message, \Cref{sm:binding}.) No comparison can judge an output kept only as a hash, one larger than $C_{\mathrm{blob}}$, unless $\tau$ is $\mathsf{T0}$, nor, at $\mathsf{T2}$, an output the closure pair disagreed on. A fake in a step that is \emph{not} replayable escapes sampling altogether, which is what the coverage fraction $n_c/n$ discloses. The chance of drawing $k$ steps all among the $n_c-m$ honest ones is $\binom{n_c-m}{k}/\binom{n_c}{k}$, so for $k\le n_c-m$ the probability of drawing at least one faked step is
\begin{equation}\label{eq:hyper}
1-\frac{\binom{n_c-m}{k}}{\binom{n_c}{k}},
\end{equation}
and it is $1$ for $k>n_c-m$, since a sample larger than the honest remainder must include a fake. A single faked step ($m=1$) is caught with probability $k/n_c$. A dataset driven by one master script has one step, $n=1$ (\Cref{sm:run}), so re-execution is then all or nothing; the arithmetic matters for deposits with several scripts.

Isolated fabrication is hard to catch, and broad fabrication is not. With $n_c=20$ replayable steps and one faked, re-running $k=4$ catches it with probability $4/20=0.2$, and even $k=10$ gives only $0.5$; certainty of drawing a single faked step requires re-running everything ($k=n_c$). With $m=5$ of $20$ faked, $k=4$ already catches at least one with probability $1-\binom{15}{4}/\binom{20}{4}\approx 0.72$. The deterrent does not depend on any particular $k$: an author deciding whether to fake a step cannot know whether anyone will draw, how many steps a reader will draw, or, if the reader draws before looking at the step durations, which ones. The durations are what an author can use against a reader who looks at them before drawing, and a target list padded with trivial steps lowers the chance that $k$ uniform draws include any one step (\Cref{sm:limits}). \Cref{eq:hyper} is the chance that the sample \emph{includes} a faked step, which is a detection only when the drawn step is judged: the reader's environment matches $\kappa$, and the compared output is one the record permits comparing. The formula is thus an upper bound on the chance that one reader's sampled re-execution detects a fake, given that the reader re-executes; it is not an estimate of how much fraud exists or of how many readers check, and with $n_c=0$ it is zero.

\subsection{The existence check: what the journal checks}\label{sm:article}

The certificate exists so that the journal need not replicate anything, and the journal's work, the existence check (the Check part of \Cref{sm:how}), is correspondingly short. This subsection gives, in order, the journal's duties, which are look-ups; the citation the article prints; and finally, the rules for comparing printed numbers with certified ones, which the copyeditor or author applies during copy editing.

\paragraph{The journal's duties.}
The journal keeps its two existing duties and gains one look-up. (1)~It accepts the paper tentatively, on condition that the author deposits the replication dataset in the archive and cites it in the article. (2)~It confirms that both were done: the published version exists, and the printed citation names it. (3)~If it requires certification, it also confirms that the cited version carries one certification metadata field, written by the archive itself, linking to a certificate that carries no exception mark (\Cref{sm:exception}) and that the archive shows as sealed ($W$ signed, $\pi$ pending) or confirmed ($\pi$ confirmed) (\Cref{sm:dataverse}). The archive must be one whose trust root the recognized certificate software carries (\Cref{sm:binding}); a look-up on any other archive's webpage does not meet this duty. Then it formally accepts the paper. The journal runs no code and chooses no tier, since the protocol fixes what a certificate must meet (\Cref{sm:exception}); why these look-ups suffice, and what catches a counterfeit webpage, is argued in the Check part of \Cref{sm:how}.

During copy editing, the copyeditor or author also places the manuscript's numbers beside the certificate's, the number comparison of the last unit of this subsection, a task of reading rather than computing; without it, the journal would be taking the author's word that the printed numbers are the certified ones. The checks of curation no re-run ever covered, such as the provenance of raw data and the rights to it, remain an editor's \citep{christian2018operationalizing}. One choice remains the journal's, with a default that needs no written policy, and one further look-up is suggested. The choice is how to read the seed sensitivity (\Cref{sm:sensitivity}). The default reads it as evidence about movement and pairing, not as a rule for sending a manuscript back: version~1 sets no threshold at which movement across seeds refuses certification or returns a manuscript. A journal that cares about movement across seeds only for the numbers the article prints finds those numbers in the number comparison and asks an editor whether the article reports them adequately. The further look-up is the landing page's counts of deposits that share bytes or group hashes with this one (\Cref{sm:dataverse}): a certificate whose deposit shares its bytes or a group hash with other deposits, the uploader's or anyone's, may be one of a series of deposits each made to draw a new seed, the paid seed shopping of \Cref{sm:limits}, and a journal may ask why.

\paragraph{Citation.}
The article prints the usual citation of the replication data, in the form the archive already generates; only the final word is new:
\begin{quote}
Smith, Jane, 2027, ``Replication Data for: The Effect of X on Y,''
\url{https://doi.org/10.7910/DVN/AB1CD2}, Harvard Dataverse, V1.
Certified.
\end{quote}
The DOI with its version, ``V1'', names one frozen version, the \emph{version of record}, which holds $P^\ast$ and the metadata field linking to $\rho$. The field also records $H(\mathrm{enc}(\rho))$, the hash that $W$ names; the check uses it, and the article need not print it. Dataverse issues one DOI per dataset, and the citation selects a version explicitly, so the check resolves the pair, DOI and version, through the archive's interface, never the bare DOI (\Cref{sm:dataverse}). The DOI alone, without the version, resolves to the latest version and is not this citation, because a later version can link a different certificate. The word ``Certified'' is a convenience for readers, not a cryptographic statement; the authoritative record is the metadata field on the cited version, which anyone can confirm by opening that version.

\paragraph{The number comparison.}
The certificate binds code and data to output files and logged estimates, not to the printed article. Authors must take the numbers they report from the certified run (\Cref{sm:run}), and the copyeditor or author checks that by placing certified values beside printed ones, number by number, citing the certified record for each match. Software may aid the comparison, a language model included, with the person who signs off responsible for it. Tables are compared as numbers, under ordinary rounding to the digits the paper printed. Figures are compared through the data that drew them, which enter $o$ when the code writes them to a file, or $Q$ through the capture command (\Cref{sm:instr}); a figure drawn from data never written has no certified counterpart. A disagreement is a question for a person, never a failed finding, and comparing pictures by a similarity score is outside this protocol.

\section{What certification guarantees}\label{sm:guarantees}

A certificate guarantees exactly what its check verifies (\Cref{sm:checking,sm:claims}), and this section states the guarantee by its limits, in three parts. \Cref{sm:limits} states what certification does not cover, from the scope of version~1 to the attacks only re-execution catches. \Cref{sm:calculus} then computes what evading the protocol costs an attacker, so that every surviving attack but one, which it names, has a stated cost, detection rate, or displayed bound. Finally, \Cref{sm:levels} names the evidence a later version may attach to a certificate, and why version~1 attaches none.

\subsection{What certification does not cover}\label{sm:limits}

The certificate does not catch every way a result can be wrong, as the main paper summarizes (\Cref{sm:m-uncovered}), and against every limit stated here a future reader can still redo the study from scratch. This subsection has five parts: first, what version~1 can certify; second, what the certificate makes no claim about; third, the residual risks; fourth, how the certificate's claims age; and finally, the falsified records that pass the record check but not re-execution.

\subsubsection{Version-1 scope}
The constants and lists of version~1 together bound what can be certified. The tree must fit in one gigabyte ($C_{\mathrm{size}}$, \Cref{sm:preflight}) and in $C_{\mathrm{files}}$ separately hashed entries, a folder of many small files counting as one entry once the author lists it as a packed folder, which may hold no code and which the run may not write into (\Cref{sm:unit}). Code is recognized by the schema's complete list of extensions, so an analysis written in a language not on the list cannot be certified; the author-supplied order of targets (\Cref{sm:entry}) accommodates an unusual \emph{structure}, not an unlisted \emph{language}. Analyses that depend on graphics processors (\Cref{sm:frozen}) will often measure $\mathsf{T2}$ and stop on the ordinary route (\Cref{sm:choices}). Capture covers the schema's lists of estimation commands, so a hand-written estimator yields few records in $Q$ unless the author routes it through the capture command (\Cref{sm:instr}), and the count of records in $\rho$ discloses how many there are (\Cref{sm:run}); a journal may ask why when that count falls far below the number of estimates the article's tables report. Data an archive may not hold are certified with the weaker guarantees of \Cref{sm:m-confidential} (\Cref{sm:confidential}). A saved intermediate result is checked by a sample only when it is a service's answers or a declared item computation; any other, such as one trained model, is certified only from the saved file forward (\Cref{sm:spot}).

Many replication datasets deposited today would not meet these bounds as they stand---most do not even run (\Cref{pfail})---but that is a fact about today's practice, not about the price of certification. Authors build replication datasets to whatever standard publication requires. Of the four levels of enforcement in \Cref{sm:problem}, the two that journals back with a check, deposit as a condition of publication and re-execution, were met once required, while the one backed by no check, data on request, was not. Many of the failures on record are fixable: automated cleaning of paths and libraries alone raised one corpus's execution rate from 25\% to 40\% \citep{trisovic2022large}. These are rates of success per file, excluding files stopped at the time limit. Closing is a diagnostic as much as a gate: preflight, the audit, and the closure report name exactly what to fix, on the author's own machine, before anyone else's time is spent. A replication dataset built to be certified is a replication dataset built to run, which is what journals already ask for; certification makes the standard mechanical and gives authors the tool that checks it.

\subsubsection{Out of scope}
The certificate makes no claim about three things, the first two of which the main paper summarizes (\Cref{sm:m-uncovered}). First, a fabricated result written into the recorded scripts, which replicates on every re-run. Second, a specification that does not answer the paper's question, including \emph{$p$-hacking}, trying many analyses and reporting only the favorable one, inside an honest run. Finally, the protocol does not provide \emph{confinement} of the run to an isolated environment, a sandbox that cannot reach undeclared files. No form of \emph{cross-platform isolation}, the same sandbox on every operating system, exists that runs without administrator rights and behaves the same on the major desktop systems, and a guarantee available to only some authors would make the strength of a certificate depend on the author's operating system.

\subsubsection{Residual risks}
The residual risks (\Cref{sm:audit}) fall in five groups: first, what recording can miss; second, what the author can still influence; third, what the clock allows; fourth, paid seed shopping; and finally, what identifiers and the ledger do not promise.

\paragraph{What recording can miss.}
The first group comes from how a run is observed. The dependency audit finds file reads by fixed text patterns, so it can miss a read, or resolve a name to a different file than the runtime does. The causes are \emph{encoding}, how characters are stored as bytes (UTF-8 and Latin-1 can make the same letters different files); \emph{case-folding}, treating ``File'' and ``file'' as one name, as some disks do; symbolic links; \emph{macros}, text that expands into other text before the program runs; and generated code, including a file path built only at run time, for which process observations and re-execution are the remaining evidence. Process observation is sampled, and a process that makes itself independent of the program that launched it, so that it no longer appears among the step's descendants, escapes it. Nothing outside the author's code can freeze the order in which a filesystem lists a folder, so an analysis that iterates over a folder without sorting it can differ between machines even at a tier the closure runs earned on one; in reproducible-builds practice this is one of the most common causes of identical code giving different output \citep{lamb2022reproducible}. The gap between searching the scripts and running them is narrowed, by recorded hashes, per-step lists of files, and a re-run of the audit, but not closed.

\paragraph{What the author can still influence.}
The second group is the author's \emph{levers}, the choices by which the author can still influence the certified result or what the certificate dates, in six parts: first, the tier; second, the seed; third, work done before deposit; fourth, the capture files; fifth, what sampling reaches; and finally, the declared environment and the target list. The certificate discloses or bounds every lever but one, and the exception is named where it arises.

First, the tier. The tier is measured on the author's own machine before deposit, so deliberately unstable code earns a looser tolerance for later re-execution. The kept pair is fixed before the seed exists, and the count of attempted pairs is a self-report (\Cref{sm:receipt,sm:cruns}). Retrying gains the author one thing: a pair that agrees by chance certifies an analysis whose stability the tier overstates, so the recorded tier is the best of the pairs attempted, and a later reader's re-execution can reveal the overstatement. The ordinary route forecloses only $\mathsf{T2}$, and the tier itself discloses what remains. At any tier looser than $\mathsf{T0}$, residual nondeterminism under the one seed also lets an author run the certified run's scripts privately more than once and submit the attempt they prefer; the certificate software does not select among attempts (\Cref{sm:run}), but no software records the author's other runs, and the disclosed tier bounds the exposure.

Second, the seed. Where the re-mapping of \Cref{sm:audit} does not reach, the author's code can override the archive's seed, making the outputs independent of $s$ and reopening seed shopping. Stata's built-in \texttt{set seed} and the seed options of its resampling commands are disclosed and marked, so for them the audit makes the override visible rather than impossible. A seeding idiom not on the schema's list is not disclosed, and it is the lever the certificate neither discloses nor bounds. Its first clue is a stochastic estimate that did not move between the two seeds, a diagnostic a person reads off the values in the closure report, which is the author's own report. The second is re-execution under $s_0$ compared with the closure report, which turns the first into evidence for a step whose inputs no earlier step wrote (\Cref{sm:reexec}). Both clues fail against code that detects which seed it was given, for example by comparing its first random draw with one stored from a closure run, and sets its own seed only when the seed is not $s_0$. Such code matches no pattern, and only reading it finds it. The audit also cannot read code that a data file carries into the run, such as an R \texttt{.rds} file or a Python pickle a script loads, or code in a compiled file. $\rho$ lists each compiled file with the registry digest it matches, or none (\Cref{sm:audit}), and a launched program the startup hook cannot reach marks the certificate (\Cref{sm:run}).

Third, work done before deposit, which the window does not date when an analysis has no random draws (\Cref{sm:window}). Even with randomness, an analysis cheap enough to run about two billion times can be computed before deposit under every one of the $2^{31}-1$ possible seeds (\Cref{sm:binding}) and run again afterward. That gives the author no choice of seed, since the seed is still the archive's, but again the window dates only the record.

Fourth, the capture files. Author code that writes into a capture file's path is disclosed by the audit's capture-path family (\Cref{sm:audit,sm:instr}), and closing stops (\Cref{sm:choices}). A write the patterns miss remains, but a counterfeit record planted that way reproduces on every re-execution, which makes it fabrication written into the code, out of scope.

Fifth, what sampling reaches. Files above $C_{\mathrm{blob}}$ are compared by hash only (\Cref{sm:reexec}), a declared dependency above $C_{\mathrm{blob}}$ is bound only by its hash, so its bytes are not in the archive's custody, and a step that is not replayable escapes sampled re-execution; the coverage fraction and the oversize list in $\rho$ disclose these.

Finally, the declared environment and the target list. $\kappa$ is declared evidence (\Cref{sm:capsule}), so a $\kappa$ naming an exotic or false environment makes every later re-execution unevaluated (environment mismatch), which lowers the number of readers who can detect a fake (\Cref{sm:calculus}), an effect the check cannot price. That is why $\kappa$'s numeric-critical fields appear on the certificate, where a journal or reader can see them and treat an unmatchable environment as they would a loose tier (\Cref{sm:reexec}). The target list is also the author's (\Cref{sm:entry}), and with it $n_c$, the denominator of every detection probability in \Cref{sm:reexec}: a single master script makes re-execution all or nothing, and a list padded with trivial steps raises $n_c$, so a reader's $k$ draws are less likely to include any one step. The per-step durations in $T$ disclose both.

\paragraph{What the clock allows.}
The third group concerns the dates. The run is no earlier than the time of the block the public random value names, and the protocol guarantees that this time is at most a day before the receipt's timestamp proof (\Cref{sm:window}); while the receipt's proof is pending, even that one-day guarantee rests on the archive's duty to confirm it in time rather than on evidence.

\paragraph{Paid seed shopping.}
The fourth group is seed shopping. The protocol stops seed shopping that costs nothing, re-running under seed after seed before choosing one, and leaves open shopping that costs a new deposit per draw (\Cref{sm:claims}). Unchanged bytes under an unchanged predecessor replay the same seed, so a fresh draw requires \emph{perturbing} the dataset, changing it slightly, or declaring a different predecessor under a later public random value (\Cref{sm:rev}); since a predecessor must be an earlier deposit of the same uploader, that route gives at most one draw per earlier deposit. Either route is a new deposit event, which the ledger records, and an author can sample perturbations until one gives a congenial draw. Each such deposit is individually honest and passes every finding, so the remaining defenses lie outside the check. The ledger records the trail of deposits that share bytes or group hashes, though no single certificate discloses the earlier deposits the author did not use. \emph{Rate limits}, how often an author may deposit, are an operational matter outside this protocol.

\paragraph{What identifiers and the ledger do not promise.}
The fifth and last group concerns identity, the archive's identifiers, its keys, and its ledger; the main paper summarizes it (\Cref{sm:m-uncovered}). Depositor identity is exactly as strong as the archive's registration of accounts. A display may show only a prefix of $\mathit{id}$ (\Cref{sm:dataverse}), but a colliding prefix forges nothing, because the signed documents bind the full $\mathit{id}$ and $H(P^\ast)$. The archive's online secrets are the private halves of its signing key and its seed key, $k_{\mathrm{priv}}$ and $k_{\mathrm{seed,priv}}$, the trust root's private key staying offline (\Cref{sm:dataverse}), and losing one has different consequences from losing the other; neither loss is a break of standard cryptography. A stolen signing key produces internally consistent certificates whose one sign is their absence from the ledger, which the ledger finding, with the key log's intervals of validity, exposes (\Cref{sm:verify,sm:dataverse}). A leaked $k_{\mathrm{seed,priv}}$ makes future seeds predictable, allowing seed shopping without a trace, at the cost of one seeded run per candidate. Rotation through the key log bounds how long either leak lasts. The worst case is a compromised signing service, the one running program that issues receipts and witnesses (\Cref{sm:binding}), that also controls the program that appends to the ledger, and so can append its forgeries to the ledger and pass the ledger finding too. Even that compromise cannot backdate: the timestamp proofs it needs can be obtained only at the real time, so its certificates carry dates from the period of compromise. Nor can it undo its own trail, since anchored entries cannot be quietly removed. Recovery is disclosure, a rotation in the key log, and a public accounting of the period of exposure, with every certificate from that period judged again against the log's intervals. The ledger's two limits, on split views and on forks abandoned within one anchoring interval, are stated with the anchor (\Cref{sm:anchor}). Finally, the DOI alone resolves to the latest version, which need not be the cited one (\Cref{sm:article}).

\subsubsection{Obsolescence}\label{sm:obsolete}
The two claims a certificate supports, record consistency and replication (\Cref{sm:checking}), age at different rates, as the main paper summarizes (\Cref{sm:m-uncovered}), and the record preserves what a later replication needs. Record consistency is arithmetic over stored bytes, so its check works the same in fifty years as today (the Check part of \Cref{sm:how}). A reader can still establish which files the run started from and wrote, which scripts ran and in what order, the environment the author declared in $\kappa$, and every logged estimate: to full precision for estimates read from memory in R and Python and for Stata estimates the capture command wrote; to the printed digits for Stata estimates read from the log. The dates rest on the public Bitcoin record rather than on anyone's memory. Replication decays, because re-running the recorded steps needs the recorded environment, and operating systems, statistical packages, and eventually processors go out of service. Two things slow the decay, one gap survives them both, and three facts about hashes, headers, and signatures complete the account.

The first is that the record changes what a late replication requires, from reconstructing what was done to translating a known computation. What defeats a replication decades later is usually not dead hardware but not knowing what was done, and a certified deposit is a detailed description of the computation together with its expected answers in machine-readable form. A later reader rewrites the analysis in whatever software then exists and compares the result with $Q$ number by number, at a tolerance the reader chooses, since the certified tier describes the recorded environment, not a reimplementation (\Cref{sm:reexec}). The certified seed does not carry over to the rewritten analysis, since seeded random streams are not portable even across versions of one runtime, so a reimplementation can match stochastic quantities in $Q$ only up to their movement across seeds (\Cref{sm:sensitivity}). The second is \emph{emulation}, software that imitates one computer on another, which outlives hardware; programs from the 1960s run in web browsers today. The frozen settings remove the dependence on core count, time zone, and item order that would make emulated replay differ, and $\kappa$ names what to emulate. The gap neither covers is a proprietary statistical package whose license and installer are decades gone, an archival and legal gap no record fills. There the safeguard \Cref{sm:limits} opens with, redoing the study from scratch, remains, and the record still serves it by preserving what the answer was.

Two duties of the archive, and one property of signatures, complete the account of aging. First, the hashes themselves age: if SHA-256 is ever retired, every binding here retires with it. So, while SHA-256 is still secure, the archive periodically re-timestamps its stored evidence under a successor hash; and if Bitcoin stops being maintained, it re-timestamps the evidence in a successor public record. That is the standard renewal procedure of the Evidence Record Syntax, an Internet standard for keeping archived timestamps valid as algorithms age \citep{rfc4998}, in the line of work begun by \citet{haber1991timestamp}. Second, storing the Bitcoin block headers beside the proofs removes the last dependence on a live service. The whole chain of headers is 80 bytes per block, under a hundred megabytes today and growing by about four megabytes a year, stored once for all certificates. A lone header's proof of work shows only that one block's work was once done, so it is the stored chain, each header naming its predecessor, that lets the fifty-year check run with no network. A stored chain is evidence only through the block hash the reader pins beside the trust root (\Cref{sm:binding}). Finally, signatures age differently from hashes: Ed25519, unlike SHA-256, can be forged with a large quantum computer, and a timestamp confirmed before the break is the mitigation, since it shows that the signed bytes existed while forgery was infeasible. For the witness, that timestamp is the anchored ledger head covering $W$'s entry (\Cref{sm:ledger}).

\subsubsection{Falsified records}\label{sm:falsified}
Short of fabrication written into the code, which is out of scope, an author who commits fraud and still passes the record check must falsify the record itself, and every way of doing so leaves untouched the stored starting files of the first step whose record is false (\Cref{sm:m-limits}). Re-executing that step from those files, with a genuine copy of the certificate software in an environment that matches $\kappa$, reproduces the honest values, so the falsification is caught whenever a reader draws and re-executes the step and the step is judged (\Cref{sm:reexec}). The ways fall in three groups, as in the main paper: first, changing the certified run while it runs; second, editing the record before the archive seals it; and finally, running a modified build of the certificate software.

First, changing the run while it runs. A value \emph{planted} in the computer's working memory during the one certified run (with a debugger, say), without changing any file, passes the record check and the number comparison, because that run really did write the planted number; to escape detection, the plant would have to recur on every reader's re-execution, on machines the attacker does not control. Code in a startup file that the launch flags do not skip (\Cref{sm:audit}) changes the run the same way. So do privileged control of the computer, meaning administrator rights used to change what the certificate software records, a malicious \emph{kernel} (the core of the operating system), or a compromised \emph{boot chain} (the software that starts the computer); and a hardware adversary, using \emph{direct memory access} (a device that reads memory without the operating system), \emph{firmware} (software built into hardware), or physical access to memory.

Second, editing the record before the seal. The author can edit the record before submitting it, and a colluding archive can sign a falsified record and add it to its ledger.

Finally, a modified build of the certificate software that reports the hash of a recognized build (\Cref{sm:deposit}) can do either: change what the run records while it runs, or write a false record afterward.

\subsection{The attacker's calculus}\label{sm:calculus}

This subsection prices each deliberate attack \Cref{sm:limits} leaves open, using one inequality from the economics of crime, and so quantifies, where it can, the risk of detection in the main paper's definition of evidence (\Cref{sm:claims}). Fraud is the rare case: the failures certification would remove most often (\Cref{pfail}) are errors, such as moved files, absolute paths, and missing libraries, the ``silly'' mistakes that make replication ``all but impossible'' \citep{gentzkow2014code}, and closing names them so the author can fix them before deposit (\Cref{sm:limits}). This subsection concerns what remains that is deliberate. The inequality says that an offense is worth committing only when its payoff exceeds its cost plus the chance of being caught times the penalty \citep{becker1968crime}; inspection games and spot-checks of volunteer computing analyze the same trade-off \citep{avenhaus2002inspection,sarmenta2002sabotage}. The protocol cannot set the payoff or the penalty, which are facts about careers and institutions, but it sets the other two terms, and they can be computed rather than described.

This subsection has three parts: first, the deterrence inequality and the claim it carries; second, the surviving attacks priced in one table, together with two facts the table cannot show, that, for attacks that record consistency catches, the yearly chance of detection does not decline with time, and that detection by re-execution accumulates across readers; and finally, what stays qualitative, and whose choice each qualitative term is.

\paragraph{The deterrence inequality.}
Write $V$ for the value to the author of the fraudulent result standing published, $S$ for the sanction if the fraud is exposed, $\mathrm{cost}(a)$ for the price of mounting attack $a$, and $\mathrm{det}(a)$ for the probability that $a$ is ever detected. The \emph{deterrence inequality} says that mounting $a$ is rational only when
\begin{equation}\label{eq:deter}
V - \mathrm{cost}(a) \;>\; \mathrm{det}(a)\,S.
\end{equation}
This subsection claims that every attack \Cref{sm:limits} leaves open short of fabrication written into the code, except one named below, meets at least one of three conditions, which we call the \emph{arms} of the claim. First, $\mathrm{cost}(a)$ has a lower bound counted in runs of the analysis, deposits, or block intervals. Second, $\mathrm{det}(a)$ rises with each reader's re-execution at a computable rate. Finally, the attack's reach is bounded by a quantity a reader can see beforehand and discount: on the certificate (the tier, the coverage fraction, the oversize list, $\kappa$'s fields), on the landing page (its counters), or in the key log (the keys' intervals of validity). One residual risk satisfies no arm: a seeding idiom not on the audit's list, which detaches the outputs from $s$ without disclosure. It is caught only by the two clues \Cref{sm:limits} names or by reading the code, none at a computable rate. With every other attack on an arm, the only unquantified terms of \Cref{eq:deter} are $V$ and $S$, which were never the protocol's to set. If exposure also forfeits $V$, as a retraction does, the condition for a rational offense becomes $(1-\mathrm{det}(a))\,V - \mathrm{cost}(a) > \mathrm{det}(a)\,S$, which is harder to satisfy than \Cref{eq:deter}; the form of \Cref{eq:deter} is thus conservative, in the attacker's favor.

\paragraph{The surviving attacks, priced.}
\Cref{tab:attacks} puts each surviving attack on its arm. The frauds the record rules out entirely are precisely the cheap ones, so they are not in the table: the edited output table, the swapped data file, and the quietly re-run script each cost nothing today and each break record consistency here, and a manuscript number retyped by hand is caught by the comparison of \Cref{sm:article}. Every surviving attack short of fabrication either costs about as much as the honest work and leaves a ledger trail, or has a detection probability that rises with re-execution, or is bounded by a quantity a reader can see beforehand, with the one residual on no arm. Two rows need comment.

The first is seed shopping, which pays per draw. If the congenial result arrives with probability $q$ under one honest seed, and each fresh seed requires a fresh deposit (\Cref{sm:limits}), the number of deposits until success is geometric with mean $1/q$. An author shopping for an outcome that honest randomness delivers one time in ten expects ten deposits, every one a permanent entry in a public ledger under the author's pseudonym (\Cref{sm:deposit,sm:anchor}). Each deposit costs one run of the part of the analysis that uses the seed, since the deterministic part can be computed once and deposited as data, plus a wait for a new public random value on the predecessor route or a perturbation on the other; the closure runs add no cost, since the closure report is a self-report. Detecting the shopping is a matter of counting, on the landing page and from the ledger snapshots (\Cref{sm:dataverse}).

The second and last is fabrication written into the code. It is outside the certificate's scope (\Cref{sm:limits}), and it must be priced honestly. The check does not catch it, so its $\mathrm{det}(a)$ under this protocol is zero, and what it saves the fabricator is the science itself, the data collection skipped. What the record adds is that the fabricated file's provenance and date are fixed on the day it ran (\Cref{sm:m-limits}), so that the later dispute is a comparison of records rather than of recollections (\Cref{sm:parties}), and that every cheaper fraud now breaks record consistency. The line the protocol draws is therefore that the only fraud the check cannot detect at all is the one that was always outside a replication check.

\begin{table}[htbp]
\centering\footnotesize
\caption{The Surviving Attacks, Priced}\label{tab:attacks}
\begin{tabular}{p{0.2\textwidth} p{0.24\textwidth} p{0.25\textwidth} p{0.2\textwidth}}
\toprule
\textbf{Attack} & \textbf{Cost floor} & \textbf{Detection} & \textbf{Displayed} \\
\midrule
Paid seed shopping: perturb the files, or declare a new predecessor
under a later public random value & one seeded run per draw, plus a perturbation or a
block interval; $1/q$ draws expected; the predecessor route gives at
most one draw per earlier deposit & every draw is a ledger entry;
counted from the snapshots & the landing page's deposit counters \\
Private selection among attempts below $\mathsf{T0}$ & extra runs &
none by the check & the tier \\
A seeding idiom not on the audit's list & none & the clue of an unmoved estimate;
re-execution under $s_0$ where the step's inputs allow; reading the
code & none: the one residual \\
A falsified record (\Cref{sm:falsified}), made by the author or signed by
a colluding archive & one run, or collusion & $1-(1-k/n_c)^N$ for one
faked step, accumulating across $N$ readers & coverage $n_c/n$ \\
An unmatchable environment $\kappa$ & none & re-execution unevaluated &
$\kappa$'s numeric-critical fields \\
Altered items in a saved intermediate file & none & sampled with probability at least $1-(1-f)^{100}$ & the spot-check's counts \\
A fake in an incomplete or oversize step & none & escapes sampling &
coverage $n_c/n$; the oversize list \\
Fabrication written into the code & saves the science & zero under the
check & out of scope; provenance and date fixed \\
A stolen signing key & subverting the signing service (\Cref{sm:binding}) & the ledger finding:
absence from the ledger, unless the ledger writer is held too & the
key log's intervals \\
Seeds computed in advance with the seed key & a seeded run per candidate; collusion or key theft & none & the key log's intervals (theft) \\
An archive choosing its public random value to steer the seed & collusion & none;
the coherence window's candidates, re-draws counted & the
count of superseded receipts \\
\bottomrule
\end{tabular}
\floatnote{For each attack, the columns give its cost floor, the least an attacker must spend, in runs, deposits, block intervals, or what must be subverted; how it can be detected; and the quantity a reader can see, on the certificate, the landing page, or the key log, that bounds it. Each attack short of fabrication written into the code, which is out of scope, has a cost floor, a detection probability that rises with readers' re-executions, or a displayed bound, except one residual risk, whose detection is by inspection rather than at a computable rate. Here $q$ is the chance a single seed gives the result sought, $k$ the steps a reader re-executes, $n_c$ the replayable steps of $n$, $N$ the readers whose environments match $\kappa$, so that their re-executions can be judged, and $f$ the share of a saved file's items an author altered.}
\end{table}

\paragraph{For record consistency, detection does not decline; for re-execution, it accumulates.}
The table cannot show two facts. Write the \emph{hazard} for the probability that a check in a given year catches a fraud not yet caught. The first fact is that, for every attack that record consistency catches, the hazard does not decline with time. Without certification, the chance that a future check catches computational fraud falls as the paper ages: environments drift, deposited code stops running \citep{trisovic2022large}, provenance fades from memory, and the dishonest author's risk declines, so waiting reduces it. The certificate stops that decline: its record and check are permanent, self-contained arithmetic (\Cref{sm:claims}), the durable half of the account of aging (\Cref{sm:obsolete}). For these attacks, such as tampering with a certificate after publication or a certificate the ledger lacks, the hazard next year is what it was this year: the risk of exposure never expires, and keeping it in force costs no reader anything. The hazard differs from the cumulative probability that some check ever catches the fraud, which cannot decrease in any system. For the attacks that only re-execution catches, the hazard holds while the recorded environment survives and then falls with it (\Cref{sm:obsolete}), so $\mathrm{det}(a)$ stops rising; rewriting the analysis in later software preserves the comparison, not the check's re-execution.

The second and last fact is that detection accumulates across readers. \Cref{eq:hyper} gives the chance that one reader's sample of $k$ among the $n_c$ replayable steps includes one of $m$ faked ones; the record is public and permanent, so one reader's sample is not the attacker's whole exposure. Write $N$ for the number of readers whose environments match $\kappa$, so that their sampled re-executions can be judged (\Cref{sm:reexec}). Their samples are drawn independently, so a fake in a single step ($m=1$) survives them all with probability $(1-k/n_c)^N$, which falls geometrically in $N$. With $n_c=20$ and $k=4$, one reader misses the fake four times in five, but ten readers together miss it with probability $0.8^{10}\approx 0.11$. Each of those checks costs its reader the step's running time and nothing else, needs no coordination with the others, and can happen at any time while an environment matching $\kappa$ can still be built. Readers whose environments match $\kappa$ are commonest in the years just after publication (\Cref{sm:obsolete}), so the accumulation comes early.

Four further points qualify the accumulation. First, the falsified records of \Cref{sm:falsified} are the surviving attacks whose detection depends on re-execution alone. A plant in memory costs one run, but the honest value returns on \emph{every} re-execution of the falsified step, so the $\mathrm{det}(a)$ of every falsified record is this same accumulating probability, and nothing the author does later can lower it. A $\kappa$ chosen to be unmatchable lowers it in advance, the lever \Cref{sm:limits} names. Second, readers are likelier to re-execute important results, so $N$ grows with a result's importance, the same importance that makes $V$ large, and under \Cref{eq:deter} raising the payoff raises the probability of paying $S$. Third, that inference is a claim about the behavior of the research community, not a property of the protocol. The certificate makes each reader's check cheap, independent, and cumulative, but $N\ge 1$ still requires someone to check, and a re-run from a deposit takes days of work today (\Cref{pcosts}); the protocol's contribution to $N$ is to reduce the cost of a check to the running time of the steps a reader draws. Finally, a public place to record the results, such as signed reports of re-execution stored beside $\rho$, which would turn private detections into public ones, is a direction for version~2 rather than a rule of version~1 (\Cref{sm:levels}).

\paragraph{What stays qualitative.}
Two kinds of term stay qualitative. The first is $V$ and $S$ (\Cref{eq:deter}): whether a caught fabrication ends a career is set by journals, universities, and funders. The second and last is one choice of the archive's: which of the day's blocks it names as the public random value in $\gamma$ (\Cref{sm:pulse}), a choice that can steer the seed among the candidates, and that no finding detects. That choice, with the seed key's power to compute a seed in advance, is the protocol's one assumption of trust in the archive (\Cref{sm:claims,sm:tools}). Three things limit that trust. First, the coherence window leaves an archive that steers the seed about 144 candidate blocks, and so 144 candidate seeds, per receipt (\Cref{sm:deposit}). A reissue for cause, which is a further draw, leaves a superseded receipt that the ledger and the landing page attribute to the archive, so re-draws beyond the window are counted rather than prevented. Second, shopping among the candidates serves only an author the archive is already colluding with, and such cooperation would buy far more than a seed (\Cref{sm:limits}). Finally, a reader who doubts a draw can re-run the recorded steps under seeds of their own, the hashes of blocks mined after the certificate's timestamp proof, and read the certified value against their spread. The seed sensitivity the certificate reports (\Cref{sm:sensitivity}) gives one measurement of how far estimates move between two seeds, a rough guide to the spread to expect; it is a diagnostic for a person, since the tier says nothing across seeds, and not a verification of the archive's derivation. The seed proof lets every reader check the derivation itself, though not the archive's choice of block within $\gamma$ or a computation the archive kept to itself (\Cref{sm:binding}). \Cref{eq:deter} divides the work between the protocol and policy: the protocol publishes lower bounds on $\mathrm{cost}(a)$ and rates for $\mathrm{det}(a)$, as formulas in $q$, $k$, $n_c$, and the ledger's counts, and policy supplies the two remaining numbers and reads off whether the inequality can hold. No claim is made that fraud is irrational, since that claim needs $V$ and $S$; the claim is only that every term the protocol controls is now a formula in published parameters, or a quantity a reader can see beforehand, rather than a description in words, except for the one residual risk that satisfies no arm.

\subsection{Evidence a later version may add}\label{sm:levels}

A certificate is the base level of assurance, the same for every author on an ordinary computer, and further evidence can be attached to it without changing what it means (the main paper's \Cref{sm:checking}). This subsection names three kinds and why version~1 adopts none: first, records of re-execution; second, attestations from protected hardware; and finally, proofs of computation.

First, records of re-execution. After a re-execution check, the certificate software could offer to send the result to the archive, which would record it on the ledger under the reader's account and show the count on the landing page: which steps were rerun, in which environment, and the verdict, with a rerun in an environment that does not match $\kappa$ recorded as unevaluated, never as a failure. A detection would then become public and the number of independent checks visible, which strengthens the deterrence of \Cref{sm:calculus}. It needs a new kind of ledger entry and a new duty of the archive, and is a direction for version~2.

Second, attestations from protected hardware. A \emph{trusted execution environment} runs code in a part of the processor that the computer's own operating system and administrator cannot read or change, and the chip signs a report naming the software it ran, into which that software can place the hash of its results. A certified run made inside one could carry such a report, and the certificate software could attach it automatically whenever it runs on such hardware, changing nothing otherwise. For those certificates it would make the falsified records that only re-execution now catches far harder to produce, for as long as the chip maker's keys and firmware hold (\Cref{sm:falsified}). Version~1 does not adopt it, for four reasons. Few authors' own computers offer it; it is available mainly in rented cloud servers. Its evidence ages faster than the rest of the certificate, because a report is worth exactly as much as the chip maker's keys and firmware, which are revised after published attacks. It adds the chip maker as a party whose word the certificate relies on. And a guarantee available to only some authors would make a certificate's strength depend on the author's computer, the reason \Cref{sm:limits} gives for declining a sandbox.

Finally, proofs of computation, which would need no one's word at all, remain far too costly for general data analysis (\Cref{sm:m-related}).

\section{Prior work and our contributions}\label{sm:related}

The main paper's \Cref{sm:m-related} credits the designs, tools, and lines of work the protocol builds on, and says what each leaves open; this section adds the detail that section omits. It has four parts: first, the designs, with TRACE, the nearest, compared in what its trusted party vouches for and in what each design costs to run; second, the borrowed tools and related work; third, the gaps that the precursors for an earlier bound in time leave; and finally, our contributions.

\paragraph{The designs, in detail.}
The main paper's one-sentence accounts of several designs omit these details. Dataverse's Universal Numerical Fingerprint is a hash of a data file's normalized content, the same whatever format stores it \citep{dataverse_unf}. RO-Crate itself records no hashes: its designers leave each file's hash to the manifest of BagIt, a separate packaging format in which an RO-Crate can be wrapped, while Croissant may record each file's SHA-256 hash \citep{soilandreyes2022rocrate,akhtar2024croissant}. Dataverse's developers have worked to connect four of the hosted platforms (Code Ocean, Whole Tale, Binder, and Renku) to Dataverse, since the platforms run code but make no commitment to keep it, while an archive keeps files but does not run them \citep{trisovic2020advancing}. The software \emph{supply chain} is the sequence of build, test, and packaging steps between a developer's source code and the program a user installs; SLSA grades signed records of it by how trustworthy the platform that generated them is, and at every level above the first the reader trusts that platform \citep{slsa}. Cascad serves confidential data through accredited staff \citep{perignon2019certify,hurlin2024role}. Cryptographic proofs of computation have been built for the training of small machine-learning models \citep{garg2023experimenting}.

TRACE and certification differ in what the trusted party vouches for. TRACE's platform is what its designers call a ``transparent research system,'' a controlled computing environment \citep{vilhuber2023trace, mcphillips2019transparent}. Because it ran the workflow itself, it signs as an eyewitness, while the archive, in signing our witness $W$, acts as a notary for the certificate it sealed. TRACE vouches for the substance, that the computation happened as recorded, and its attestation is the evidence, with the hashes, signatures, and timestamps of that attestation exposed for anyone to check. The record's assurance is therefore the platform's continued integrity, and no reader can re-derive the substance without trusting the platform. Here no party vouches for the substance, which rests instead on the open-source certificate software that any reader can test (\Cref{sm:claims}). The archive is a notary, never an observer (\Cref{sm:parties}; \Cref{sm:tools} states the trust placed in the seed, and \Cref{sm:limits} the exposure if the archive's keys are stolen). The seed bounds the run from below in time, which no platform's word supplies.

The two designs also place the cost of running the analysis on different parties. A platform that attests by executing provisions hardware for whatever authors bring, so its costs, like a data editor's, grow with the computation authors choose to run. An archive in our protocol hashes, signs, stores, and timestamps, so it pays per certificate the cost of hashing and storing what was deposited and recorded, and of running the findings it can evaluate at seal. When the timestamp calendars lag (\Cref{sm:binding}), it pays one transaction fee shared across that day's receipts. The runs themselves are paid for by the author, on the author's own machine (\Cref{sm:parties}).

\paragraph{The borrowed tools and related work, in detail.}
Two parts of the design come from Certificate Transparency \citep{laurie2013certificate}: the form of the hash tree, whose \texttt{0x00} and \texttt{0x01} tags make a leaf's hash equal an interior node's only through a collision of $H$, and the Merkle log the ledger uses. In Sigstore, the deployed system closest to the archive's role, independent parties cosign the heads of the \emph{transparency log}, an append-only log like the ledger \citep{sigstore2022,rekor}. A certificate could be exported, for supply-chain tooling, into the format of the internet standard that the main paper cites \citep{scitt2025architecture}, but version~1 defines no such export, and the canonical record is the flat $\mathrm{enc}$ record. The nonce is the output of the verifiable random function of RFC~9381 \citep{rfc9381}, the salts of a restricted deposit are made with HMAC \citep{bellare1996keying}, and the signatures are Ed25519 \citep{bernstein2012high}. Timestamping backed by Bitcoin was built for researchers a decade ago \citep{gipp2015decentralized}, and a consortium of archives now records hashes of deposited research data on a blockchain as a service to its members \citep{wittek2020bloxberg}. The signed record of a step in in-toto, a framework for securing how software is built and distributed, also carries its console output and exit code, as the certificate's does, and its designers call a signed record made for a step that never ran a \emph{fake-check}, which they treat as a consequence of a compromised signing key \citep{torresarias2019intoto}; a signature over a record of work is only evidence that someone with the key said the work happened. The workflow engine Snakemake tracks the code, parameters, inputs, and software of every job it executes \citep{molder2021sustainable}, and Two Sigma's Python framework Memento transforms data incrementally through code while recording its provenance \citep{memento}. The tool noWorkflow records the provenance of an ordinary Python script without a workflow engine, by reading the script's syntax and profiling it as it runs \citep{murta2015noworkflow}. Accountable virtual machines, which replay random segments of a signed, hash-chained log of a machine's execution, build on PeerReview \citep{haeberlen2010avm,haeberlen2007peerreview}, and a check of which data trained a model draws its seed from a hash of the training data and re-runs sampled segments of the training \citep{choi2023tools}. ReproZip and SciUnit record every request a run makes to the operating system, such as each file it opens, then pack the files the run used, programs included, for replay \citep{chirigati2016reprozip,tonthat2017sciunits}; \Cref{sm:audit} says why the static audit does without that completeness. A cryptographic near-namesake, ``proof of replication,'' is an unrelated primitive for storage, which proves that data are held in dedicated copies, as in Filecoin \citep{fisch2018porep}.

\paragraph{The gaps.}
The precursors \Cref{sm:m-related} names for an earlier bound in time each leave one of three gaps: (a) the unpredictable value is tied to named materials only if the user first publishes them; (b) whoever commits also computes the seed, so the author can change the plan and derive the value again; or (c) the bound dates a signature, not a run's draws. Election audits, including their statistical theory and a version with a cryptographic commitment \citep{stark2008conservative,benaloh2011soba}, the committee selection of RFC~3797, Saa and Stern's registration, and the randomness beacon of \Cref{sm:pulse} leave gap (a); a beacon's value is tied to named materials by hashing it together with a description of them, as the guidance for the beacon of the National Institute of Standards and Technology describes \citep{kelsey2019beacons}, and distributed beacons are now deployed at scale \citep{syta2017scalable,drand}. Seeding from a hash of a preregistration leaves gap (b), because nothing stops the author from adjusting the plan and deriving a new seed \citep{mazor2019novel}, and blind analysis leaves it as well, since only the analysts' own discipline enforces the blinding \citep{maccoun2015blind}. The postdated signatures of \citet{abadi2020timed} leave gap (c). Hardware enforcement closes gaps (a) and (b) by confining the analysis inside a \emph{hardware enclave}, a protected region of a processor, which differs from the secure facility of \Cref{sm:confidential}, a place where data may be analyzed but not copied out \citep{meissner2021peqes}.

\paragraph{Contributions.}
The protocol makes ten contributions (\Cref{sm:m-related}). None is new cryptography; each is a new use of standard methods. The last three, (8) to (10), are what make acceptance by an existence check sound.

\emph{(1) Acceptance by existence.} A journal can accept a paper by confirming that a certificate exists (\Cref{sm:checking}; \Cref{sm:article}), rather than by re-running the analysis; \Cref{sm:m-related} cites two recent proposals with the same aim. The certified run already did that work, on the terms of \Cref{sm:claims}: the manuscript must take its numbers from the certified run, anyone can compare them, and anyone can re-run the record. Certification removes outright the data editor's re-execution, with the months of delay and the staff costs of \Cref{pcosts}, and moves the repair work that re-execution triggers to the author, before final acceptance, where it is paid once and most cheaply. A data editor's elapsed months include the authors' own repairs: at the \emph{American Journal of Political Science}, verification takes about seventy-five days on average, the first report alone about twenty-five, and the authors' responses account for part of the rest \citep{ajps2018verification}. The saving extends past the journal: a later researcher, and especially anyone replicating at scale across many papers, can settle record consistency in seconds of hashing instead of re-running every analysis (\Cref{sm:parties}). What certification does not remove is the comparison of the manuscript with the certified record; it makes that comparison mechanical and leaves it to the copyeditor or author during copy editing, and the checks of curation to an editor (\Cref{sm:article}).

\emph{(2) A shared record.} The certificate supplies the checkable record of the original computation whose absence, \Cref{pcosts} showed, turns failed replications into personal disputes; with it, a failed replication becomes a comparison of computations rather than a contest of reputations (\Cref{sm:parties}).

\emph{(3) A record that outlives its software.} A verifying data editor leaves behind a verdict; the certificate leaves the computation itself, still checkable, and still a reference for reimplementation, after the software that ran it has gone out of service (\Cref{sm:m-uncovered}; \Cref{sm:obsolete}).

\emph{(4) Computable deterrence.} The security argument is computed rather than asserted: every term of cost and detection the protocol controls is a formula in published parameters, or a quantity a reader can see beforehand, with one residual risk named as neither. The deterrence it computes requires that some reader checks, which the protocol makes cheap and cannot compel, and it leaves to policy only the payoff and the penalty, $V$ and $S$ (\Cref{sm:calculus}).

\emph{(5) The estimates themselves are bound.} In R and Python, capture code writes every estimate a listed command produces, read from the runtime's memory, into the step's capture file, and in Stata every such estimate the log shows; in every language, a quantity the author routes through the capture command is written there as well (\Cref{sm:instr,sm:run}). The certificate binds those files through the hash of $Q$, so it binds the estimates themselves, not only the files that produced them. The article's numbers are tied to those estimates by the rule that authors report certified values, and by whoever compares the two (\Cref{sm:certify}; \Cref{sm:article}).

\emph{(6) A closed specification.} Nothing that affects what a check computes is left to an implementer's choice, so independent implementations must agree---on the hashes, the seed, and the tier---and must accept each other's certificates (\Cref{sm:version}).

\emph{(7) Evidence of a split view stored with the certificates.} Certificate Transparency catches a log that shows different parties different histories through dedicated monitors and, in its design literature, through gossip among clients \citep{chuat2015efficient,meiklejohn2020think}. Here each certificate's bundle stores its own inclusion proofs under signed, timestamped heads, so two certificates whose heads of equal size conflict prove a split view with no help from the archive, and heads of different sizes oblige the archive to produce a consistency proof (\Cref{sm:anchor}).

\emph{(8) Seed shopping made visible.} Seed shopping (\Cref{sm:claims}) is a specification search preregistration does not prevent, because the analysis never changes. Certification makes it visible. Election audits and seeds drawn after a public registration use the same order of commitment, then randomness (\Cref{sm:m-related}); here the commitment is the archive's receipt of the deposit. A fresh draw therefore requires a fresh deposit, which the ledger records (\Cref{sm:claims}; \Cref{sm:window,sm:rev}). The author's own resets of the seed are re-mapped through $s$ where the language allows and disclosed where it does not, for every reset in the text the audit reads (\Cref{sm:audit}); a seeding idiom not on the audit's list makes the outputs independent of $s$ without disclosure, and that residual risk is stated (\Cref{sm:limits}). Three ingredients, the seed's three inputs (\Cref{sm:certify}), do different jobs. The archive's seed key makes the seed unpredictable to the author and its proof makes it checkable by anyone, and the hash of the deposit inside the nonce context ties it to these bytes. The public random value beside it, which the check confines to the day before the receipt's timestamp proof, makes the receipt impossible to date to any earlier moment, so the earlier bound in time holds even against the archive (\Cref{sm:window}).

\emph{(9) The earned tier.} The test a later re-run must pass needs a threshold that no one chose for their own benefit, since an author who set it would be choosing the standard against which re-runs of the author's own work are later judged. Measurement supplies it: two closure runs before deposit earn the tier (\Cref{sm:ready}; \Cref{sm:cruns}), the closure report is stored and hashed, and the check re-derives the tier from the report rather than reading it off the certificate (\Cref{sm:verify}). Like the capture code, the tier relies on the certificate software being unmodified: a modified build that reports the expected hash is not caught by the record check, though re-execution can expose what it falsified (\Cref{sm:m-limits}; \Cref{sm:deposit,sm:falsified,sm:bytes-capture}). What the author keeps is influence rather than choice: code written to be unstable earns a looser tier. But that trade is visible on the certificate: $\mathsf{T2}$ on the ordinary route stops certification (\Cref{sm:choices}), and a $\mathsf{T1}$-reported certificate states that the analysis reproduces only to within half a unit in the fourth significant figure. Badging schemes, in which a journal or conference prints a badge on a paper whose materials passed a review, by contrast record a mark without a recorded, measured tolerance. The recommended practice that standardizes them places ``the process for verifying results'' outside its scope \citep{niso2021badging}, and the largest deployed regime, the artifact review of the Association for Computing Machinery, vests verification in conference committees \citep{acm2020artifact}.

\emph{(10) The open check.} Acceptance by an existence check is safe only if the check itself is open to everyone, and here it is: a reader who doubts a certificate runs the same calculation as the archive that signed it, needing only the deposited files and the two pinned values (\Cref{sm:checking}; \Cref{sm:binding}). The open check also lets an archive offer certification without running the analysis, which makes adoption by an existing archive practical (\Cref{sm:dataverse}). The property is standard in software supply-chain systems; what is new is having it in a research certificate, since prior systems that verify a published analysis vest the check in a designated person, agency, or witnessing platform \citep{nust2021codecheck,perignon2019certify,vilhuber2023trace}, a category that includes the trusted hardware and signing service of two recent proposals \citep{li2026tee,keita2026nonrepudiable}, and hand the reader that party's verdict instead. It holds for every reader of a public deposit and for every approved reader of a restricted one (\Cref{sm:confidential}).

\renewcommand{\newline}{\unskip\space}
\section{Cryptographic tools used throughout}\label{sm:tools}

This section defines exactly each tool that \Cref{sm:toolkit} named in a sentence, with its byte spellings and the security property it supplies; the main paper credits the tools' sources (\Cref{sm:m-related}). Producing a certificate (\Cref{sm:produce}) and checking one (\Cref{sm:check}) use the same small toolkit, in three groups: the tools that name bytes (\Cref{sm:naming}), the tools that bind a person or a time to bytes (\Cref{sm:binding}), and the tool that folds a list into one hash (\Cref{sm:merkle}). Three preliminaries come first, because they govern every tool and every section of this specification: the schema, which fixes byte-level spellings; the protocol version, which names the edition of the rules under which every signed object was built, together with what lies outside the protocol; and the assumptions on which every security claim rests.

\paragraph{Schema.}\phantomsection\label{sm:schema}
The \emph{schema} is the machine-readable half of this specification: one small tree of files that fixes every byte-level spelling whose meaning this specification explains. It fixes the key names of each signed object (the canonical encoding fixes their order), the grammars of the result capture and of the closure report, the byte rules of normalization, the dependency audit's patterns, the capture code itself, and the per-language lists of estimation commands. The objects themselves are defined in \Cref{sm:close,sm:produce}. Spellings must live in a file because implementations consume them exactly, and a PDF is typeset, not parsed: whether the arrival time's key is spelled \texttt{t\_arr} or \texttt{arrival\_unix} changes every hash in the protocol, since the canonical encoding (\Cref{sm:naming}) writes fields sorted by key name. The schema is named the way every object here is named, by its hash, $h_{\mathrm{schema}}=H(\mathrm{pack}(\cdot))$ over its tree (\Cref{sm:pack}), and it is deposited in the archive under a DOI and version, preserved by the same infrastructure the protocol relies on. The published version of this specification prints that hash here: $h_{\mathrm{schema}} = \langle\text{printed at publication}\rangle$.

\paragraph{Protocol version $v$.}\phantomsection\label{sm:version}
A version number says which edition of these rules a certificate was built under, as the edition of a codebook does; later rules do not apply retroactively. This unit has three parts: first, how the pair of version and schema hash binds every object; second, what it means that version~1 is closed; and finally, what lies outside the protocol.

First, the binding. Every signed object carries its \emph{protocol context}: the pair $(v,\,h_{\mathrm{schema}})$, with the key id of the key that signed it. A reader applies the rules of the declared $v$, and the check requires the pair to match across the receipt, the witness, and the certificate that name one deposit (\Cref{sm:verify}). A certificate that claims $v=1$ with any other schema hash is not a version-1 certificate. Each artifact is thus fixed by the level above it: the schema by $h_{\mathrm{schema}}$ (\Cref{sm:schema}), this specification by its version of record, and every certificate by the pair it names.

Second, closure. This specification is version~1, and version~1 is closed in this text. Every rule that affects what a check computes is fixed here or in the schema, whose hash is printed at publication (\Cref{sm:schema}), and no value is left for an implementer to pick. Until that hash is printed the schema is a draft, and what follows is a claim about the published pair, tested by the conformance vectors the schema supplies. Two independent implementations of version~1 must then compute the same hashes, the same roots, the same seed, and the same tier from the same replication dataset, and each must accept the other's certificates. The current prototype does not yet implement all of version~1; \Cref{tab:profiles} lists its execution profiles beside the full version-1 target. Where this specification names a constant, that value \emph{is} the rule, not a suggested default: $C_{\mathrm{size}}$, $C_{\mathrm{blob}}$, $C_{\mathrm{files}}$, $C_{\mathrm{time}}$, the anchoring interval, $\Delta$, the domain strings, the seed range, the tolerance at $\mathsf{T1}$, the two-hour widening, the depth of six confirmations, and the overlap in recognition of software builds. The rest are the attestation deadline of \Cref{sm:receipt}, the four significant figures of $\mathsf{T1}$-reported (\Cref{sm:cruns}), the seed sensitivity's schema name and pairing-count fields (\Cref{sm:sensitivity}), the fixed build date of \Cref{sm:bytes-frozen}, and the spot-check's sample size of 100 (\Cref{sm:spot}). The domain strings include $\Lambda$, the per-step and per-reseed strings of \Cref{sm:binding}, and $\Lambda_{\mathrm{spot}}$ of \Cref{sm:spot}, and the strings $\Sigma$ of the trace, the manifests, and the ledger (\Cref{sm:merkle}); the schema fixes the complete list, one per construction. Where the specification names a fixed list (the tiers; the fields of $D$, $W$, and $\kappa$), the list is complete as given, and every byte-level spelling those lists leave open is fixed in the version-1 schema (\Cref{sm:schema}, which lists what it fixes). And where it names an outside service (the timestamp service of \Cref{sm:binding}; the source of the public random value of \Cref{sm:pulse}), the named service is the one a version-1 certificate is issued and checked against. Changing any of these is a new $v$, not a local option. Each is defined where its object is introduced, and \Cref{fig:map} maps the objects to their sections.

\begin{table}[htbp]
\centering
\caption{Execution Profiles of the Current Prototype Certificate Software, \texttt{reproai-cert-v5}}\label{tab:profiles}
\begin{tabular}{p{0.28\textwidth} p{0.62\textwidth}}
\toprule
\textbf{Profile} & \textbf{Status in \texttt{reproai-cert-v5}} \\
\midrule
\texttt{run-seed-v1} & Default profile. The run seed is seated at each
target, and the current continuous sensitivity diagnostic uses the first
closure run as its baseline. \\
\texttt{step-seed-q-v1} & Explicit opt-in profile for per-step seeds and
per-step $Q$ artifacts. It is not a default migration. \\
\texttt{python-child-hooks-v1} & Explicit opt-in historical Python
child-hook profile. \\
\texttt{mixed-child-hooks-v1} & Explicit opt-in bounded profile for
Python, R, and Stata child hooks and process-level $Q$ capture. \\
Full version-1 target & Human-readable target specification described in
this paper; not a real implementation profile identifier and not fully
implemented by the current prototype. \\
\bottomrule
\end{tabular}
\floatnote{A profile is a named configuration under which the prototype runs, and a profile name's suffix \texttt{-v1} is the prototype's own numbering, not protocol version~1. Only the last row is the full version-1 specification.}
\end{table}

Finally, what lies outside. Three things a check might be expected to inspect lie outside the protocol, and a check never inspects any of them: how often an archive lets an account deposit (rate limits, \Cref{sm:limits}); the archive's backups, succession arrangements, and landing-page design (\Cref{sm:dataverse}); and any aid a person uses when comparing an output with the printed article (\Cref{sm:article}). Each is named as outside so that silence is not mistaken for latitude. The archive's terms of deposit, the list of recognized builds, and the payoff and sanction of \Cref{sm:calculus} are outside the protocol too, and each is named where it arises.

\paragraph{Assumptions.}
Apart from three limits stated where they arise, the archive's word on account facts (\Cref{sm:claims}), the integrity of a recognized build of the certificate software (\Cref{sm:deposit}), and, for held files, a facility's custodian statement or the author's word (\Cref{sm:confidential}), every security claim in the main paper and this specification reduces to five standard assumptions, two operational premises, and one assumption of trust, collected here. No one can find two inputs with one SHA-256 hash (a collision), or an input with a given hash (a preimage) (\Cref{sm:naming}). The outputs of HMAC-SHA-256, which derives the salts of a restricted deposit, are unpredictable without the key, the property of being a pseudorandom function (\Cref{sm:bytes-salt}). The outputs of the verifiable random function are unpredictable without its private key, and for each input only one output verifies, under the assumptions RFC~9381 states (\Cref{sm:binding}). Ed25519 signatures cannot be forged without the private key (\Cref{sm:binding}). And Bitcoin's blockchain is immutable at depth, meaning that a block buried deep enough in the chain cannot feasibly be altered; its block times are approximately honest; and its block headers are available, since the archive stores them beside the proofs (\Cref{sm:obsolete,sm:binding,sm:pulse}). The two operational premises sit beneath all five: the archive's keys were drawn from a well-seeded, cryptographically secure generator of random numbers, one an outsider cannot predict; and the reader pinned the genuine trust root and one Bitcoin block hash (\Cref{sm:binding}). The one assumption of trust is that the archive does not use its seed key to steer a seed, either by choosing the one field of $\gamma$ it controls or by computing in advance, for an author, the seeds of deposits not yet made (\Cref{sm:deposit}); \Cref{sm:calculus} states what that trust admits and what bounds it. The argument is informal throughout: each claim is reduced to these assumptions in prose, with no formal security games. Where a claim needs less, the text says so, and where an assumption ages, \Cref{sm:obsolete} says what the archive does about it.

\subsection{Naming bytes: hash and canonical encoding}\label{sm:naming}

Three tools name bytes: the hash, which names any run of bytes; the canonical encoding, which turns a record with named fields into one run of bytes so that the hash can name it; and $\mathrm{pack}$, which does the same for a file tree and is defined with the closed copy it lays out (\Cref{sm:pack}). This subsection defines the first two.

\paragraph{Hash $H$.}
A hash is a function that computes a fingerprint of a file's bytes: change even one character, and the hash changes completely. The file cannot be recovered from its hash except by guessing the file and checking the guess, a channel \Cref{sm:confidential} closes for restricted files. No one can find a file that hashes to a given value. And no one can usefully construct \emph{two} different files that share one hash, the property called \emph{collision resistance}, which carries the most weight here, because the author, who supplies the bytes being hashed, is also the party the record must bind. Let $H$ be SHA-256, a standard cryptographic hash, which takes any run of bytes and always returns 32 bytes, written as 64 lowercase \emph{hexadecimal} characters (the digits 0--9 and letters a--f, two characters per byte). $H(z)$ is the hash of a run of bytes $z$; a record $z$ is hashed as $H(\mathrm{enc}(z))$, and a file tree as $H(\mathrm{pack}(z))$, which $H(P^\ast)$ abbreviates (\Cref{sm:pack}). One use recurs throughout: a hash stored inside a later signed or hashed object, so that swapping the hashed bytes makes the stored hash fail to match. The receipt, the witness, and the commitment $x$ (\Cref{sm:objects}) all bind bytes this way.

\paragraph{Canonical encoding $\mathrm{enc}(\cdot)$.}
The \emph{canonical encoding} turns a record with named fields into one run of bytes that every implementation spells identically. This unit has three parts: first, why a rule is needed, and what $\mathrm{enc}$ is; second, what may not appear in a signed object; and finally, how this encoding relates to the standard one.

First, the rule. $H$ takes one run of bytes, from a first byte to a last, as a single file already is. A record with named fields is not a run of bytes until someone writes down a rule for laying it out in one order and spelling every part the same way each time, and two implementations that pick different rules compute different hashes for identical material. $\mathrm{enc}(\cdot)$ converts a structured object to bytes as JSON, a text format of named fields such as \texttt{\{"seed":42,"tier":"T0"\}}. Keys are sorted by Unicode code point, the standard numbering of characters, compared as numbers, so that uppercase letters precede lowercase ones and no locale's order is used. No extra whitespace appears: a colon and a comma have nothing around them. And any character outside ASCII (the basic English letters, digits, and punctuation) is written as a \texttt{\textbackslash u} code, so every implementation spells it the same way.

Second, what may not appear. A \emph{floating-point} value is a number stored in binary scientific notation. No floating-point value may appear in any signed object or in any \emph{hash chain}, a list in which each item carries the hash of the previous one. A signature covers bytes, not the number those bytes were meant to represent, and two implementations that agree on a number may still disagree on the shortest spelling of a float. So times are integer \emph{Unix seconds}, whole seconds since 1970-01-01 00:00:00 UTC, a fixed origin, as Stata's dates count days from 1 January 1960; durations are integer \emph{milliseconds}; and any ratio is stored as the two integers from which it is computed. The order of an array, unlike the order of keys, is part of its value and is never sorted: the commitment's five hashes (\Cref{sm:objects}), the nonce context's four fields (\Cref{sm:deposit}), and the ledger entry's four fields (\Cref{sm:ledger}) mean what they mean by position. The finished document is pure ASCII, so its byte encoding needs no separate rule. Every finer detail of spelling (the escape rules, the case of hexadecimal, the format of integers) is the schema's (\Cref{sm:schema}).

Finally, the standard. This encoding is a deliberately stricter sibling of the standard JSON Canonicalization Scheme \citep{rfc8785}: that scheme admits floating-point numbers, spelled by a rule that picks the shortest spelling that reads back to the same value, and this protocol forbids them outright. Because every key in a version-1 object is ASCII fixed by the schema, the two schemes' different treatments of keys outside ASCII never arise. An implementer must build $\mathrm{enc}$ to this unit and the schema, not to an off-the-shelf canonicalizer.

\subsection{Binding a person or a time: signatures, nonce, seed, timestamp}\label{sm:binding}

Four tools bind a person or a time to bytes: the signature binds the archive to what it signs; the nonce and the seed bind the run's random draws, and so the record, to a moment after the deposit; and the timestamp proof binds bytes to a moment by which they existed. This subsection defines them in that order; the main paper describes their use (\Cref{sm:certify}) and what they establish together (\Cref{sm:claims}).

\paragraph{Signatures $\mathrm{Sign}$ and $\mathrm{Vrfy}$.}
A \emph{digital signature} is a mark that only the holder of a private key can make and that anyone holding the matching public key can check, like a specimen signature on file, except that the mathematics makes forgery infeasible. This unit has three parts: first, the two operations; second, where the archive's keys live; and finally, how a reader obtains the public key.

First, the operations. Under Ed25519, the standard signature scheme used here, a signature is a number of 64 bytes, ordinarily written as 128 hexadecimal characters. Two operations define it. $\mathrm{Sign}(k_{\mathrm{priv}}, m)$ returns the signature of a message $m$, the bytes being signed, under the private key $k_{\mathrm{priv}}$. $\mathrm{Vrfy}(k_{\mathrm{pub}}, m, \sigma)=1$ exactly when the signature $\sigma$ is valid for $m$ under the public key $k_{\mathrm{pub}}$, and $0$ otherwise.\footnote{Ed25519 verifiers are not all alike: published libraries disagree on edge cases (non-canonical encodings of points; verification with or without the cofactor). The disagreement is enough that one library can accept a signature another rejects \citep{chalkias2020eddsas}. The schema therefore pins the criteria of verification (\Cref{sm:schema}), so that two version-1 implementations return the same verdict on the same bytes.}

Second, the keys. The archive's \emph{signing service}, the one running copy of its software, on the archive's own server, that issues receipts and witnesses, generates a \emph{key pair}, a private half $k_{\mathrm{priv}}$ and a matching public half $k_{\mathrm{pub}}$, the first time it starts and again at each rotation (\Cref{sm:dataverse}). It stores $k_{\mathrm{priv}}$, and the private half $k_{\mathrm{seed,priv}}$ of its seed key (the next unit), where only it can read them and never copies either anywhere else; the root key is made and kept offline (the last part of this unit). Producing the archive's signatures then requires subverting that one service or stealing its stored key, not merely working at the archive.

Finally, the trust root. The reader must obtain the archive's key from somewhere the certificate cannot influence. Anyone checking a signature \emph{pins} the archive's trust root: settles on it in advance, obtains it apart from the files being checked, and holds it fixed. The trust root is a root key, kept offline and used for nothing else, in the discipline of key rotation used by software-update systems \citep{samuel2010survivable}. The private half of that key signs the archive's key log (\Cref{sm:receipt,sm:dataverse}). Each recognized build of the certificate software carries the trust roots of the archives it supports and is itself checked against its published hash. For a reader who runs no such software, the root's public key is deposited under a DOI and version beside the key log and the ledger snapshots. A copy of a key shipped alongside the files being checked does not suffice, because a substituted archive would ship the substitute's key, and its forged signatures would then appear valid. One more value is pinned the same way, beside the root: the hash of one Bitcoin block, as recent as the certificate software that carries it. The stored chain of block headers must pass through that block and, beyond it, meet the chain's own difficulty rule header by header (\Cref{sm:obsolete}), so that a forged chain would cost real proof of work rather than the \emph{minimum difficulty}, the least work per block Bitcoin's rules allow, which a forger could otherwise choose. Beyond the pinned block, the difficulty rule proves that work was spent, not that the headers are the public chain, so the pin should postdate the proofs being checked. Each recognized build carries a recent pin. A reader checking proofs newer than their pin gets a passing dates finding (finding~9), which its reason code marks provisional until the online check, or a later build's pin, confirms those later headers (\Cref{sm:verify}).

\paragraph{Nonce $\nu$ and seed proof $\phi$.}
A \emph{nonce} here is a number the author could not have known in advance: the archive makes it with a private key of its own, and the run's seed is then computed from it. The term ordinarily stresses that such a number is used once; here $\nu$ is unique to one combination of bytes, key, and public random value in the nonce context $\gamma$ (\Cref{sm:deposit}). The property that does the work is unpredictability: any document or result that contains $\nu$, or was computed from it, must have come after the archive issued it. The tool that makes it is a \emph{verifiable random function}: a keyed function whose output only the holder of a private key can compute, and which comes with a short proof that anyone holding the matching public key can check. Version~1 uses the suite ECVRF-EDWARDS25519-SHA512-TAI of RFC~9381 \citep{rfc9381}, on the same kind of key as the archive's signatures. The signing service generates the archive's \emph{seed key}, a private half $k_{\mathrm{seed,priv}}$ and a public half $k_{\mathrm{seed,pub}}$, from a cryptographically secure generator of random numbers (\Cref{sm:tools}) when it first starts, and again at each rotation the key log records (\Cref{sm:dataverse}); it uses the key for nothing else, so no other use of the key exists to collide with it. Write $\mathrm{VRF}(k_{\mathrm{seed,priv}}, m)$ for the function's output and proof on a message $m$. The nonce and its proof are
\begin{equation}
(\nu,\,\phi)=\mathrm{VRF}(k_{\mathrm{seed,priv}},\,\mathrm{enc}(\gamma)),
\end{equation}
where $\gamma$ is the four-field record of \Cref{sm:deposit}, the nonce $\nu$ is the function's 64-byte output, and the \emph{seed proof} $\phi$ is 80 bytes; each is written in lowercase hexadecimal when stored in a JSON document. A reader cannot compute $\nu$, lacking $k_{\mathrm{seed,priv}}$, but can check it: a check that also validates $k_{\mathrm{seed,pub}}$, as RFC~9381 provides, accepts exactly one output for a given $\gamma$, even for a key the archive generated dishonestly, so the archive cannot report any other. The archive keeps two freedoms, both stated as the one assumption of trust in it (\Cref{sm:tools}): which recent block the public random value names (\Cref{sm:pulse}), and, since it holds the key, computing in advance the seed a deposit not yet made would receive.

\paragraph{Seed $s$.}
The seed is the number that starts the random draws in an analysis: the simulations, bootstraps, and random splits whose results depend on where the draws begin. The author does not choose it, because choosing would allow seed shopping (\Cref{sm:claims}), which contribution~(8) of \Cref{sm:related} is built to make visible. This unit has three parts: first, who fixes the seed and who can check it; second, the range it must fall in; and finally, the public rule that produces it.

First, who fixes it. The archive's nonce $\nu$ fixes it, and the arrangement has a secret part and a public part, which do different jobs. Making $\nu$ takes the private half $k_{\mathrm{seed,priv}}$ of the archive's seed key, which is why the author could not have known the seed before depositing. Checking $\nu$ takes only the public half, through the seed proof, and turning $\nu$ into $s$ takes no key at all: the rule is public, so any reader holding $\nu$ can recompute $s$ and check it against the seed the certificate reports.

Second, the range. The rule must produce an integer $s$ in $\{1,2,\ldots,2^{31}-1\}$, a range that every runtime running a file type on the schema's extension list accepts (\Cref{sm:dataset}).

Finally, the rule. It hashes the nonce and reduces the result to that range. Write $\Lambda$ for the fixed ASCII domain string (\Cref{sm:audit}) \texttt{certified-runtime-seed-v1|}, of lowercase letters, digits, hyphens, and a final vertical bar. Write $a \,\|\, b$ for the bytes of $a$ followed immediately by the bytes of $b$, and $a \bmod n$ for the remainder after dividing $a$ by $n$, an integer in $\{0,1,\ldots,n-1\}$. Concatenate $\Lambda$ and $\nu$, with $\Lambda$ as ASCII bytes and $\nu$ as the 64 raw bytes of the function's output; hash the result; and take the first eight of the 32 output bytes, written $[0..7]$. Write $\mathrm{BE}_{64}$ for reading eight bytes as a \emph{big-endian unsigned} integer: a non-negative whole number whose first byte carries the highest-order 8 bits, the same convention as writing 1234 with the thousands digit first. Then \[ s \;=\; 1 + \bigl(\mathrm{BE}_{64}\bigl(H(\Lambda \,\|\, \nu)[0..7]\bigr) \bmod (2^{31}-1)\bigr). \] The remainder maps the 64-bit integer into $\{0,1,\ldots,2^{31}-2\}$, and adding 1 shifts it into $\{1,2,\ldots,2^{31}-1\}$. Reducing a uniform 64-bit integer modulo $2^{31}-1$ leaves a deviation from uniformity below one part in $2^{33}$, negligible for any use made of the seed. A reader recomputes $s$ from the signed $\nu$ and requires it to equal the seed recorded in $\rho$; a seed that is merely reported is not accepted. Derived seeds use the same reduction. A step's seed $s_i$ (\Cref{sm:run}) and an author's re-mapped seed (\Cref{sm:audit}) are \[ 1+\bigl(\mathrm{BE}_{64}\bigl(H(\Lambda' \,\|\, \mathrm{BE}_{64}(s) \,\|\, \mathrm{BE}_{64}(i))[0..7]\bigr) \bmod (2^{31}-1)\bigr), \] where $\Lambda'$ is the schema's per-step or per-reseed domain string, $\mathrm{BE}_{64}(\cdot)$ applied to an integer writes it as eight big-endian bytes, and $i$ is the step's index or the seed the author requested. Under the per-step string this is the step seed $s_i=\mathrm{derive}(s,i)$ of \Cref{sm:run}; under the per-reseed string, with $i$ the requested seed $j$ (\Cref{sm:audit}), it is the re-mapped seed.

\paragraph{Timestamp proof $\pi$.}
The timestamp proof stops one particular fraud: manufacturing a certificate after the fact and passing it off as contemporaneous with the research. Why stopping that fraud takes a public record no party controls, and the two-edged window the proof closes around the certified run, are stated with the certificate (\Cref{sm:window}); this unit gives the mechanism. It has three parts: first, the proof and its three outcomes; second, the service that supplies it, and why no one need trust that service; and finally, how precise its times are.

First, the proof. A \emph{timestamp proof} is a public mark that these exact bytes already existed by a known moment; it says nothing about how much earlier they may have existed. The proof $\pi$ binds the exact bytes of $\mathrm{enc}(\rho)$ at issue. Checking $\pi$ returns one of three outcomes. $\mathsf{pending}$: the mark is recorded but not yet confirmed on the public record, which in version~1 is Bitcoin's blockchain. $\mathsf{confirmed}(t_{\pi})$: the mark is in a block whose time is $t_{\pi}$, so $\mathrm{enc}(\rho)$ existed no later than $t_{\pi}$; a proof is confirmed only once that block is six confirmations deep. Or $\mathsf{fail}$. The depth rule is this specification's, added on top of the timestamp service's own rules; it is the same rule as the public random value's (\Cref{sm:pulse}), so that a proof cannot be confirmed one hour and \emph{orphaned}, displaced from the chain by a reorganization, the next. Pending is not failure; a hash of $\rho$ that disagrees with $\pi$ is always failure. Confirmation must not require trusting the archive or the author.

Second, the service. The three outcomes are an interface, and version~1 fills it with one service: OpenTimestamps, which timestamps into Bitcoin \citep{opentimestamps}. Timestamping itself is much older \citep{haber1991timestamp}. The commercial arrangement is a \emph{timestamping authority}, a company that signs a statement saying it saw a document at a stated moment \citep{rfc3161}, which only moves the trust, since such a signature is worth exactly what the company is. Bitcoin replaces the company with a public record no one owns, but space in that record must be bought, and the network accepts only a few entries a second. OpenTimestamps closes that gap by batching strangers together. The archive sends $H(\mathrm{enc}(\rho))$ to a public \emph{calendar} server, which collects the hashes reaching it from many unrelated users over a short interval and folds them into one Merkle tree of its own (a list of hash operations rather than the RFC~6962 tree of \Cref{sm:merkle}). Periodically the calendar writes that tree's root into a Bitcoin transaction. The root then belongs to a block of which thousands of independent computers keep a copy. Altering that block would mean redoing the \emph{proof of work}, the deliberately expensive computation that had to be performed to publish a block, for that block and every later one, faster than the rest of the network combined. The proof $\pi$ is a small file holding the chain of hash operations that leads from $H(\mathrm{enc}(\rho))$ through the calendar's root and the transaction that carries it to the root of the block's own tree of transactions (Bitcoin's, not the Merkle root of \Cref{sm:merkle}), with the block's height. Checking it means recomputing that chain, reading the named block's header from the stored chain of headers (\Cref{sm:obsolete}) or from any Bitcoin node, confirming that the chain ends at the transaction root the header records, and reading the block's recorded time. No part of this check consults the archive or the author.

Sharing the tree is what makes the service affordable and harmless. One fee and one transaction per interval cover every document submitted in it, since the calendar writes one 32-byte root whatever the number of documents, so a document's share is a fraction of a cent. The public calendars charge nothing to submit, and an archive running a calendar of its own would pay once per interval rather than once per certificate. All that $\pi$ keeps of the other documents is a handful of hashes from the tree, which disclose nothing about them and demand nothing of whoever submitted them. The proof grows by about one hash each time the number of documents doubles, so a million documents in one interval add about twenty hashes. The calendar never becomes a party anyone must trust. It cannot backdate a document, since a chain of operations leading to a root already in Bitcoin cannot be constructed without breaking $H$. It cannot revoke one, since $\pi$ is self-contained and never consults the calendar again, so its proofs survive the calendar's shutdown. The worst it can do is decline to include a document, which is obvious at once. A calendar that shuts down before writing its root to Bitcoin would leave a proof permanently pending, or \emph{stranded}, so the archive submits every hash to several independent calendars, as the standard client does, and one calendar's shutdown strands nothing. When the archive records a receipt's hash in a transaction of its own at the attestation deadline (\Cref{sm:receipt}), the proof's chain ends in that transaction's block instead.

Finally, precision. Every timestamp proof the protocol uses---the receipt's, the certificate's $\pi$, and each head's---has the three outcomes and the depth rule above; this unit writes $\pi$ for the certificate's. A proof normally stays $\mathsf{pending}$ for a matter of hours. And Bitcoin block times are only approximate, because the network's rules on time are asymmetric. A node rejects a block stamped more than two hours ahead of its own clock, a rule of node policy, not of consensus. And a block's stamp need only exceed the median of the previous eleven blocks' stamps, an original consensus rule (the \emph{median time past}, the quantity \citealp{bip113} defines for the lock-times of transactions), which trails real time by roughly an hour. Neither rule ties a stamp to real time by itself; the median lags real time by more than two hours for about one block in fifty \citep{szalachowski2018towards}. So $t_{\pi}$, and every block time the protocol uses, is read as a moment good to an hour or two, not to the second. The one finding that sets a block time beside a clock reading, finding~9, allows two hours at each end for this imprecision (\Cref{sm:verify}). The imprecision does not matter for the use made of block times here, where the question is whether a certificate existed before some later event, not exactly when.

\subsection{Folding a list into one hash: the Merkle root}\label{sm:merkle}

\paragraph{Merkle root.}
A \emph{Merkle tree} \citep{merkle1980protocols} folds a long ordered list into one hash, so that changing any item, or the order of the items, changes it: pair the items, hash each pair, and repeat until one value, the top hash $t$, remains. The \emph{Merkle root}, defined at the end of this unit, wraps $t$ with the list's domain string and length. The protocol folds three kinds of list this way: the trace (root $R_T$, \Cref{sm:run}), each manifest (root $R_{M_i}$), and the ledger (\Cref{sm:ledger}). For an ordered list of records $r_0,\ldots,r_{n-1}$, form the \emph{leaves}, one hash per record, as $H(\texttt{0x00}\,\|\,\mathrm{enc}(r_i))$, where \texttt{0x00} is a single byte of value 0. Then, as long as more than one hash remains, pair them from left to right: each pair $(a,b)$ becomes an \emph{interior node} $H(\texttt{0x01}\,\|\,a\,\|\,b)$, where \texttt{0x01} is a single byte of value 1. A leftover hash with no partner is \emph{promoted}, kept as it is into the next round, never copied to fill a pair. The one-byte tags are \emph{domain-separation} prefixes taken from RFC~6962. They keep a leaf from being mistaken for an interior node: a match across the two prefixes would require two different inputs with one hash, which no one can usefully find (\Cref{sm:naming}).

The tree is exactly RFC~6962's for every non-empty list. Pairing from left to right and promoting the leftover builds, for every length of list, the same tree the standard defines by splitting a list at the largest power of two below its length. So a Certificate Transparency library (\Cref{sm:related}) computes the same top hash, and the standard's inclusion and consistency proofs apply unchanged to every tree in this protocol. What this specification adds is a final wrap. Write $t$ for the tree's top hash after pairing is finished (a bare $t$ always means this top hash; a $t$ with a subscript is a time), $n$ for the number of leaves, written in decimal digits as the schema fixes (\Cref{sm:schema}), and $\Sigma$ for the fixed ASCII domain string of the list being folded: \texttt{certified-merkle-v1|} for the trace, and the manifest's and the ledger's own strings for those trees (\Cref{sm:run,sm:ledger}). Like $\Lambda$, every such $\Sigma$ ends in a vertical bar that appears nowhere else in any domain string, so that no domain string can be the first bytes of another; it does for whole constructions what the one-byte tags do for the levels of a tree. The Merkle root is $H$ of three pieces concatenated, the bytes of $\Sigma$, then those digits, then the 32 bytes of $t$: \[ H(\Sigma \,\|\, n \,\|\, t). \] An empty list has no $t$, and its root is $H(\Sigma \,\|\, \texttt{0})$, defined for completeness only, since preflight's requirement of a code file (\Cref{sm:preflight}) keeps a certified run's list of steps (\Cref{sm:run}) from ever being empty. Promotion and the one-byte tags together prevent a list and a padded twin (the same list with a dummy extra item made by copying the last leaf) from sharing a root. That is not a hypothetical hazard: Bitcoin's rule, still in force, of duplicating the last leaf let distinct lists of transactions share a root, a disclosed flaw its implementation still guards against \citep{cve2012merkle}.

\section{Requirements of the archive}\label{sm:dataverse}

This section collects every duty of the archive role, so that an archive can be built from it; the main paper describes the archive's part (\Cref{sm:certify}) and its cost (\Cref{sm:parties}). It is a checklist: it names objects rather than redefining them, and each object is defined where it was introduced (\Cref{sm:tools,sm:close,sm:deposit,sm:seal}). Dataverse already stores replication datasets, issues a DOI for each dataset with a citation for each version, and freezes a version when it is published; certification uses those functions, and everything else on the list is what Dataverse, or any archive taking the role, adds. The duties come in the order they are met. First come the keys, once, at installation. Then, for each deposit, come the draft, the duties after the certified run, storage, the certification metadata field, the ledger with its proofs and snapshots, and the counters on the landing page. Then come the standing duties: renewal, recovery from a crash, the interface, and change of operator with the key log. Last come restricted files and revisions.

\paragraph{Keys and the trust root.}
Generate one Ed25519 signing key pair and one seed key pair for the verifiable random function (\Cref{sm:binding}) the first time the signing service starts, and again at each rotation the key log records. Generate the archive's root key once, offline, for signing key-log entries and nothing else. Never copy a private half of a key off the device that holds it. Keep both private halves in the most hardened storage the deployment has (a hardware security module where one exists), since one compromise of the signing service takes both at once; \Cref{sm:limits} states how each failure shows. Publish the root through the recognized certificate software, and deposit the root's public key, the currently pinned block hash, and the key log under DOIs and versions (\emph{Operator change and the key log}, below); readers pin the root, never a key that arrives with files.

\paragraph{Draft until the certificate is linked.}
Accept a closed copy $P^\ast$ into an unpublished draft, and refuse an unrecognized or lapsed build of the certificate software (\Cref{sm:deposit}). Hash the bytes received, $P^\ast$ and the closure report, never accepting a hash the author computed except the hiding commitment of a held file (\Cref{sm:confidential}), and run the dependency audit on $P^\ast$, since the group hashes require the audit's division of the files into groups (\Cref{sm:dataset}). Refuse a $P^\ast$ larger than $C_{\mathrm{size}}$ or with more than $C_{\mathrm{files}}$ entries, or one whose packed-folder list breaks the rules of \Cref{sm:unit}. Read the public random value $\beta$ (\Cref{sm:pulse}), and record it in $\gamma$; refuse the deposit if no Bitcoin node can be reached. If the declared predecessor is in this archive, refuse the deposit unless the predecessor belongs to the uploader; if it is not in this archive, sign it as an unresolved link (\Cref{sm:rev}); refuse a deposit without its closure report (\Cref{sm:evidence}). Only then sign the receipt $D$, including $\nu$, $s$, $\mathit{identity\_assurance}$ from the uploader's account, and the ledger position its entry takes. Append that entry under the ledger's lock (\Cref{sm:ledger}), submit $H(\mathrm{enc}(D))$ to the timestamp service, store the returned proof beside $D$, and serve the two together. At the attestation deadline, if no Bitcoin block yet records the receipt's hash, broadcast a transaction of the archive's own, or a purchased one (\Cref{sm:receipt}). Watch the proof until it confirms, raising the fee if the transaction lags. Reissue for cause any receipt whose proof confirms outside the coherence window, signing the replacement's supersedes field and storing the superseded receipt and its proof beside it (\Cref{sm:receipt}). Keep the draft unpublished.

\paragraph{After the certified run.}
Accept one $\rho$ per receipt, and refuse a second $\rho$ citing a $D$ already witnessed (\Cref{sm:run}). Accept a $\rho$ only from the account that deposited its $D$, or from one that account has authorized for this deposit, such as a coauthor's, or that of a reader who certifies public bytes; $W$ records the submitter. Before signing, run every finding that can be evaluated at that moment, and refuse on any failure. Those are findings~1 through~6, over the deposited files and the submitted $\rho$, and finding~9 whole, which means waiting for $D$'s proof to confirm inside the coherence window. The wait is the calendar's time to confirmation, ordinarily some hours; a run of that length or longer absorbs it, a shorter one does not, and no deadline runs against the author meanwhile. For held files a secure facility keeps, also require its custodian statement, verify it under the key the key log lists for that facility, and refuse if it reports a failed finding or names another $H(\mathrm{enc}(\rho))$; for held files the author keeps, the findings over them stay unevaluated (\Cref{sm:confidential}). Recompute the exception mark as finding~5 does, and write that value, never the one the certificate software recorded, onto the metadata field. Findings~7, 8, and~10 concern the witness, the timestamp proof, and the ledger proofs, which do not yet exist; the later pass that gathers $\pi$'s confirmation evaluates them and records the result beside $\rho$. Sign $W$, and place the timestamp proof $\pi$ on $\mathrm{enc}(\rho)$. Then store, link, and publish as \emph{Storage}, next, specifies; the linked $\rho$ must be the one whose hash $W$ names.

A submission is therefore in one of five states, which the archive shows on the metadata field: \emph{submitted}; \emph{refused}, with the failing finding; \emph{sealed}, $W$ signed and $\pi$ pending; \emph{confirmed}, which reports $\pi$'s own outcome of that name; and \emph{failed}, naming the finding, a state the later pass writes when a proof or a ledger proof fails. For an unchanged submission that is retried, the archive returns the stored original response and evidence for the same $W$; it does not issue a successor witness as a matter of routine. If the archive itself made the original evidence unusable, the separate mechanism of reissue for cause may issue a distinct receipt or witness that names the failed archival object and keeps both objects in the bundle. That recovery is for failures the archive caused, not for an author's ordinary retransmission, and it is not a general promise that every interrupted case is reconstructed automatically. The witness $W$ is kept as durably as $D$ (\Cref{sm:receipt}; \emph{Crash recovery}, below).

\paragraph{Storage.}
The archive keeps the replication dataset under content addressing (\Cref{sm:pack}): the file's name on disk is its hash, so a later swap would carry the wrong name. Store $P^\ast$ under $H(P^\ast)$. Write to a temporary file, \emph{flush} it to durable storage (force the disk to finish writing), read it back and hash it again, and then move it into place, read-only. A mismatch is a failure of storage integrity and stops the deposit. Before serving a file, hash it again; a file whose bytes no longer match the name it is stored under is \emph{quarantined}, moved out of the live store and kept for inspection. Serve $P^\ast$ as the packed run of bytes itself, one opaque file exempt from ingest (\Cref{sm:pack}). Store a deposited container image beside $P^\ast$, addressed by its digest and outside $C_{\mathrm{size}}$, and serve it to readers who re-execute (\Cref{sm:container}).

After $W$ is signed, store $\rho$ in the archive's store of certificates, addressed by $H(\mathrm{enc}(\rho))$, and not as a file in the replication dataset. Store beside it, addressed the same way, the certificate bundle (\Cref{sm:seal}). The blobs are the archive's storage exposure: many steps, each writing files up to $C_{\mathrm{blob}}$, can together far exceed $C_{\mathrm{size}}$. An archive prices that exposure in its terms of deposit, which are outside the protocol.

\paragraph{The certification metadata field.}
The certification metadata field is written by the signing service and cannot be changed by depositors or curators. On the version of the replication dataset about to be published, it links to the stored $\rho$ and records $H(\mathrm{enc}(\rho))$ as 64 hexadecimal characters. It also records the tier; the exception mark; the seed-sensitivity summary, which gives the state, the pairing counts, and the largest movement between the two seeds in baseline standard errors where that scale is available (\Cref{sm:sensitivity}); the count of records in $Q$ (\Cref{sm:run}); the counts of records excluded from the summary as loops and of records the capture command emitted (\Cref{sm:bytes-sensitivity}); the count of compiled files that match no registry digest (\Cref{sm:audit}); the re-execution coverage $n_c/n$ and the count of oversize outputs (\Cref{sm:run}); the spot-check's sample sizes and counts (\Cref{sm:spot}); the submission's state; and, for held files, the custodian and the date its statement names (\Cref{sm:confidential}). The version is published only after the archive has checked that the linked $\rho$'s $D$ names this version's $H(P^\ast)$, a match the landing page states, so that a passing certificate cannot be pointed at a version whose files it did not cover. Publish only after $P^\ast$, the stored $\rho$, and the field are in place, so that the citation, a DOI with its version selector (\Cref{sm:article}), names a frozen set.

\paragraph{Ledger, proofs, and snapshots.}
Append $D$'s entry when the receipt is issued, $W$'s when the certificate is sealed, and the publication entry when the version is published, each a leaf of the tree of \Cref{sm:merkle}; the publication entry's document is the metadata field's contents at publication, excluding the state that changes later, under statement type \texttt{publication}. At each anchoring interval (\Cref{sm:anchor}), sign a head with the fields of \Cref{sm:anchor}, submit its hash to the timestamp service, and keep the proof. Serve signed heads with their proofs, an inclusion proof for any entry, and a consistency proof between any two heads, and attach each certificate's own proofs to its bundle once the covering head's proof has confirmed. At least daily, deposit a snapshot of the whole ledger, signed by the signing key with the protocol context, as an ordinary archival object under its own DOI and version. The snapshot holds every leaf with the document its hash names (a receipt, witness, key-log entry, or publication, each with its signature), every signed head with its proof, and the key log. Other archives keep copies of the snapshot, and a copy held elsewhere, not this archive's own, is the source this archive cannot later revise, so that the reference head of finding~10 survives this archive.

A repeat deposit of the same bytes under the same predecessor is served the stored $D$, or the replacement of a superseded one, and appends no second $D$ (\Cref{sm:deposit}). A certificate signed under a different $v$ is treated as absent, so the replication dataset is deposited again under the current $v$, naming the earlier deposit as its predecessor; the new certificate is then a dated link rather than a fresh draw with no trail. Do not rewrite a published version, and never reduce a withdrawn version that carries a certification field to a bare notice of withdrawal: its page keeps serving the certificate bundle and states the withdrawal, so that a doctored successor cannot hide the version it replaced.

\paragraph{Landing-page counters.}
A landing page should also show four counts from the ledger: the uploader's deposits that name the same predecessor; the uploader's receipts never followed by a witness; the receipts the archive superseded for cause; and the uploader's witnessed certificates never followed by a publication entry. The superseded receipts are shown, not excluded, so that an archive's re-draws are counted. Beside them, from the snapshot, the page shows the number of receipts, across all uploaders, that share this deposit's bytes or any of its group hashes, and the days between chained versions. The trail of seed shopping that \Cref{sm:limits} describes is then something to read off a page rather than to investigate. The counters attach to one account at one archive; only an identity assurance through ORCID ties accounts to a person. An honest run that failed under $s$ and was then fixed leaves the same abandoned receipt a shopper does, so a count is read against an author's total, never as a verdict.

\paragraph{Renewal.}
Two duties, argued in \Cref{sm:obsolete}, keep old certificates checkable as the cryptography ages. Periodically timestamp the stored evidence again under the successor hash while SHA-256 still stands, and into a successor public record should Bitcoin's stop being maintained. And store the Bitcoin block headers, the whole chain of them, once, beside the proofs, so that the check runs with no network at all.

\paragraph{Crash recovery.}
If the archive's computer stops halfway through taking a deposit, the archive finishes what was completed and discards what was not. Every deposit writes a work-log entry, a note of work in progress, before making the record permanent, and writes the finished record so that it survives a crash. On startup, rebuild the lookup table in working memory from the finished records, then compare the unfinished notes with the ledger. A note whose deposit the ledger already has is left over and is cleared; a note with no ledger entry is unfinished, and its upload is discarded. Return $D$ to the author only after its ledger entry is durable, so that a receipt the author holds is never one a crash could lose; and refuse a $\rho$ citing a $D$ the ledger does not hold.

\paragraph{Interface.}
The interface is the small set of requests the archive answers. Two writes require a login: accept a deposit and return $D$; accept a finished $\rho$ and return $W$. The public reads are the issued $D$ with its timestamp proof; the certificate bundle, by the DOI and version; the signed heads with their proofs; an inclusion proof for any entry; a consistency proof between any two heads; the key log; and \emph{liveness}, whether the archive is responding. The certificate bundle (\Cref{sm:seal}) is returned with $\rho$ in one read, so that a reader never assembles it by hand. Uploads use a secret of at least sixteen characters that resolves to an uploader's account; what the archive knows about that account is written into $D$ as $\mathit{identity\_assurance}$. Because the reads need no login, the archive cannot use credentials to decide who may run the check on a published deposit; for restricted deposits, access control governs who holds the files (\Cref{sm:confidential}), never what the check computes from them.

\paragraph{Operator change and the key log.}
When a new archive takes over from an old one, nothing already signed is signed again. A certificate is judged under the key that signed it, and never under another key. What a reader pins across changes of key is the archive's trust root, under which the key log's signature is checked (\Cref{sm:binding}): the key-log entries on the ledger (\Cref{sm:ledger}). Each entry carries a public signing key with its key id, $H(k_{\mathrm{pub}})$ (the hash alone would not reconstruct the key an offline reader needs), and the public seed key $k_{\mathrm{seed,pub}}$ in force with it; its interval of validity; its successor; the recognized builds of the certificate software, with their intervals of recognition (\Cref{sm:deposit}); and the recognized custodians, each a secure facility with its public key and interval of recognition (\Cref{sm:confidential}). The key id stamped into every $D$ and $W$ names which entry applies. Validity is judged at the public dates the check reads, never at the archive's own asserted times: for $D$, the time of the block carrying the receipt's proof (\Cref{sm:receipt}); for a witness, the confirmed time of the head covering $W$'s entry; and for a head, its own confirmed time. A certificate whose dates fall outside its key's interval is judged by the interval, not by the signature alone. A rotation is itself a ledger entry, anchored and timestamped like every other, so a thief cannot backdate one. The seed key rotates with the signing key on the same terms rather than living forever, and each receipt's seed proof is checked under the public seed key the log lists for that receipt's key id. A retiring key's interval closes no earlier than the latest confirmed public date of any statement signed under it (a witness's covering head, a receipt's proof, a head's own proof), so that a calendar's delay never fails an honest certificate. A snapshot's weight comes from its deposit under a DOI and the anchored heads inside it, not from the date of its signature. A rotation cross-signed by the old key is one a thief holding that key could sign too, so the root, not the cross-signature, is what a reader trusts. The root key's own succession is a new recognized release of the certificate software and a new deposit under a DOI. Every verdict is relative to the newest key log the reader holds: an interval closed after that log was taken cannot be seen offline, and the online check fetches the current log first. The ledger either continues under one key or begins a new tree whose first entry names the preceding head. A \emph{renderer}, a program that displays $\rho$, shows the archive's identity, which archive now holds the files, and $H(\mathrm{enc}(\rho))$; it must produce the same page from the same fields, so that anyone can regenerate the page from the downloaded files.

\paragraph{Restricted files.}
Files the archive may hold but not publish form a deposit under restricted access (\Cref{sm:confidential}): store $P^\ast$ and the evidence set and serve them to approved readers only; publish $\rho$, $D$, $W$, $\pi$, and the ledger proofs as usual; and publish the standalone or redacted $Q$ beside $\rho$ only after the author's disclosure review. Files the archive may not hold at all are held files (\Cref{sm:confidential}): accept their declarations without the bytes; verify a facility's custodian statement before signing $W$; and store under restricted access the salt file of held files the author keeps, serving it to a reader who shows the same license.

\paragraph{Revisions.}
A later, different replication dataset is a new draft, with a new $D$, a new seed, a new $\rho$, a new metadata link, and a new published version. The predecessor link is the earlier version's $\mathit{id}$, signed into the new $D$. Files beside the packed $P^\ast$ in the same version other than the closure report and the evidence set, such as a README or the article, are outside the certificate: the renderer lists them as uncertified, and changing them is not a revision.

\section{Byte-level rules of closing and deposit}\label{sm:bytes}

This section collects the byte-level rules whose purpose \Cref{sm:close,sm:produce} state and whose spelling only an implementer needs. Each unit names the unit of those sections it serves, and the version-1 schema (\Cref{sm:schema}) holds the same spellings in machine-readable form. The units follow those sections' order: paths; the environment capsule's fields; the preflight list; the frozen settings by name; the mechanism of the capture code; the normalization list; and the mechanism of the dependency audit. Then come the receipt's fields; the spelling of numbers in the capture files; the grouping rules of the seed sensitivity; and finally, salting for restricted deposits.

\paragraph{Paths.}\phantomsection\label{sm:bytes-paths}
These are the byte-level rules for paths (\Cref{sm:paths}). Paths are written as UTF-8, the standard byte encoding of Unicode text. They are first put in Unicode normalization form~C, the standard single spelling of accented characters, so that the same name typed on macOS and on Linux gives the same bytes; otherwise an accented character can be stored either as one character or as a letter plus a combining mark. They are then compared as bytes, which is why letter case matters. No component may be \texttt{.}, \texttt{..}, or empty. No path may contain a control character, a byte below \texttt{0x20}, the range that includes tab and newline: the group hash (\Cref{sm:group}) writes one path per line, and a path holding a newline could make two different sets of files hash alike. Ordinary spaces are allowed, since every line ends in a hash of fixed width. Finally, a path names a regular file: symbolic links, names that point at another file like a shortcut, are neither followed nor listed. A packed folder's entry is its path, which obeys every rule here, followed by one \texttt{/} (\Cref{sm:unit}).

\paragraph{Environment capsule fields.}\phantomsection\label{sm:bytes-capsule}
The capsule $\kappa$ (\Cref{sm:capsule}) contains the operating system's name and version and the processor architecture; and, for every language the targets will run (\Cref{sm:entry}), the language's name and its exact version string. For every add-on library the dependency audit (\Cref{sm:audit}) finds the scripts loading, whether vendored inside $P^\ast$ or in the deposited image, it contains the library's name and exact version. It contains the numerical library that performs the linear algebra, that is, the implementation and version of the code that multiplies and inverts matrices. It contains the C mathematics library the runtime links. When the author declares at closing that the analysis uses a graphics processor, it contains that processor's model, its driver's version, and the version of its numerical software, such as NVIDIA's CUDA. And it contains the frozen settings actually in force (\Cref{sm:frozen}), the container image's content digest when one is deposited (\Cref{sm:container}), and the certificate software's own version and source hash, the pair $(v_{\mathrm{tool}},\,h_{\mathrm{tool}})$ of \Cref{sm:deposit}.

\paragraph{Preflight list.}\phantomsection\label{sm:bytes-preflight}
Before closing (\Cref{sm:preflight}), the software rejects the tree on any of the following grounds. The replication dataset is not a directory (a folder). There is no empty, writable place for the fresh temporary copy (\Cref{sm:run}). The code group has no file, so there is nothing to run. Two paths differ only in letter case, so the tree cannot be unpacked on the common filesystems that ignore case. A directory is empty: packing carries only files, so a run that writes into a pre-existing empty folder would fail on re-execution. A runtime required by the scripts present is missing (R is missing when the replication dataset has \texttt{.R} files, and likewise for Stata and Python). The tree is larger than the size ceiling $C_{\mathrm{size}}$, or has more than $C_{\mathrm{files}}$ entries, counting each packed folder as one, or its packed-folder list breaks the rules of \Cref{sm:unit}. The archive cannot be reached.

\paragraph{Frozen settings, by name.}\phantomsection\label{sm:bytes-frozen}
These are the settings of \Cref{sm:frozen}, as the schema names them. For clocks and formatting, the process time zone on $P^\ast$ is set to $\mathtt{UTC}$ (\Cref{sm:frozen}) and the process locale to $\mathtt{C}$. $\mathtt{SOURCE\_DATE\_EPOCH}$, the build-date setting, is set to the schema's fixed value, the Unix second for 2024-06-01 00:00:00 UTC. For cores, every setting on the schema's threading list is set to $1$, and, where a numerical library runs different code for different processors, its reproducibility switch (Intel's \texttt{MKL\_CBWR}) is set to its fixed code path. For table order, the per-language setting (\texttt{PYTHONHASHSEED} for Python) is on the schema's list of frozen settings and is frozen to the step's seed $s_i$ (\Cref{sm:run}). The list also disables Python's cache of compiled bytecode (\texttt{PYTHONDONTWRITEBYTECODE}) and R's saved workspace and history, whose files would otherwise enter $o$ carrying the time they were copied.

\paragraph{Capture code: mechanism.}\phantomsection\label{sm:bytes-capture}
This unit gives the mechanism behind \Cref{sm:instr}, in four parts: first, how the code arrives and what its loading line does; second, how the capture files are protected; third, what capture code may and may not do; and finally, the startup hook that extends all of this to processes a target launches.

First, how the code arrives. Capture code arrives as files, not as edits to the author's statements: it writes its own files into the code group and inserts a single loading line at the top of each target (\Cref{sm:entry}). The line goes at the first position the language permits, after any prefix the language requires to come first, as the schema's per-language rule fixes. Such a prefix is a \emph{shebang} (the \texttt{\#!} first line that names a script's interpreter), an encoding declaration, or Python's \texttt{\_\_future\_\_} imports. The loading line performs the four duties of \Cref{sm:instr} before any of the author's code executes. The settings it fixes that no process setting reaches are the schema's per-language list, and only Stata needs any: \texttt{set processors 1}, since Stata/MP, the multiprocessor edition, ignores the core-count settings and the command exists only there; and \texttt{set sortseed} to $s_i$, since Stata's \texttt{sort} otherwise breaks ties at random. No statement of the author's is changed, which keeps the difference between the author's code and the deposited code small enough to audit. The text of the loading line and the contents of the added files are fixed in the version-1 schema (\Cref{sm:schema}), not composed afresh for each deposit. They have to be, since otherwise two implementations would build different closed copies from the same $P$ and compute different $h_c$ for it.

Second, how the capture files are protected. A capture file is written by code running in the author's process, so author code could write into it too, which is why the audit's capture-path family matches its path (\Cref{sm:audit}); closing stops before deposit, while the author can still rename the output, rather than at a reader's check years later. Only the seed banner and the lines recording re-mapped seeds, one framed line per re-mapped reset, naming the requested seed and the seed seated, travel in the transcript, each framed by a schema-fixed delimiter, so that the delivery of the seed stays checkable from the log (finding~2, \Cref{sm:verify}). The check also verifies that the deposited capture code is the schema's: finding~1 compares every added file with the schema's canonical file for its language, and checks that removing the loading line from each touched target yields the recorded before-hash (\Cref{sm:verify}), so a doctored hook cannot enter the record. The schema also fixes one capture command that the author may call, a registered call that emits a record to $Q$ and that a hand-written estimator may invoke; its records are labeled as emitted by the author in $Q$, so that an analysis whose estimator matches no listed command can still bind its estimates instead of leaving $Q$ thin (\Cref{sm:limits}).

Third, what capture code may and may not do. After the loading line's four duties, capture code may only read, print, and write its capture files. It must not change what the author's code computes, since a run of altered code would not be a run of the author's analysis, and the determinism tier would be measuring something else. This rule states the goal, not the mechanism, which differs by language and which the schema specifies for each. R and Python register hooks (a masked function binding in R; an import hook in Python), which alter interpreter state that the audit exempts by rule. Stata runs no hook after a command and resolves its built-in commands ahead of any user path, so there the capture code parses the log, a parse the closure runs exercise, with the limit and remedy \Cref{sm:run} states. Beyond finding~1's comparison of bytes, nothing in the check inspects what the added lines do. What enforces the constraint is the build of the certificate software: the archive signs only for a build it recognizes by $(v_{\mathrm{tool}},\,h_{\mathrm{tool}})$ (\Cref{sm:deposit}), so the code that does the inserting is published and held fixed rather than assembled for each deposit. That defense has the limit stated with the build check (\Cref{sm:deposit}).

Finally, the startup hook. The startup hook of \Cref{sm:run} repeats the loading line's duties for a process a target launches, through \texttt{system}, \texttt{shell}, \texttt{subprocess}, and kin. It is installed outside the tree, as R's profile, Python's site hook, or a file on Stata's system ado-path, and reached through the run's process settings. The hook acts only when the schema's process setting shows that a loading line has already run: in the target's own process it fires before the loading line and does nothing, and every process launched after the loading line inherits the setting and is seeded, masked, and captured. The hook's banner always carries the delimiters reserved for a launched process, and finding~2 counts only banners between the delimiters that the loading line alone prints, so the one banner it counts is the target's own. Every banner prints the seed value actually seated, not one computed separately. A launch that suppresses the hook or clears the process settings would leave the launched process unseeded and uncaptured, so closing names such a line and stops; the cases are R's \texttt{--vanilla} and \texttt{--no-init-file}, Python's \texttt{-I}, \texttt{-S}, and \texttt{-E}, and any launch with fresh process settings.

The hook does not take its seed from the process settings, which author code controls. Instead, the certificate software writes the hook file for each run with $s$, the fresh temporary copy's top folder, and the path of a record baked in, and the hook appends one record per process (the process id, the parent's process id, and the seed seated) to a schema-named file outside the tree, which is kept in the evidence set (\Cref{sm:evidence}). Every observed process is recorded in the step's process observations with its executable's path and hash. An executable is exempt from the mark in four cases: it is a \emph{hooked interpreter}, an interpreter on the schema's list of runtimes the hook serves, with a hook record; it is a text file the audit read; its hash matches an entry of the artifact manifest that has a registry digest; or it is a file under a runtime root recorded in $\kappa$. From every other observed executable, finding~5 derives the exception mark unhooked-process (\Cref{sm:exception}). $\rho$ lists each observed executable with its status in the manifest, so a utility the run called, such as a document converter or a compressor, is exempt by its registry digest, while an author's own binary, vendored or built into a custom layer of the image, is marked. Finding~5 also requires, for every observed process whose executable is a hooked interpreter, a hook record naming the enclosing step's $s_i$; an interpreter observed with no record, or with another seed, yields the unhooked-process mark. The read family also matches any reference in executed code to the two seed settings or to the variables that carry the hook, and a match marks the certificate rather than being disclosed. Every target, and, through the process settings that carry the hook, every launched process, is launched with the flags that skip the machine's user startup files: Python's \texttt{-s}; R's \texttt{--no-environ}, with \texttt{R\_ENVIRON\_USER} and \texttt{R\_LIBS\_USER} empty; and, for Stata, \texttt{S\_ADO} pointing its personal and plus directories at empty ones. Because the hook lives outside $P^\ast$, no finding compares it; the recognized build that writes it holds it fixed.

\paragraph{Normalization list.}\phantomsection\label{sm:bytes-norm}
Before comparison (\Cref{sm:norm}), the software strips the following, and nothing else. It strips the creation timestamps and producer strings of PDF files (the producer string names the program that wrote the file), the stamps in log headers, the date stamps in documents, and the timestamp in the header of every Stata \texttt{.dta} file. It strips the per-entry times inside office formats built on zip (\texttt{.xlsx}, \texttt{.docx}), and PDF dates wherever the writing program places them, compressed object streams and metadata packets included. And it strips lines in logs that report runtime metrics (elapsed time, memory) and the uninitialized padding in the metadata of binary data files, leftover bytes a program did not set. Whether a stripped region is removed or filled with zeros is fixed per category in the schema.

\paragraph{Dependency audit: mechanism.}\phantomsection\label{sm:bytes-audit}
This unit gives the mechanism behind \Cref{sm:audit}, in three parts: first, the fetch layer and the library root; second, the masks on set-seed calls; and finally, what is audited where.

First, the fetch layer and the library root. A reference is resolved from the fresh temporary copy's top folder, the working directory of every step (\Cref{sm:run}). The fetch layer (\Cref{sm:audit}) is the only channel through which a locator is resolved, and it does no resolution of versions of its own. It fetches bytes and requires them to match the declared hash. It places each file at the path the declaration names under the dependency root, the schema-named directory inside the fresh temporary copy that holds fetched files. And it writes into $\rho$ the record of what was fetched and its hash, so that the record the check reads (finding~5, \Cref{sm:verify}) has a defined producer. A declared file must not carry a code extension, no target may invoke a file under the dependency root, and $P^\ast$ holds nothing under that root. Which files are the runtime, exempt from the audit and recorded in $\kappa$, and which are add-ons is fixed per language by the schema: the runtime's own installation roots are the runtime, and everything under a user library path is an add-on. A vendored library is a project-local library under the schema-named library root, and a \emph{lockfile} (\texttt{renv}'s, or the one Python's hash-checking install uses) is how an author reproduces one exactly. It costs space against $C_{\mathrm{size}}$, a price stated rather than hidden. Its files are other inputs, never targets, and exempt from the read family (\Cref{sm:dataset,sm:entry}) but not from the reseed and capture-path families. A reset of the seed inside a vendored package is disclosed like one in a script. A vendored file whose bytes differ from the public registry's record for that package is named in $\rho$, and a vendored file that names a capture file's path stops certification. When a container image is deposited, its text files are searched by the reseed and capture-path families, as the tree's are (\Cref{sm:audit}), and the image is part of the evidence set (\Cref{sm:evidence}). The artifact manifest of \Cref{sm:audit} covers the compiled files, including byte-compiled ones, of the image and of the tree. The registry is the schema's per-language package index (CRAN, PyPI, the image registry). A match means that the file's hash equals the one that index publishes for the same package, version, and platform, and a file the index does not list by hash is recorded as matching none. Finding~5 recomputes the manifest from the image in hand, and reports the parts of that finding that read the image unevaluated when the image is not in hand.

Second, the masks. The mask on the set-seed call is installed where package code resolves the call---R's base namespace binding and Python's module attributes---not only in the script's own scope, so a reset inside a package is re-mapped too. The masks also cover the call that every seed passed as a keyword argument reaches (Python's \texttt{RandomState}, which scikit-learn's and SciPy's \texttt{random\_state} arguments construct) and R's direct assignment to the generator's state. Two requested seeds collide in the range of seeds with probability about $B^2/2^{32}$ for $B$ distinct requests, which is negligible below a few thousand requests; an argument that is not an integer, or is absent, is seated as the schema fixes and recorded as such. The capture code's files carry their own implementation of the hash and load nothing from the author's tree, so the arithmetic of re-mapping is the schema's, not the author's.

Finally, what is audited where. The read family runs over the scripts; compiled files are unaudited and listed in the artifact manifest. The launch of a separate process is an arrow of the invocation graph of \Cref{sm:entry}, so the launched file is reachable for the families and is not itself a target (\Cref{sm:bytes-capture}).

\paragraph{The receipt's fields.}\phantomsection\label{sm:bytes-receipt}
For the receipt $D$ (\Cref{sm:receipt}), the archive signs $\mathrm{enc}(D)$, where $D$ is an object with the following fields; the canonical encoding orders them by key name, and the order below is for reading only. First come the statement type \texttt{deposit}; the archive's name; $\mathit{id}$; $H(P^\ast)$; $h_c,h_d,h_u,h_e$; $t_{\mathrm{arr}}$; and the uploader's account identifier, a stable pseudonym the archive assigns. The pseudonym is resolved to a person only on the landing page of a published version, so the daily snapshot names no one who has not yet published. The pseudonym is stable because the landing page's counters need it, so once any version is published, every deposit under it is attributable. That a deposit exists, with its predecessor link, is public from the day of the receipt, whether or not the deposit is ever published. Next come the identity assurance, $\mathit{identity\_assurance}\in\{\mathsf{shared},\,\mathsf{account},\,\mathsf{orcid}\}$; the predecessor $\mathit{id}^-$ when the author declared one, and otherwise a recorded absence (\Cref{sm:rev}); and the protocol context, $v$, the schema hash $h_{\mathrm{schema}}$, and the archive key id (\Cref{sm:deposit}). Last come the recognized build of the certificate software, $(v_{\mathrm{tool}},h_{\mathrm{tool}})$; the ledger position the receipt's entry takes; the public random value $\beta$ (\Cref{sm:pulse}); $\nu$; the seed proof $\phi$; $s$; the closure hash (\Cref{sm:cruns}); and, for a receipt reissued for cause, the superseded receipt's ledger document hash and confirmed time, and otherwise a recorded absence (\Cref{sm:receipt}).

\paragraph{The capture file's number spelling.}\phantomsection\label{sm:bytes-q}
Each record of the result capture $Q$ (\Cref{sm:run}) is a single JSON object per estimation, and it carries no floating-point values. Each number is written in one canonical hexadecimal spelling of its floating-point value, or as a pair of integers if it is a ratio. The spelling is the form Python's \texttt{float.hex()} prints: a minus sign if the value is negative, \texttt{0x1.}, thirteen hexadecimal digits, \texttt{p}, and a signed decimal exponent, with the forms it prints for zero and for subnormal values. The capture code converts into that form what C's \texttt{\%a} format and Stata's \texttt{\%21x} format print, since the three formatters spell one value three ways. The form is an exact spelling, fixed to the byte, of the binary value, so no formatter's rounding enters. Non-finite values are spelled as the schema fixes. The per-estimate values and the summary's movement in $\rho$ are spelled the same way.

\paragraph{Seed sensitivity: grouping and pairing.}\phantomsection\label{sm:bytes-sensitivity}
These are the rules behind the sensitivity summary of \Cref{sm:sensitivity}. The first closure run is the baseline. The schema pairs only logged estimates whose command identities are unique, in the scope it initially supports. A candidate identity consists of the step, the script, the source line or recorded call, the command, the dependent variable, and the set of coefficient names. An identity must occur exactly once in the baseline and exactly once in the certified run, and it must contain enough information about meaning to distinguish the estimation. Repeated identities are ambiguous: they are not paired by position and not reported as a measured zero movement. Records present under one seed only are unpaired. Coefficients with invalid numbers, non-finite movement, missing comparison values, or no positive, finite baseline standard error are disclosed through the counts of invalid, paired, and standardized coefficients, rather than folded into a convention that records a flip or zero movement.

\paragraph{Hiding commitments for restricted deposits.}\phantomsection\label{sm:bytes-salt}
For a restricted deposit (\Cref{sm:confidential}), every per-file hash that a public document carries is a hiding commitment. Those hashes are the per-file lines over which the group hashes are computed (\Cref{sm:group}), a packed folder's line among them, whose hash is $H(\mathrm{salt}\,\|\,\mathrm{pack}(U))$ under the salt derived from its entry, the trailing \texttt{/} included; the per-file hashes of outputs; the hashes of written files, transcripts, and capture files in the step records of $\rho$; the hashes of declared dependencies; and each value of a redacted $Q$. A value of a redacted $Q$ is committed to under its path, position, and field, so that equal withheld values hash differently. A restricted deposit's group hashes, and the $D$ that carries them, are therefore computed over lines that carry hiding commitments. Each published hash of a file is $H(\mathrm{salt}\,\|\,\mathrm{bytes})$ under a random 32-byte salt for that file. The salt is HMAC-SHA-256 \citep{bellare1996keying} of the file's path, keyed by the deposit's random file $R$, described next. One secret, made where the hashing happens, thus yields an independent salt for each file and each certificate. The salts are stored with the restricted evidence set, which the findings that recompute hashes over the files consume for approved readers. The whole-object hashes cannot take a salt, because $H(P^\ast)$, the group hashes, and $h_x$ are the protocol's own equations. A restricted deposit therefore adds one schema-named file $R$, of 32 random bytes, to the other-inputs group $u$ before closing, so that every whole-object hash covers bytes no outsider can guess and so confirms no guess. (The letter is reused: $R_T$ and $R_{M_i}$ are Merkle roots.) Because $R$ appears in every manifest, it also keeps every manifest root from confirming a guess. Held files (\Cref{sm:confidential}) are committed to by the same rule under a second random file of the same form, kept with them rather than in $P^\ast$: by a secure facility, inside it, and for licensed files the author holds, by the archive under restricted access; $P^\ast$ declares that file's hash.

\end{document}